%% file: draft9.tex
\documentstyle[12pt]{article}
 \makeatletter
 \def\leqq{\mathrel{\mathpalette\gl@align<}}
 \def\geqq{\mathrel{\mathpalette\gl@align>}}
 \def\gl@align#1#2{\lower.6ex\vbox{\baselineskip\z@skip\lineskip\z@
     \ialign{$\m@th#1\hfil##\hfil$\crcr#2\crcr=\crcr}}}
 \makeatother

 \makeatletter
 \def\sileqq{\mathrel{\mathpalette\gs@align<}}
 \def\sigeqq{\mathrel{\mathpalette\gs@align>}}
 \def\gs@align#1#2{\lower.6ex\vbox{\baselineskip\z@skip\lineskip\z@
     \ialign{$\m@th#1\hfil##\hfil$\crcr#2\crcr\sim\crcr}}}
 \makeatother
\begin{document}
\hbadness=10000
\hbadness=10000
\begin{titlepage}
\nopagebreak
\begin{flushright}
{\normalsize
HIP-1999-52/TH\\
DPSU-99-6\\
August, 1999}\\
\end{flushright}
\vspace{0.5cm}
\begin{center}

{\large \bf  Generic Gravitational Corrections 
to Gauge Couplings in SUSY $SU(5)$ GUTs}

\vspace{0.8cm}
 
{ Katri Huitu$^{{a}}$, Yoshiharu Kawamura$^{{c}}$, 
Tatsuo Kobayashi$^{{a,b}}$,   
Kai~Puolam\"{a}ki$^{{a}}$
}

\vspace{0.5cm}
$^a$ Helsinki Institute of Physics, 
FIN-00014 University of Helsinki, Finland \\
$^b$ Department of Physics, 
FIN-00014 University of Helsinki, Finland \\
and\\
$^c$ Department of Physics, Shinshu University,
Matsumoto 390-0802, Japan\\

\end{center}
\vspace{0.5cm}

\nopagebreak

\begin{abstract}
We study non-universal corrections to the gauge couplings due to 
higher dimensional operators in supersymmetric $SU(5)$ grand 
unified theories.
The corrections are, in general, 
parametrized by three components originating from
{\bf 24}, {\bf 75} and {\bf 200} representations.
We consider the prediction of $\alpha_3(M_Z)$ 
along each {\bf 24}, {\bf 75} and {\bf 200} direction, 
and their linear combinations.
The magnitude of GUT scale and its effects on proton decay 
are discussed.
Non-SUSY case is also examined.

\end{abstract}
\vfill
\end{titlepage}
\pagestyle{plain}
\newpage
\def\thefootnote{\fnsymbol{footnote}}

1. The prediction of $\alpha_3(M_Z)$ from the 
precision data $\alpha$ and $\sin \theta_W$ 
is one of the strong motivations for supersymmetric 
grand unified theories (SUSY GUTs) \cite{SUSY-GUT}.
On the analysis of gauge coupling constants, several corrections
have been considered, e.g. 
threshold corrections \cite{guni,GUT-th}
due to superparticles at the weak scale 
and heavy particles around the GUT scale.
In addition, non-renormalizable interactions in gauge kinetic terms
can give corrections suppressed by the reduced Planck mass $M$
as an effect of quantum gravity \cite{g-nonuni} \footnote{Such 
corrections are important for the gauge coupling unification 
with extra dimensions, too \cite{HK}.}.
We call them gravitational corrections.
The gravitational corrections are not universal but 
proportional to group theoretical factors.
If the $F$-component of the Higgs field
has a non-vanishing vacuum expectation value (VEV), 
gauginos also receive a non-universal 
correction to their masses \cite{EKN}.

In the SUSY $SU(5)$ GUT, gravitational corrections are 
parametrized by three components originating from
{\bf 24}, {\bf 75} and {\bf 200} representations
because these (elementary and/or composite) fields can couple to 
the gauge multiplets in gauge kinetic terms.
Non-zero VEVs of $F$-component of these
fields lead to proper types of non-universal gaugino masses.
Recently, several phenomenological aspects of models with
such non-universal 
gaugino masses have been studied and interesting 
difference among models have been shown \cite{nonunigm} \footnote{
See also Ref.~\cite{GLMR}.}.
In our previous analysis, 
it is assumed that 
non-universal corrections to gauge couplings $\alpha_i$ $(i=1,2,3)$
are small enough for 
the $SU(5)$ breaking scale $M_U$ to be the ordinary unification scale 
$M_X = 2.0 \times 10^{16}$GeV.

In this paper, we study gravitational corrections to gauge couplings
based on the SUSY $SU(5)$ GUT.
We consider the prediction of $\alpha_3(M_Z)$ 
along each {\bf 24}, {\bf 75} and {\bf 200} direction, 
and their linear combination.
We discuss which value is allowed as $M_U$
in the presence of gravitational corrections 
and its phenomenological implication.
The non-SUSY case is also examined.

The gauge kinetic function is given by
\begin{eqnarray}
{\cal L}_{g.k.} &=& \sum_{\alpha, \beta} \int d^2\theta 
f_{\alpha\beta}(\Phi^I) W^\alpha W^\beta   +   H.c.
\nonumber \\
&=& -{1 \over 4} \sum_{\alpha, \beta}
Re f_{\alpha\beta}(\phi^I) F^\alpha_{\mu\nu} F^{\beta \mu\nu} 
\nonumber \\
&~& 
+ \sum_{\alpha, \beta, \alpha', \beta'} \sum_I F^I_{\alpha'\beta'} 
{\partial f_{\alpha\beta}(\phi^I) \over \partial \phi^I_{\alpha'\beta'}}
\lambda^\alpha \lambda^\beta   +   H.c. + \cdots
\label{gaugekinetic}
\end{eqnarray}
where $\alpha, \beta$ are indices related to gauge generators,
$\Phi^I$'s are chiral superfields and 
$\lambda^\alpha$'s are the $SU(5)$ gaugino fields.
The scalar and $F$-components of $\Phi^I$ are denoted by
$\phi^I$ and $F^I$, respectively.
The gauge multiplet is in the adjoint representation and 
the symmetric product of ${\bf 24} \times {\bf 24}$ is decomposed as 
\begin{equation}
( {\bf 24} \times {\bf 24} )_s = {\bf 1} + {\bf 24}+{\bf 75}+{\bf 200}.
\end{equation}
Hence the gauge kinetic function $f_{\alpha\beta}(\Phi^I)$ is
also decomposed as
\begin{eqnarray}
f_{\alpha\beta}(\Phi^I) &=& \sum_R f^{R}_{\alpha \beta}(\Phi^I) 
\label{f2}
\end{eqnarray}
where $f^{R}_{\alpha \beta}(\Phi^I)$ is a part of gauge kinetic functions 
which transforms as $R$-representation ($R={\bf 1},{\bf 24},{\bf 75},
{\bf 200}$).

After a breakdown of $SU(5)$ at $M_U$, a boundary condition (BC) 
of $\alpha_i$ is given by 
\begin{eqnarray}
\alpha_i^{-1}(M_U)=\alpha_U^{-1}(1+C_i)
\label{BC}
\end{eqnarray}
where $C_i$'s are non-universal factors which parametrize generic 
gravitational corrections such as
\footnote{The factors $1/2\sqrt{15}$, $1/6$ and $1/2\sqrt{21}$
come from that the normalization $Tr (T^a T^b) = \delta^{ab}/2$ of 
the $5 \times 5$, $10 \times 10$ and $15 \times 15$ matrices
representing ${\bf 24}$, ${\bf 75}$ and ${\bf 200}$.}
\begin{eqnarray}
(C_1,C_2,C_3)&=& {x_{\bf 24} \over 2\sqrt{15}}(-1,-3,2) 
+ {x_{\bf 75} \over 6}(-5,3,1) 
+ {x_{\bf 200} \over 2\sqrt{21}}(10,2,1) \nonumber \\
&=& x'_{\bf 24}(-1,-3,2) + x'_{\bf 75}(-5,3,1) 
+ x'_{\bf 200}(10,2,1) .
\label{Ci}
\end{eqnarray}
Here $x_R$'s are model-dependent quantities including 
the VEV of Higgs fields, and their order is 
supposed to be $O(M_U/M)$ or less.\\

2. Let us predict $\alpha_3(M_Z)$ using the experimental values 
$\alpha_1^{-1}(M_Z)=59.98$ and $\alpha_2^{-1}(M_Z)=29.57$ 
based on the assumption that 
the minimal supersymmetric standard model (MSSM) holds on
below $M_U$ and the SUSY $SU(5)$ GUT is realized above $M_U$.

In the case without gravitational corrections,
the following value is obtained
\begin{equation}
\alpha_3^{(0)}(M_Z) = 0.127
\label{alpha3-cal}
\end{equation}
based on solutions of one-loop renormalization group (RG) equations
with a common SUSY threshold $m_{SUSY} = M_Z$
\begin{eqnarray}
\alpha_i^{-1}(M_Z) = \alpha_U^{-1}  + 
{b_i \over 2\pi}\ln{M_U \over M_Z}
\label{RGE}
\end{eqnarray}
where $(b_1, b_2, b_3) = (33/5, 1, -3)$.
The experimental value of $\alpha_3(M_Z)$ is \cite{p-data}
\begin{equation}
\alpha_3(M_Z) = 0.119 \pm 0.002.
\label{alpha3-exp}
\end{equation}

There are several possibilities to explain the difference between values
(\ref{alpha3-cal}) and (\ref{alpha3-exp}), e.g.,
threshold corrections due to superparticles at the weak scale,
heavy particles around the GUT scale and
gravitational corrections.
Here we neglect non-universal threshold corrections
and pay attention to gravitational corrections
by the ${\bf 24}$, ${\bf 75}$ or
${\bf 200}$ Higgs field.
Before numerical calculations of two-loop RG equations, 
we estimate the magnitude of $x_R$ using analytical results
of one-loop RG equations with BC (\ref{BC}).
For small $x_R$, $\alpha_3(M_Z)$ receives the following 
correction,
\begin{equation}
\alpha_3(M_Z)= \alpha_3^{(0)}(M_Z)+\sum_i a_iC_i,
\label{1-loop1}
\end{equation}
where $a_1=-0.28$, $a_2=0.67$ and $a_3=-0.39$.
Using Eq. (\ref{Ci}), allowed regions for $x_R$ are estimated as
\begin{eqnarray}
&~& 0.019 \sileqq x_{\bf 24} \sileqq 0.031 , \quad 
-0.020 \sileqq x_{\bf 75} \sileqq -0.012 , \quad  
\nonumber \\
&~& 0.030 \sileqq x_{\bf 200} \sileqq 0.050
\label{xR}
\end{eqnarray}
for each contribution.

Now let us study corrections solving the two-loop RG equations 
numerically.
We take the top quark mass $m_t=175$ GeV and $\tan \beta =3$,
and assume the universal SUSY threshold $m_{SUSY}=1$ TeV.
The three curves in Fig. 1 correspond to predictions of $\alpha_3(M_Z)$
along the pure {\bf 24},  {\bf 75} and  {\bf 200} directions, respectively.
Here we use a notation $x$ instead of $x_R$.
Fig. 2 shows the magnitude of GUT scale for each case.
Thus we obtain a good agreement with the experiment in the region 
with $|x|\sileqq O(0.01)$ for the three pure directions.
This value is consistent with $x_R \sileqq O(M_U/M)$.
For small $x_R$, $a_i$'s in eq.(\ref{1-loop1}) are obtained 
as $a_1=-0.27$, $a_2=0.61$ and $a_3=-0.36$ at the two-loop level.
In addition, $\alpha_3(M_Z)$ receives the SUSY thereshold correction, 
$0.0039 \times (m_{SUSY}/1$ TeV).

\begin{center}
\input fig1.tex

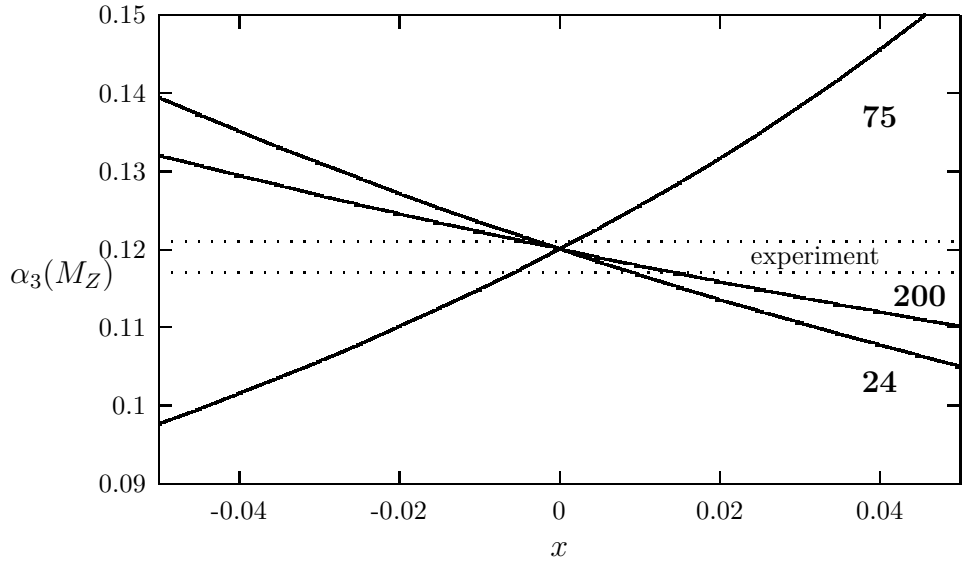
Fig.1: $\alpha_3(M_Z)$ along the {\bf 24}, {\bf 75} and  {\bf 200} 
directions. 
\end{center}

\newpage
\begin{center}
\input fig2.tex

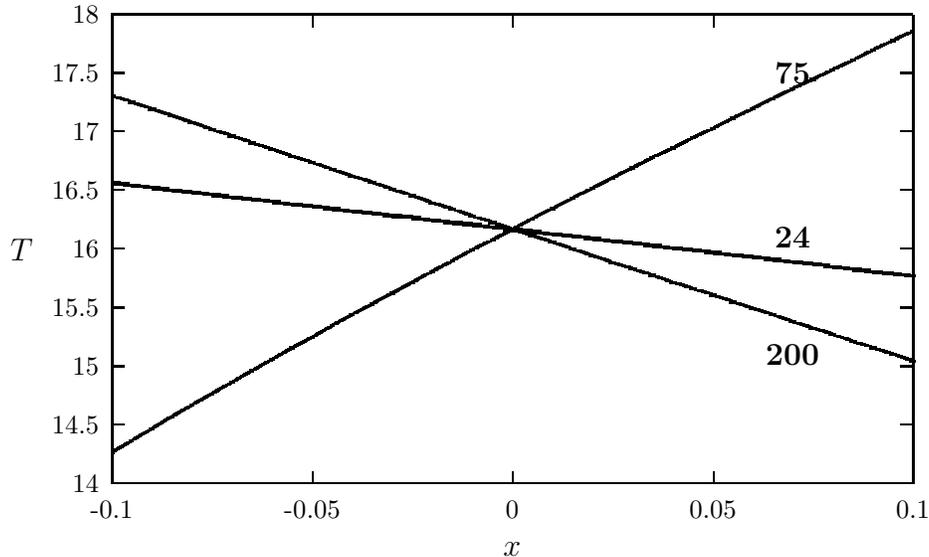
Fig.2: The GUT scale along the {\bf 24}, {\bf 75} and  {\bf 200} 
directions, where $T$ denotes $T=\log_{10}M_U$ [GeV]. 
\end{center}

We give comments on non-universal SUSY threshold corrections.
Under the assumption that the dominant contribution 
to gaugino masses comes from the VEV of
$F$-component of {\bf 24}, {\bf 75} or {\bf 200} Higgs field,
the ratio of gaugino mass magnitudes at the weak scale 
is given by \cite{EKN,nonunigm}
\begin{eqnarray}
M_1 : M_2 : M_3 &=& 0.4 : 0.8 : 2.9 ~~~~ (R={\bf 1}) , \nonumber \\
                &=& 0.2 : 1.2 : 2.9 ~~~~ (R={\bf 24}) , \nonumber \\
                &=& 2.1 : 2.5 : 2.9 ~~~~~~ (R={\bf 75}) , \nonumber \\
                &=& 4.1 : 1.6 : 2.9 ~~~~~~ (R={\bf 200}) . \nonumber 
\end{eqnarray}
This difference among gaugino masses leads to a small correction 
compared with the universal SUSY threshold.
For example,
$\alpha_3^{(0)}(M_Z)$ is raised by $0.001$ for $R={\bf 24}$.
Here we have assumed all soft scalar masses in the MSSM are equal to $M_3$.
Similarly, the case with ${\bf 75}$ Higgs condensation leads to a
tiny correction compared with the other two.

Detailed analysis shows that SUSY threshold corrections due to 
scalar masses could lead to sizable corrections \cite{guni}.
It is possible to deviate from $|x|\sileqq O(0.01)$ with a good agreement 
with the experimental value 
even in each pure direction case, although
the width of the good parameter region 
$\Delta x$ would be as narrow as those in Fig.1, 
i.e. $\Delta x = O(0.01)$.

The GUT scale threshold corrections due to heavy particles
are also important to the precise prediction of $\alpha_3(M_Z)$.
Since they depend on details of a GUT model, it would be difficult to
derive model-independent predictions.
We will discuss a model with {\bf 24} and {\bf 75} later.\\

3. We find that $|x|\sileqq O(0.01)$ and $M_U = 10^{16.1 \sim 16.2}$ GeV 
for the three pure directions 
without sizable non-universal threshold corrections.
Next let us explore a parameter region with a higher breaking scale.
It is expected that a higher GUT scale is realized
by some linear combination of the 
{\bf 24},  {\bf 75} and  {\bf 200} directions 
as we see from Figs.1 and 2.
Hereafter we consider only contributions from {\bf 24} and {\bf 75} Higgs
fields for simplicity.

First we estimate which value of $M_U$ is allowed
in the presence of gravitational corrections
based on one-loop RG analysis.
By using solutions of RG equations with the BC (\ref{BC}), 
the formula of $M_U$ is given by
\begin{eqnarray}
M_U = M_Z \cdot \exp \left( 2\pi 
((\alpha^{-1}_1 + \alpha^{-1}_2 + 2 \alpha^{-1}_3)(M_Z)
- 4 \alpha^{-1}_U) \over b_1 + b_2 + 2 b_3 \right)
\label{MU}
\end{eqnarray}
independent of $x'_{\bf 24}$ and $x'_{\bf 75}$.
The $x'_R$ $(R = {\bf 24}, {\bf 75})$ are written in terms of 
$\alpha_U$ and $M_U$ by
\begin{eqnarray}
x'_{R} = \alpha_U \sum_{i=1}^3 K_R^i (\alpha_i^{-1}(M_Z) - 
{b_i \over 2\pi}\ln{M_U \over M_Z})
\label{x'R}
\end{eqnarray}
where the matrix $K_R^i$ is given by
\begin{eqnarray}
K_R^i = \left(
\begin{array}{ccc}
-{1 / 18} & -{1 / 6} & {2 / 9}\\
-{5 / 36} & {1 / 12} & {1 / 18}
\end{array}
\right) .
\label{K}
\end{eqnarray}
Typical values of $\alpha_U^{-1}$, $x'_{\bf 24}$
and $x'_{\bf 75}$ are given in Table 1.
Here we use $\alpha_3^{-1}(M_Z) = 8.40$.
This result suggests that the GUT symmetry can be broken
down to the SM one $G_{SM}$ anywhere between $M_X$ and $M$.

\begin{table}

\caption{SUSY case}
\begin{center}
\begin{tabular}{|c|c|l|l|}
\hline
$M_U$ (GeV) &  $\alpha_U^{-1}$ & $x'_{\bf 24}$ 
& $x'_{\bf 75}$ \\
\hline\hline
$2 \times 10^{18}$ &  24.19 & 0.0322 & 0.0235 \\ \hline
$2 \times 10^{17}$ &  24.34 & 0.0139 & 0.0083 \\ \hline
$2 \times 10^{16}$ &  24.48 & $-0.0041$ & $-0.0067$ \\ \hline
\end{tabular}
\end{center}

\end{table}

Next we solve two-loop RG equations numerically with a common SUSY
threshold $m_{SUSY} = 1$TeV 
and calculate corrections for the linear combination of 
the {\bf 24} and {\bf 75} directions, i.e.,
$\langle {\bf 75} \rangle \cos \theta 
+ \langle {\bf 24} \rangle \sin \theta$.
Here GUT scale threshold corrections are not considered for simplicity.
Fig. 3 shows $\alpha_3(M_Z)$ against $\theta/\pi$ 
for $x=0.01, 0.05$ and 0.1 where we set $x=x_{\bf 24}=x_{\bf 75}$ and
Fig. 4 shows $M_U$ against $\theta/\pi$.
The correction to $\alpha_3(M_Z)$ is very small
at $\tan \theta \cong 1.4$ because of a cancellation between
contributions from ${\bf 24}$ and ${\bf 75}$.
The points with $\tan \theta = 1.4$ are denoted by 
the dotted vertical lines in Fig. 4. 
For $\sin \theta >0$, $M_U$ increases as $x$ does 
at $\tan \theta \cong 1.4$, although $\alpha_3(M_Z)$ does not change.
For example, $x=0.01, 0.05$ and $0.1$ correspond to $M_U=10^{16.2}$, 
$M_U=10^{16.5}$ and $M_U=10^{16.9}$ [GeV] at $\tan \theta =1.4$ 
and $\sin \theta >0$, while these values of $x$ 
correspond to $M_U=10^{16.1}$, $M_U=10^{15.9}$ and $M_U=10^{15.7}$ [GeV]
at $\tan \theta = 1.4$ and $\sin \theta <0$.
Such a relatively lower GUT scale is not realized naturally 
for $x > O(0.01)$
because the magnitude of $x$ is expected to be equal to or
smaller than $O(M_U/M)$.

\begin{center}
\input newfig3.tex

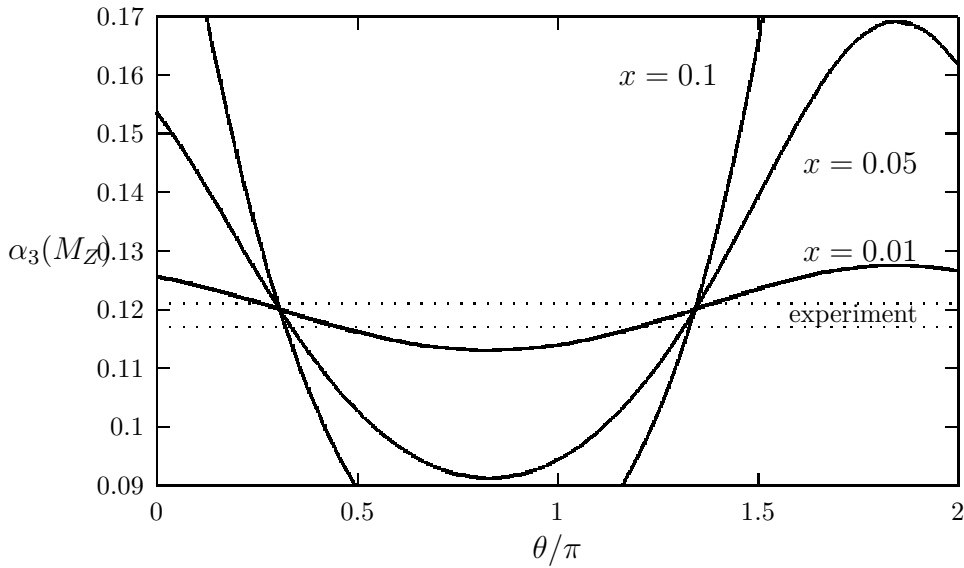
Fig.3: $\alpha_3(M_Z)$ along linear combinations of {\bf 24} and {\bf 75}
directions. 
\end{center}

\begin{center}
\input newfig4.tex

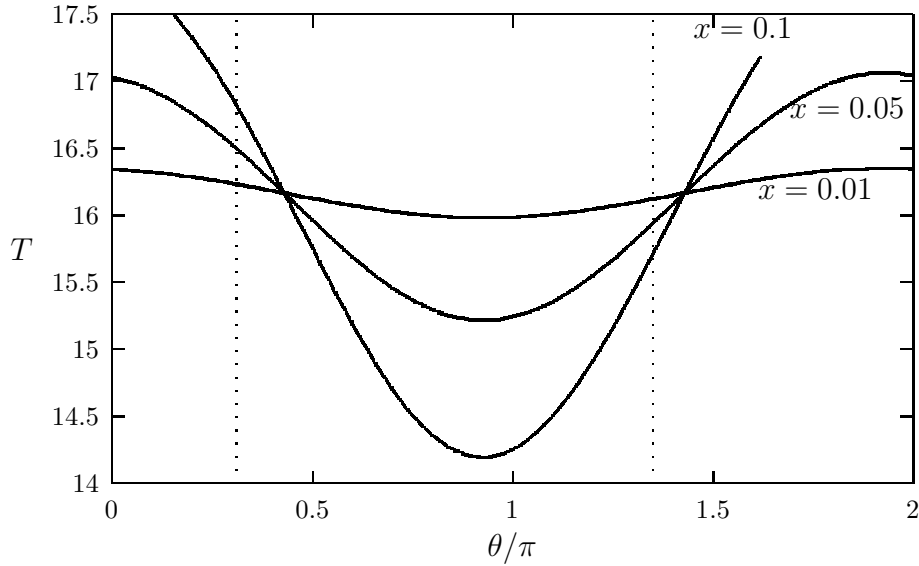
Fig.4: The GUT scale along linear combinations of {\bf 24} and {\bf 75}
directions, where $T$ denotes $T=\log_{10}M_X$ [GeV]. 
\end{center}

Similarly we can discuss the case with generic linear combination 
including a contribution from the ${\bf 200}$ Higgs field.
However, the model with the fundamental {\bf 200} Higgs particle has 
so large $\beta$-function coefficient that
the $SU(5)$ gauge coupling blows up near to 
the GUT scale.
For example, in a model with the fundamental {\bf 200} and {\bf 24}
Higgs particle and the minimal matter content,
the blowing-up energy scale $M_Y$ is obtained 
$M_Y/M_U=10^{0.74}=5.5$.
Thus, this model is not connected perturbatively with a theory 
at $M$.
There is a possibility that the {\bf 200} Higgs field is a composite one
made from fields with smaller representations.\\

4. Finally we discuss proton decay 
in the presence of gravitational corrections.
Here our purpose is to show qualitative features, how much 
gravitational corrections are important for discussions of proton decay.
Thus, we use only one-loop RG equations.
In order to carry out a precise analysis, 
it is necessary to consider two-loop effects, because 
two-loop effects of RG flows can be comparable to  
gravitational corrections.

For simplicity, we assume that particle contents of SUSY $SU(5)$ GUT
are $SU(5)$ gauge multiplet ${\bf 24}$, Higgs multiplets
${\bf 24}$ and ${\bf 75}$, and matter multiplets 
$N_g (\bar{\bf 5} + {\bf 10})$, ${\bf 5} + \bar{\bf 5}$
and extra matter multiplets.
Note that, in our usage, Higgs doublets in the MSSM belong to 
${\bf 5} + \bar{\bf 5}$
in matter multiplets and 
we assume that all extra matter multiplets acquire heavy masses
much bigger than $m_{SUSY}$.
The gauge symmetry $SU(5)$ is assumed to be broken down to $G_{SM}$ 
by a combination of 
VEVs of Higgs bosons ${\bf 24}$ and ${\bf 75}$.
In this case, would-be Nambu-Goldstone multiplets are a combination
of $({\bf 3}, {\bf 2})$ in ${\bf 24}$ and ${\bf 75}$
and a combination
of $(\bar{\bf 3}, {\bf 2})$ in ${\bf 24}$ and ${\bf 75}$.

Following the procedure in Ref.\cite{GUT-th}, we get the following relations
at one-loop level,
\begin{eqnarray}
&~& (3\alpha_2^{-1} - 2\alpha_3^{-1} - \alpha_1^{-1})(M_Z)
+ 12 \alpha_U^{-1}(x'_{\bf 24}-x'_{\bf 75}) \nonumber \\
&~& ~~~ = {1 \over 2\pi}\left({12 \over 5} \ln {\hat{M}_C \over M_Z}
 -2 \ln {m_{SUSY} \over M_Z} - {12 \over 5} \Delta_1\right) ,
\label{RGE-gr-th-1} \\
&~& (5\alpha_1^{-1} - 3\alpha_2^{-1} - 2\alpha_3^{-1})(M_Z)
+ 36 \alpha_U^{-1} x'_{\bf 75} \nonumber \\
&~& ~~~ = {1 \over 2\pi}\left(36 \ln {\hat{M}_U \over M_Z}
 +8 \ln {m_{SUSY} \over M_Z} + 36 \Delta_2 \right) 
\label{RGE-gr-th-2}
\end{eqnarray}
where we use solutions of RG equations of gauge couplings including 
a universal SUSY threshold and a GUT scale threshold correction.
Here $\hat{M}_C$ and $\hat{M}_U$ are an effective colored Higgs mass 
and GUT scale, respectively.
For example, in the minimal model $\hat{M}_C$ is 
the colored Higgs mass $M_{H_C}$ itself.
The effective GUT scale is given by
\begin{eqnarray}
\hat{M}_U = \left({M_V^2 M_{\bf 24} M_{\bf 75} \over M'} \right)^{1/3}
\label{GUT-scale}
\end{eqnarray}
where $M_V$ is $X$, $Y$ gauge boson mass, 
$M_{\bf 24}$ heavy ${\bf 24}$ Higgs mass, 
$M_{\bf 75}$ heavy ${\bf 75}$ Higgs mass
and $M'$ mass of orthogonal components to Nambu-Goldstone
multiplets.
The corrections $\Delta_{1,2}$ come from a mass splitting
among Higgs multiplets.
In a missing partner model, it is known that 
$\Delta_{1}$ is sizable,
e.g., $\Delta_{1} = \ln (1.7 \times 10^{4})$
and it can relax a constraint from proton decay\footnote{See the third
and fourth papers in Ref.\cite{GUT-th}.}.

Using relations (\ref{RGE-gr-th-1}) and (\ref{RGE-gr-th-2}),
we can estimate the magnitude of $\hat{M}_C$ and $\hat{M}_U$
as follows,
\begin{eqnarray}
\hat{M}_C &=& M_Z \cdot \left({m_{SUSY} \over M_Z}\right)^{5/6}
\nonumber \\
&~&
\times \exp \left({5\pi \over 6}((3\alpha_2^{-1} - 2\alpha_3^{-1} 
- \alpha_1^{-1})(M_Z) 
 + 12 \alpha_U^{-1}(x'_{\bf 24}-x'_{\bf 75})) 
+ \Delta_1 \right) \nonumber \\
&\sim& 3.2 \times 10^{15} \cdot \left({m_{SUSY} \over M_Z}\right)^{5/6}
\cdot \exp \left( 10 \pi \alpha_U^{-1}(x'_{\bf 24}-x'_{\bf 75}) 
+ \Delta_1 \right) 
\label{hatMC} \\
\hat{M}_U &=& M_Z \cdot \left({M_Z \over m_{SUSY}}\right)^{2/9}
\nonumber \\
&~& \times \exp \left({\pi \over 18}((5\alpha_1^{-1} - 3\alpha_2^{-1} 
- 2\alpha_3^{-1})(M_Z) + 36 \alpha_U^{-1} x'_{\bf 75}) 
- \Delta_2 \right) \nonumber \\
 &\sim& 4.8 \times 10^{16} \cdot \left({M_Z \over m_{SUSY}}\right)^{2/9}
\cdot \exp \left( 2\pi \alpha_U^{-1} x'_{\bf 75} - \Delta_2 \right)
\label{hatMU} 
\end{eqnarray}
where we use experimental data of $\alpha_i$.

SUSY-GUT models, in general, possess dangerous dimension-five and 
six operators to induce a rapid proton decay \cite{proton-decay}.
The dimension-five operators due to the exchange of colored Higgs boson
are suppressed only by a single power of
$\hat{M}_C$ in most cases.
The present lower bound of proton decay experiment suggests that
$\hat{M}_C$ is heavier than $O(10^{16})$GeV.
On the other hand, the dimension-six operators due to the exchange 
of $X$ and $Y$ gauge bosons
are suppressed by power of ${M}_V^2$.
Hence if $M_V$ is $O(10^{16})$GeV, nucleon lifetime can be longer than
$10^{34}$ years.
As we can see from Eqs. (\ref{hatMC}) and (\ref{hatMU}),
the magnitude of $\hat{M}_C$ and $M_V$ is sensitive to that of 
$\Delta_{1,2}$ and $\alpha_U^{-1} x'_R$.
It is important to get values or a relation between $\alpha_U^{-1} x'_R$
from other analysis.
Then we can obtain useful information on GUT scale mass spectrum
and scales such as $\hat{M}_C$ and $\hat{M}_U$
from precision measurement of sparticle masses
and RG analysis.

In the same way, we have analyzed a non-SUSY case with 
gravitational correction.
Under the assumption that particle contents of $SU(5)$ GUT
are $SU(5)$ gauge boson ${\bf 24}$, Higgs bosons
${\bf 24}$ and ${\bf 75}$, matter fermions 
$N_g (\bar{\bf 5} + {\bf 10})$, a fundamental representation Higgs 
${\bf 5}$ and extra matter fields, we get the following relations
at one-loop level,
\begin{eqnarray}
&~& (3\alpha_2^{-1} - 2\alpha_3^{-1} - \alpha_1^{-1})(M_Z)
+ 12 \alpha_U^{-1}(x'_{\bf 24}-x'_{\bf 75}) \nonumber \\
&~& ~~~~~~~~ 
= {1 \over 5\pi}\left( \ln {\hat{M}_C \over M_Z} - \Delta_1\right) ,
\label{RGE-gr-th-1-non} \\
&~& (5\alpha_1^{-1} - 3\alpha_2^{-1} - 2\alpha_3^{-1})(M_Z)
+ 36 \alpha_U^{-1} x'_{\bf 75} 
= {22 \over \pi}\left(\ln {\hat{M}_U \over M_Z}
+ \Delta_2 \right)  .
\label{RGE-gr-th-2-non}
\end{eqnarray}
Using relations (\ref{RGE-gr-th-1-non}) and (\ref{RGE-gr-th-2-non}),
we can estimate the magnitude of $\hat{M}_C$ and $\hat{M}_U$
as follows,
\begin{eqnarray}
\hat{M}_C &=& M_Z \cdot \exp \left(5\pi (3\alpha_2^{-1} - 2\alpha_3^{-1} 
- \alpha_1^{-1})(M_Z) 
 + 60 \pi \alpha_U^{-1}(x'_{\bf 24}-x'_{\bf 75}) 
+ \Delta_1 \right) \nonumber \\
&\sim& 10^{83} \cdot
\exp \left( 60 \pi \alpha_U^{-1}(x'_{\bf 24}-x'_{\bf 75}) 
+ \Delta_1 \right) 
\label{hatMC-non} \\
\hat{M}_U &=& M_Z \cdot 
 \exp \left({\pi \over 22}(5\alpha_1^{-1} - 3\alpha_2^{-1} 
- 2\alpha_3^{-1})(M_Z) + {18 \pi \over 11} \alpha_U^{-1} x'_{\bf 75} 
- \Delta_2 \right) \nonumber \\
 &\sim&  10^{14} \cdot 
 \exp \left( {18 \pi \over 11} \alpha_U^{-1} x'_{\bf 75} - \Delta_2 \right)
\label{hatMU-non} 
\end{eqnarray}
where we use experimental data of $\alpha_i$.

We require that the magnitude of $\hat{M}_U$ is bigger than $O(10^{16})$
to suppress a rapid proton decay by the exchange of $X$ and $Y$ 
gauge bosons.
Hence the magnitude of $x'_{\bf 75}$ is estimated as 
$x'_{\bf 75} = 0.024 \sim 0.044$ for $\hat{M}_U = 10^{16 \sim 18}$ GeV,
$\Delta_2 = 1$ and $\alpha_U = 1/45$.
Further we obtain a reasonable value as
$x'_{\bf 75} - x'_{\bf 24} \sim 0.019$
for $\hat{M}_C = 10^{16 \sim 18}$ GeV,
$\Delta_1 = 10$ and $\alpha_U = 1/45$.
Hence the non-SUSY $SU(5)$ GUT can revive in the presence of generic
gravitational corrections.\\

5. To summarize, we have studied non-universal corrections 
to gauge couplings due to higher dimensional operators 
along the three independent 
directions, {\bf 24}, {\bf 75} and {\bf 200} directions, and their 
linear combinations based on the SUSY $SU(5)$ GUT.
We have obtained a good agreement with the experimental values of 
$\alpha_i$ with $|x| \sileqq O(0.01)$ and $M_U = O(10^{16.1 \sim 16.2})$
GeV for the three pure directions.
A higher energy scale can be allowed as a breaking scale of $SU(5)$
in the presence of gravitational corrections by
a certain linear combination of contribution from 
{\bf 24} and {\bf 75} Higgs fields.
The constraints from the suppression of rapid proton decay
is sensitive to magnitude of $\Delta_{1,2}$ and $\alpha_U^{-1} x'_R$.
It is important to get values or a relation between $\alpha_U^{-1} x'_R$
from other analysis.
Then we can obtain useful information on GUT scale mass spectrum
and scales such as $\hat{M}_C$ and $\hat{M}_U$
from precision measurement of sparticle masses
and RG analysis.
The non-SUSY $SU(5)$ GUT can be revived in the presence of
gravitational corrections.

\section*{Acknowledgments}
This work was partially supported by the Academy of Finland under 
Project no. 163394. 
Y.K. acknowledges support by the Japanese Grant-in-Aid for Scientific 
Research ($\sharp$10740111) from the Ministry of
Education, Science and Culture.

\end{document}

%% file: fig1.tex
\setlength{\unitlength}{0.240900pt}
\ifx\plotpoint\undefined\newsavebox{\plotpoint}\fi
\sbox{\plotpoint}{\rule[-0.200pt]{0.400pt}{0.400pt}}%
\begin{picture}(1500,900)(0,0)
\font\gnuplot=cmr10 at 10pt
\gnuplot
\sbox{\plotpoint}{\rule[-0.200pt]{0.400pt}{0.400pt}}%
\put(181.0,123.0){\rule[-0.200pt]{4.818pt}{0.400pt}}
\put(161,123){\makebox(0,0)[r]{0.09}}
\put(1419.0,123.0){\rule[-0.200pt]{4.818pt}{0.400pt}}
\put(181.0,246.0){\rule[-0.200pt]{4.818pt}{0.400pt}}
\put(161,246){\makebox(0,0)[r]{0.1}}
\put(1419.0,246.0){\rule[-0.200pt]{4.818pt}{0.400pt}}
\put(181.0,369.0){\rule[-0.200pt]{4.818pt}{0.400pt}}
\put(161,369){\makebox(0,0)[r]{0.11}}
\put(1419.0,369.0){\rule[-0.200pt]{4.818pt}{0.400pt}}
\put(181.0,491.0){\rule[-0.200pt]{4.818pt}{0.400pt}}
\put(161,491){\makebox(0,0)[r]{0.12}}
\put(1419.0,491.0){\rule[-0.200pt]{4.818pt}{0.400pt}}
\put(181.0,614.0){\rule[-0.200pt]{4.818pt}{0.400pt}}
\put(161,614){\makebox(0,0)[r]{0.13}}
\put(1419.0,614.0){\rule[-0.200pt]{4.818pt}{0.400pt}}
\put(181.0,737.0){\rule[-0.200pt]{4.818pt}{0.400pt}}
\put(161,737){\makebox(0,0)[r]{0.14}}
\put(1419.0,737.0){\rule[-0.200pt]{4.818pt}{0.400pt}}
\put(181.0,860.0){\rule[-0.200pt]{4.818pt}{0.400pt}}
\put(161,860){\makebox(0,0)[r]{0.15}}
\put(1419.0,860.0){\rule[-0.200pt]{4.818pt}{0.400pt}}
\put(307.0,123.0){\rule[-0.200pt]{0.400pt}{4.818pt}}
\put(307,82){\makebox(0,0){-0.04}}
\put(307.0,840.0){\rule[-0.200pt]{0.400pt}{4.818pt}}
\put(558.0,123.0){\rule[-0.200pt]{0.400pt}{4.818pt}}
\put(558,82){\makebox(0,0){-0.02}}
\put(558.0,840.0){\rule[-0.200pt]{0.400pt}{4.818pt}}
\put(810.0,123.0){\rule[-0.200pt]{0.400pt}{4.818pt}}
\put(810,82){\makebox(0,0){0}}
\put(810.0,840.0){\rule[-0.200pt]{0.400pt}{4.818pt}}
\put(1062.0,123.0){\rule[-0.200pt]{0.400pt}{4.818pt}}
\put(1062,82){\makebox(0,0){0.02}}
\put(1062.0,840.0){\rule[-0.200pt]{0.400pt}{4.818pt}}
\put(1313.0,123.0){\rule[-0.200pt]{0.400pt}{4.818pt}}
\put(1313,82){\makebox(0,0){0.04}}
\put(1313.0,840.0){\rule[-0.200pt]{0.400pt}{4.818pt}}
\put(181.0,123.0){\rule[-0.200pt]{303.052pt}{0.400pt}}
\put(1439.0,123.0){\rule[-0.200pt]{0.400pt}{177.543pt}}
\put(181.0,860.0){\rule[-0.200pt]{303.052pt}{0.400pt}}
\put(30,451){\makebox(0,0){$\alpha_3(M_Z)$}}
\put(810,21){\makebox(0,0){$x$}}
\put(1313,283){\makebox(0,0){\bf 24}}
\put(1313,479){\makebox(0,0)[r]{experiment}}
\put(1376,418){\makebox(0,0){\bf 200}}
\put(1313,700){\makebox(0,0){\bf 75}}
\put(181.0,123.0){\rule[-0.200pt]{0.400pt}{177.543pt}}
\sbox{\plotpoint}{\rule[-0.400pt]{0.800pt}{0.800pt}}%
\put(181,730){\usebox{\plotpoint}}
\multiput(181.00,728.09)(1.183,-0.504){47}{\rule{2.067pt}{0.121pt}}
\multiput(181.00,728.34)(58.711,-27.000){2}{\rule{1.033pt}{0.800pt}}
\multiput(244.00,701.09)(1.230,-0.504){45}{\rule{2.138pt}{0.121pt}}
\multiput(244.00,701.34)(58.562,-26.000){2}{\rule{1.069pt}{0.800pt}}
\multiput(307.00,675.09)(1.230,-0.504){45}{\rule{2.138pt}{0.121pt}}
\multiput(307.00,675.34)(58.562,-26.000){2}{\rule{1.069pt}{0.800pt}}
\multiput(370.00,649.09)(1.336,-0.504){41}{\rule{2.300pt}{0.122pt}}
\multiput(370.00,649.34)(58.226,-24.000){2}{\rule{1.150pt}{0.800pt}}
\multiput(433.00,625.09)(1.336,-0.504){41}{\rule{2.300pt}{0.122pt}}
\multiput(433.00,625.34)(58.226,-24.000){2}{\rule{1.150pt}{0.800pt}}
\multiput(496.00,601.09)(1.315,-0.504){41}{\rule{2.267pt}{0.122pt}}
\multiput(496.00,601.34)(57.295,-24.000){2}{\rule{1.133pt}{0.800pt}}
\multiput(558.00,577.09)(1.397,-0.505){39}{\rule{2.391pt}{0.122pt}}
\multiput(558.00,577.34)(58.037,-23.000){2}{\rule{1.196pt}{0.800pt}}
\multiput(621.00,554.09)(1.463,-0.505){37}{\rule{2.491pt}{0.122pt}}
\multiput(621.00,554.34)(57.830,-22.000){2}{\rule{1.245pt}{0.800pt}}
\multiput(684.00,532.09)(1.536,-0.505){35}{\rule{2.600pt}{0.122pt}}
\multiput(684.00,532.34)(57.604,-21.000){2}{\rule{1.300pt}{0.800pt}}
\multiput(747.00,511.09)(1.536,-0.505){35}{\rule{2.600pt}{0.122pt}}
\multiput(747.00,511.34)(57.604,-21.000){2}{\rule{1.300pt}{0.800pt}}
\multiput(810.00,490.09)(1.536,-0.505){35}{\rule{2.600pt}{0.122pt}}
\multiput(810.00,490.34)(57.604,-21.000){2}{\rule{1.300pt}{0.800pt}}
\multiput(873.00,469.09)(1.616,-0.505){33}{\rule{2.720pt}{0.122pt}}
\multiput(873.00,469.34)(57.355,-20.000){2}{\rule{1.360pt}{0.800pt}}
\multiput(936.00,449.09)(1.616,-0.505){33}{\rule{2.720pt}{0.122pt}}
\multiput(936.00,449.34)(57.355,-20.000){2}{\rule{1.360pt}{0.800pt}}
\multiput(999.00,429.09)(1.705,-0.506){31}{\rule{2.853pt}{0.122pt}}
\multiput(999.00,429.34)(57.079,-19.000){2}{\rule{1.426pt}{0.800pt}}
\multiput(1062.00,410.09)(1.805,-0.506){29}{\rule{3.000pt}{0.122pt}}
\multiput(1062.00,410.34)(56.773,-18.000){2}{\rule{1.500pt}{0.800pt}}
\multiput(1125.00,392.09)(1.776,-0.506){29}{\rule{2.956pt}{0.122pt}}
\multiput(1125.00,392.34)(55.866,-18.000){2}{\rule{1.478pt}{0.800pt}}
\multiput(1187.00,374.09)(1.805,-0.506){29}{\rule{3.000pt}{0.122pt}}
\multiput(1187.00,374.34)(56.773,-18.000){2}{\rule{1.500pt}{0.800pt}}
\multiput(1250.00,356.09)(1.918,-0.507){27}{\rule{3.165pt}{0.122pt}}
\multiput(1250.00,356.34)(56.431,-17.000){2}{\rule{1.582pt}{0.800pt}}
\multiput(1313.00,339.09)(1.918,-0.507){27}{\rule{3.165pt}{0.122pt}}
\multiput(1313.00,339.34)(56.431,-17.000){2}{\rule{1.582pt}{0.800pt}}
\multiput(1376.00,322.09)(1.918,-0.507){27}{\rule{3.165pt}{0.122pt}}
\multiput(1376.00,322.34)(56.431,-17.000){2}{\rule{1.582pt}{0.800pt}}
\put(181,217){\usebox{\plotpoint}}
\multiput(181.00,218.41)(1.397,0.505){39}{\rule{2.391pt}{0.122pt}}
\multiput(181.00,215.34)(58.037,23.000){2}{\rule{1.196pt}{0.800pt}}
\multiput(244.00,241.41)(1.281,0.504){43}{\rule{2.216pt}{0.121pt}}
\multiput(244.00,238.34)(58.401,25.000){2}{\rule{1.108pt}{0.800pt}}
\multiput(307.00,266.41)(1.281,0.504){43}{\rule{2.216pt}{0.121pt}}
\multiput(307.00,263.34)(58.401,25.000){2}{\rule{1.108pt}{0.800pt}}
\multiput(370.00,291.41)(1.281,0.504){43}{\rule{2.216pt}{0.121pt}}
\multiput(370.00,288.34)(58.401,25.000){2}{\rule{1.108pt}{0.800pt}}
\multiput(433.00,316.41)(1.183,0.504){47}{\rule{2.067pt}{0.121pt}}
\multiput(433.00,313.34)(58.711,27.000){2}{\rule{1.033pt}{0.800pt}}
\multiput(496.00,343.41)(1.121,0.504){49}{\rule{1.971pt}{0.121pt}}
\multiput(496.00,340.34)(57.908,28.000){2}{\rule{0.986pt}{0.800pt}}
\multiput(558.00,371.41)(1.099,0.504){51}{\rule{1.938pt}{0.121pt}}
\multiput(558.00,368.34)(58.978,29.000){2}{\rule{0.969pt}{0.800pt}}
\multiput(621.00,400.41)(1.099,0.504){51}{\rule{1.938pt}{0.121pt}}
\multiput(621.00,397.34)(58.978,29.000){2}{\rule{0.969pt}{0.800pt}}
\multiput(684.00,429.41)(1.026,0.503){55}{\rule{1.826pt}{0.121pt}}
\multiput(684.00,426.34)(59.210,31.000){2}{\rule{0.913pt}{0.800pt}}
\multiput(747.00,460.41)(0.963,0.503){59}{\rule{1.727pt}{0.121pt}}
\multiput(747.00,457.34)(59.415,33.000){2}{\rule{0.864pt}{0.800pt}}
\multiput(810.00,493.41)(0.963,0.503){59}{\rule{1.727pt}{0.121pt}}
\multiput(810.00,490.34)(59.415,33.000){2}{\rule{0.864pt}{0.800pt}}
\multiput(873.00,526.41)(0.906,0.503){63}{\rule{1.640pt}{0.121pt}}
\multiput(873.00,523.34)(59.596,35.000){2}{\rule{0.820pt}{0.800pt}}
\multiput(936.00,561.41)(0.881,0.503){65}{\rule{1.600pt}{0.121pt}}
\multiput(936.00,558.34)(59.679,36.000){2}{\rule{0.800pt}{0.800pt}}
\multiput(999.00,597.41)(0.833,0.503){69}{\rule{1.526pt}{0.121pt}}
\multiput(999.00,594.34)(59.832,38.000){2}{\rule{0.763pt}{0.800pt}}
\multiput(1062.00,635.41)(0.791,0.502){73}{\rule{1.460pt}{0.121pt}}
\multiput(1062.00,632.34)(59.970,40.000){2}{\rule{0.730pt}{0.800pt}}
\multiput(1125.00,675.41)(0.740,0.502){77}{\rule{1.381pt}{0.121pt}}
\multiput(1125.00,672.34)(59.134,42.000){2}{\rule{0.690pt}{0.800pt}}
\multiput(1187.00,717.41)(0.735,0.502){79}{\rule{1.372pt}{0.121pt}}
\multiput(1187.00,714.34)(60.152,43.000){2}{\rule{0.686pt}{0.800pt}}
\multiput(1250.00,760.41)(0.686,0.502){85}{\rule{1.296pt}{0.121pt}}
\multiput(1250.00,757.34)(60.311,46.000){2}{\rule{0.648pt}{0.800pt}}
\multiput(1313.00,806.41)(0.657,0.502){89}{\rule{1.250pt}{0.121pt}}
\multiput(1313.00,803.34)(60.406,48.000){2}{\rule{0.625pt}{0.800pt}}
\multiput(1376.00,854.40)(0.650,0.526){7}{\rule{1.229pt}{0.127pt}}
\multiput(1376.00,851.34)(6.450,7.000){2}{\rule{0.614pt}{0.800pt}}
\put(181,639){\usebox{\plotpoint}}
\multiput(181.00,637.09)(2.046,-0.507){25}{\rule{3.350pt}{0.122pt}}
\multiput(181.00,637.34)(56.047,-16.000){2}{\rule{1.675pt}{0.800pt}}
\multiput(244.00,621.09)(2.046,-0.507){25}{\rule{3.350pt}{0.122pt}}
\multiput(244.00,621.34)(56.047,-16.000){2}{\rule{1.675pt}{0.800pt}}
\multiput(307.00,605.09)(2.192,-0.508){23}{\rule{3.560pt}{0.122pt}}
\multiput(307.00,605.34)(55.611,-15.000){2}{\rule{1.780pt}{0.800pt}}
\multiput(370.00,590.09)(2.046,-0.507){25}{\rule{3.350pt}{0.122pt}}
\multiput(370.00,590.34)(56.047,-16.000){2}{\rule{1.675pt}{0.800pt}}
\multiput(433.00,574.09)(2.192,-0.508){23}{\rule{3.560pt}{0.122pt}}
\multiput(433.00,574.34)(55.611,-15.000){2}{\rule{1.780pt}{0.800pt}}
\multiput(496.00,559.09)(2.323,-0.509){21}{\rule{3.743pt}{0.123pt}}
\multiput(496.00,559.34)(54.232,-14.000){2}{\rule{1.871pt}{0.800pt}}
\multiput(558.00,545.09)(2.361,-0.509){21}{\rule{3.800pt}{0.123pt}}
\multiput(558.00,545.34)(55.113,-14.000){2}{\rule{1.900pt}{0.800pt}}
\multiput(621.00,531.09)(2.361,-0.509){21}{\rule{3.800pt}{0.123pt}}
\multiput(621.00,531.34)(55.113,-14.000){2}{\rule{1.900pt}{0.800pt}}
\multiput(684.00,517.09)(2.361,-0.509){21}{\rule{3.800pt}{0.123pt}}
\multiput(684.00,517.34)(55.113,-14.000){2}{\rule{1.900pt}{0.800pt}}
\multiput(747.00,503.08)(2.560,-0.509){19}{\rule{4.077pt}{0.123pt}}
\multiput(747.00,503.34)(54.538,-13.000){2}{\rule{2.038pt}{0.800pt}}
\multiput(810.00,490.09)(2.361,-0.509){21}{\rule{3.800pt}{0.123pt}}
\multiput(810.00,490.34)(55.113,-14.000){2}{\rule{1.900pt}{0.800pt}}
\multiput(873.00,476.08)(2.560,-0.509){19}{\rule{4.077pt}{0.123pt}}
\multiput(873.00,476.34)(54.538,-13.000){2}{\rule{2.038pt}{0.800pt}}
\multiput(936.00,463.08)(2.796,-0.511){17}{\rule{4.400pt}{0.123pt}}
\multiput(936.00,463.34)(53.868,-12.000){2}{\rule{2.200pt}{0.800pt}}
\multiput(999.00,451.08)(2.560,-0.509){19}{\rule{4.077pt}{0.123pt}}
\multiput(999.00,451.34)(54.538,-13.000){2}{\rule{2.038pt}{0.800pt}}
\multiput(1062.00,438.08)(2.796,-0.511){17}{\rule{4.400pt}{0.123pt}}
\multiput(1062.00,438.34)(53.868,-12.000){2}{\rule{2.200pt}{0.800pt}}
\multiput(1125.00,426.08)(2.751,-0.511){17}{\rule{4.333pt}{0.123pt}}
\multiput(1125.00,426.34)(53.006,-12.000){2}{\rule{2.167pt}{0.800pt}}
\multiput(1187.00,414.08)(2.796,-0.511){17}{\rule{4.400pt}{0.123pt}}
\multiput(1187.00,414.34)(53.868,-12.000){2}{\rule{2.200pt}{0.800pt}}
\multiput(1250.00,402.08)(3.082,-0.512){15}{\rule{4.782pt}{0.123pt}}
\multiput(1250.00,402.34)(53.075,-11.000){2}{\rule{2.391pt}{0.800pt}}
\multiput(1313.00,391.08)(2.796,-0.511){17}{\rule{4.400pt}{0.123pt}}
\multiput(1313.00,391.34)(53.868,-12.000){2}{\rule{2.200pt}{0.800pt}}
\multiput(1376.00,379.08)(3.082,-0.512){15}{\rule{4.782pt}{0.123pt}}
\multiput(1376.00,379.34)(53.075,-11.000){2}{\rule{2.391pt}{0.800pt}}
\sbox{\plotpoint}{\rule[-0.200pt]{0.400pt}{0.400pt}}%
\put(181,455){\usebox{\plotpoint}}
\put(181.00,455.00){\usebox{\plotpoint}}
\put(201.76,455.00){\usebox{\plotpoint}}
\put(222.51,455.00){\usebox{\plotpoint}}
\put(243.27,455.00){\usebox{\plotpoint}}
\put(264.02,455.00){\usebox{\plotpoint}}
\put(284.78,455.00){\usebox{\plotpoint}}
\put(305.53,455.00){\usebox{\plotpoint}}
\put(326.29,455.00){\usebox{\plotpoint}}
\put(347.04,455.00){\usebox{\plotpoint}}
\put(367.80,455.00){\usebox{\plotpoint}}
\put(388.55,455.00){\usebox{\plotpoint}}
\put(409.31,455.00){\usebox{\plotpoint}}
\put(430.07,455.00){\usebox{\plotpoint}}
\put(450.82,455.00){\usebox{\plotpoint}}
\put(471.58,455.00){\usebox{\plotpoint}}
\put(492.33,455.00){\usebox{\plotpoint}}
\put(513.09,455.00){\usebox{\plotpoint}}
\put(533.84,455.00){\usebox{\plotpoint}}
\put(554.60,455.00){\usebox{\plotpoint}}
\put(575.35,455.00){\usebox{\plotpoint}}
\put(596.11,455.00){\usebox{\plotpoint}}
\put(616.87,455.00){\usebox{\plotpoint}}
\put(637.62,455.00){\usebox{\plotpoint}}
\put(658.38,455.00){\usebox{\plotpoint}}
\put(679.13,455.00){\usebox{\plotpoint}}
\put(699.89,455.00){\usebox{\plotpoint}}
\put(720.64,455.00){\usebox{\plotpoint}}
\put(741.40,455.00){\usebox{\plotpoint}}
\put(762.15,455.00){\usebox{\plotpoint}}
\put(782.91,455.00){\usebox{\plotpoint}}
\put(803.66,455.00){\usebox{\plotpoint}}
\put(824.42,455.00){\usebox{\plotpoint}}
\put(845.18,455.00){\usebox{\plotpoint}}
\put(865.93,455.00){\usebox{\plotpoint}}
\put(886.69,455.00){\usebox{\plotpoint}}
\put(907.44,455.00){\usebox{\plotpoint}}
\put(928.20,455.00){\usebox{\plotpoint}}
\put(948.95,455.00){\usebox{\plotpoint}}
\put(969.71,455.00){\usebox{\plotpoint}}
\put(990.46,455.00){\usebox{\plotpoint}}
\put(1011.22,455.00){\usebox{\plotpoint}}
\put(1031.98,455.00){\usebox{\plotpoint}}
\put(1052.73,455.00){\usebox{\plotpoint}}
\put(1073.49,455.00){\usebox{\plotpoint}}
\put(1094.24,455.00){\usebox{\plotpoint}}
\put(1115.00,455.00){\usebox{\plotpoint}}
\put(1135.75,455.00){\usebox{\plotpoint}}
\put(1156.51,455.00){\usebox{\plotpoint}}
\put(1177.26,455.00){\usebox{\plotpoint}}
\put(1198.02,455.00){\usebox{\plotpoint}}
\put(1218.77,455.00){\usebox{\plotpoint}}
\put(1239.53,455.00){\usebox{\plotpoint}}
\put(1260.29,455.00){\usebox{\plotpoint}}
\put(1281.04,455.00){\usebox{\plotpoint}}
\put(1301.80,455.00){\usebox{\plotpoint}}
\put(1322.55,455.00){\usebox{\plotpoint}}
\put(1343.31,455.00){\usebox{\plotpoint}}
\put(1364.06,455.00){\usebox{\plotpoint}}
\put(1384.82,455.00){\usebox{\plotpoint}}
\put(1405.57,455.00){\usebox{\plotpoint}}
\put(1426.33,455.00){\usebox{\plotpoint}}
\put(1439,455){\usebox{\plotpoint}}
\put(181,504){\usebox{\plotpoint}}
\put(181.00,504.00){\usebox{\plotpoint}}
\put(201.76,504.00){\usebox{\plotpoint}}
\put(222.51,504.00){\usebox{\plotpoint}}
\put(243.27,504.00){\usebox{\plotpoint}}
\put(264.02,504.00){\usebox{\plotpoint}}
\put(284.78,504.00){\usebox{\plotpoint}}
\put(305.53,504.00){\usebox{\plotpoint}}
\put(326.29,504.00){\usebox{\plotpoint}}
\put(347.04,504.00){\usebox{\plotpoint}}
\put(367.80,504.00){\usebox{\plotpoint}}
\put(388.55,504.00){\usebox{\plotpoint}}
\put(409.31,504.00){\usebox{\plotpoint}}
\put(430.07,504.00){\usebox{\plotpoint}}
\put(450.82,504.00){\usebox{\plotpoint}}
\put(471.58,504.00){\usebox{\plotpoint}}
\put(492.33,504.00){\usebox{\plotpoint}}
\put(513.09,504.00){\usebox{\plotpoint}}
\put(533.84,504.00){\usebox{\plotpoint}}
\put(554.60,504.00){\usebox{\plotpoint}}
\put(575.35,504.00){\usebox{\plotpoint}}
\put(596.11,504.00){\usebox{\plotpoint}}
\put(616.87,504.00){\usebox{\plotpoint}}
\put(637.62,504.00){\usebox{\plotpoint}}
\put(658.38,504.00){\usebox{\plotpoint}}
\put(679.13,504.00){\usebox{\plotpoint}}
\put(699.89,504.00){\usebox{\plotpoint}}
\put(720.64,504.00){\usebox{\plotpoint}}
\put(741.40,504.00){\usebox{\plotpoint}}
\put(762.15,504.00){\usebox{\plotpoint}}
\put(782.91,504.00){\usebox{\plotpoint}}
\put(803.66,504.00){\usebox{\plotpoint}}
\put(824.42,504.00){\usebox{\plotpoint}}
\put(845.18,504.00){\usebox{\plotpoint}}
\put(865.93,504.00){\usebox{\plotpoint}}
\put(886.69,504.00){\usebox{\plotpoint}}
\put(907.44,504.00){\usebox{\plotpoint}}
\put(928.20,504.00){\usebox{\plotpoint}}
\put(948.95,504.00){\usebox{\plotpoint}}
\put(969.71,504.00){\usebox{\plotpoint}}
\put(990.46,504.00){\usebox{\plotpoint}}
\put(1011.22,504.00){\usebox{\plotpoint}}
\put(1031.98,504.00){\usebox{\plotpoint}}
\put(1052.73,504.00){\usebox{\plotpoint}}
\put(1073.49,504.00){\usebox{\plotpoint}}
\put(1094.24,504.00){\usebox{\plotpoint}}
\put(1115.00,504.00){\usebox{\plotpoint}}
\put(1135.75,504.00){\usebox{\plotpoint}}
\put(1156.51,504.00){\usebox{\plotpoint}}
\put(1177.26,504.00){\usebox{\plotpoint}}
\put(1198.02,504.00){\usebox{\plotpoint}}
\put(1218.77,504.00){\usebox{\plotpoint}}
\put(1239.53,504.00){\usebox{\plotpoint}}
\put(1260.29,504.00){\usebox{\plotpoint}}
\put(1281.04,504.00){\usebox{\plotpoint}}
\put(1301.80,504.00){\usebox{\plotpoint}}
\put(1322.55,504.00){\usebox{\plotpoint}}
\put(1343.31,504.00){\usebox{\plotpoint}}
\put(1364.06,504.00){\usebox{\plotpoint}}
\put(1384.82,504.00){\usebox{\plotpoint}}
\put(1405.57,504.00){\usebox{\plotpoint}}
\put(1426.33,504.00){\usebox{\plotpoint}}
\put(1439,504){\usebox{\plotpoint}}
\end{picture}

%% file: fig2.tex
\setlength{\unitlength}{0.240900pt}
\ifx\plotpoint\undefined\newsavebox{\plotpoint}\fi
\sbox{\plotpoint}{\rule[-0.200pt]{0.400pt}{0.400pt}}%
\begin{picture}(1500,900)(0,0)
\font\gnuplot=cmr10 at 10pt
\gnuplot
\sbox{\plotpoint}{\rule[-0.200pt]{0.400pt}{0.400pt}}%
\put(181.0,123.0){\rule[-0.200pt]{4.818pt}{0.400pt}}
\put(161,123){\makebox(0,0)[r]{14}}
\put(1419.0,123.0){\rule[-0.200pt]{4.818pt}{0.400pt}}
\put(181.0,215.0){\rule[-0.200pt]{4.818pt}{0.400pt}}
\put(161,215){\makebox(0,0)[r]{14.5}}
\put(1419.0,215.0){\rule[-0.200pt]{4.818pt}{0.400pt}}
\put(181.0,307.0){\rule[-0.200pt]{4.818pt}{0.400pt}}
\put(161,307){\makebox(0,0)[r]{15}}
\put(1419.0,307.0){\rule[-0.200pt]{4.818pt}{0.400pt}}
\put(181.0,399.0){\rule[-0.200pt]{4.818pt}{0.400pt}}
\put(161,399){\makebox(0,0)[r]{15.5}}
\put(1419.0,399.0){\rule[-0.200pt]{4.818pt}{0.400pt}}
\put(181.0,492.0){\rule[-0.200pt]{4.818pt}{0.400pt}}
\put(161,492){\makebox(0,0)[r]{16}}
\put(1419.0,492.0){\rule[-0.200pt]{4.818pt}{0.400pt}}
\put(181.0,584.0){\rule[-0.200pt]{4.818pt}{0.400pt}}
\put(161,584){\makebox(0,0)[r]{16.5}}
\put(1419.0,584.0){\rule[-0.200pt]{4.818pt}{0.400pt}}
\put(181.0,676.0){\rule[-0.200pt]{4.818pt}{0.400pt}}
\put(161,676){\makebox(0,0)[r]{17}}
\put(1419.0,676.0){\rule[-0.200pt]{4.818pt}{0.400pt}}
\put(181.0,768.0){\rule[-0.200pt]{4.818pt}{0.400pt}}
\put(161,768){\makebox(0,0)[r]{17.5}}
\put(1419.0,768.0){\rule[-0.200pt]{4.818pt}{0.400pt}}
\put(181.0,860.0){\rule[-0.200pt]{4.818pt}{0.400pt}}
\put(161,860){\makebox(0,0)[r]{18}}
\put(1419.0,860.0){\rule[-0.200pt]{4.818pt}{0.400pt}}
\put(181.0,123.0){\rule[-0.200pt]{0.400pt}{4.818pt}}
\put(181,82){\makebox(0,0){-0.1}}
\put(181.0,840.0){\rule[-0.200pt]{0.400pt}{4.818pt}}
\put(496.0,123.0){\rule[-0.200pt]{0.400pt}{4.818pt}}
\put(496,82){\makebox(0,0){-0.05}}
\put(496.0,840.0){\rule[-0.200pt]{0.400pt}{4.818pt}}
\put(810.0,123.0){\rule[-0.200pt]{0.400pt}{4.818pt}}
\put(810,82){\makebox(0,0){0}}
\put(810.0,840.0){\rule[-0.200pt]{0.400pt}{4.818pt}}
\put(1125.0,123.0){\rule[-0.200pt]{0.400pt}{4.818pt}}
\put(1125,82){\makebox(0,0){0.05}}
\put(1125.0,840.0){\rule[-0.200pt]{0.400pt}{4.818pt}}
\put(1439.0,123.0){\rule[-0.200pt]{0.400pt}{4.818pt}}
\put(1439,82){\makebox(0,0){0.1}}
\put(1439.0,840.0){\rule[-0.200pt]{0.400pt}{4.818pt}}
\put(181.0,123.0){\rule[-0.200pt]{303.052pt}{0.400pt}}
\put(1439.0,123.0){\rule[-0.200pt]{0.400pt}{177.543pt}}
\put(181.0,860.0){\rule[-0.200pt]{303.052pt}{0.400pt}}
\put(40,491){\makebox(0,0){$T$}}
\put(810,21){\makebox(0,0){$x$}}
\put(1250,510){\makebox(0,0){\bf 24}}
\put(1250,326){\makebox(0,0){\bf 200}}
\put(1250,768){\makebox(0,0){\bf 75}}
\put(181.0,123.0){\rule[-0.200pt]{0.400pt}{177.543pt}}
\sbox{\plotpoint}{\rule[-0.400pt]{0.800pt}{0.800pt}}%
\put(181,596){\usebox{\plotpoint}}
\put(181,592.34){\rule{6.400pt}{0.800pt}}
\multiput(181.00,594.34)(17.716,-4.000){2}{\rule{3.200pt}{0.800pt}}
\put(212,588.34){\rule{6.600pt}{0.800pt}}
\multiput(212.00,590.34)(18.301,-4.000){2}{\rule{3.300pt}{0.800pt}}
\put(244,584.34){\rule{6.400pt}{0.800pt}}
\multiput(244.00,586.34)(17.716,-4.000){2}{\rule{3.200pt}{0.800pt}}
\put(275,580.84){\rule{7.709pt}{0.800pt}}
\multiput(275.00,582.34)(16.000,-3.000){2}{\rule{3.854pt}{0.800pt}}
\put(307,577.34){\rule{6.400pt}{0.800pt}}
\multiput(307.00,579.34)(17.716,-4.000){2}{\rule{3.200pt}{0.800pt}}
\put(338,573.34){\rule{6.600pt}{0.800pt}}
\multiput(338.00,575.34)(18.301,-4.000){2}{\rule{3.300pt}{0.800pt}}
\put(370,569.84){\rule{7.468pt}{0.800pt}}
\multiput(370.00,571.34)(15.500,-3.000){2}{\rule{3.734pt}{0.800pt}}
\put(401,566.34){\rule{6.600pt}{0.800pt}}
\multiput(401.00,568.34)(18.301,-4.000){2}{\rule{3.300pt}{0.800pt}}
\put(433,562.34){\rule{6.400pt}{0.800pt}}
\multiput(433.00,564.34)(17.716,-4.000){2}{\rule{3.200pt}{0.800pt}}
\put(464,558.84){\rule{7.709pt}{0.800pt}}
\multiput(464.00,560.34)(16.000,-3.000){2}{\rule{3.854pt}{0.800pt}}
\put(496,555.34){\rule{6.400pt}{0.800pt}}
\multiput(496.00,557.34)(17.716,-4.000){2}{\rule{3.200pt}{0.800pt}}
\put(527,551.34){\rule{6.400pt}{0.800pt}}
\multiput(527.00,553.34)(17.716,-4.000){2}{\rule{3.200pt}{0.800pt}}
\put(558,547.84){\rule{7.709pt}{0.800pt}}
\multiput(558.00,549.34)(16.000,-3.000){2}{\rule{3.854pt}{0.800pt}}
\put(590,544.34){\rule{6.400pt}{0.800pt}}
\multiput(590.00,546.34)(17.716,-4.000){2}{\rule{3.200pt}{0.800pt}}
\put(621,540.84){\rule{7.709pt}{0.800pt}}
\multiput(621.00,542.34)(16.000,-3.000){2}{\rule{3.854pt}{0.800pt}}
\put(653,537.34){\rule{6.400pt}{0.800pt}}
\multiput(653.00,539.34)(17.716,-4.000){2}{\rule{3.200pt}{0.800pt}}
\put(684,533.34){\rule{6.600pt}{0.800pt}}
\multiput(684.00,535.34)(18.301,-4.000){2}{\rule{3.300pt}{0.800pt}}
\put(716,529.84){\rule{7.468pt}{0.800pt}}
\multiput(716.00,531.34)(15.500,-3.000){2}{\rule{3.734pt}{0.800pt}}
\put(747,526.34){\rule{6.600pt}{0.800pt}}
\multiput(747.00,528.34)(18.301,-4.000){2}{\rule{3.300pt}{0.800pt}}
\put(779,522.34){\rule{6.400pt}{0.800pt}}
\multiput(779.00,524.34)(17.716,-4.000){2}{\rule{3.200pt}{0.800pt}}
\put(810,518.84){\rule{7.468pt}{0.800pt}}
\multiput(810.00,520.34)(15.500,-3.000){2}{\rule{3.734pt}{0.800pt}}
\put(841,515.34){\rule{6.600pt}{0.800pt}}
\multiput(841.00,517.34)(18.301,-4.000){2}{\rule{3.300pt}{0.800pt}}
\put(873,511.84){\rule{7.468pt}{0.800pt}}
\multiput(873.00,513.34)(15.500,-3.000){2}{\rule{3.734pt}{0.800pt}}
\put(904,508.34){\rule{6.600pt}{0.800pt}}
\multiput(904.00,510.34)(18.301,-4.000){2}{\rule{3.300pt}{0.800pt}}
\put(936,504.34){\rule{6.400pt}{0.800pt}}
\multiput(936.00,506.34)(17.716,-4.000){2}{\rule{3.200pt}{0.800pt}}
\put(967,500.84){\rule{7.709pt}{0.800pt}}
\multiput(967.00,502.34)(16.000,-3.000){2}{\rule{3.854pt}{0.800pt}}
\put(999,497.34){\rule{6.400pt}{0.800pt}}
\multiput(999.00,499.34)(17.716,-4.000){2}{\rule{3.200pt}{0.800pt}}
\put(1030,493.34){\rule{6.600pt}{0.800pt}}
\multiput(1030.00,495.34)(18.301,-4.000){2}{\rule{3.300pt}{0.800pt}}
\put(1062,489.84){\rule{7.468pt}{0.800pt}}
\multiput(1062.00,491.34)(15.500,-3.000){2}{\rule{3.734pt}{0.800pt}}
\put(1093,486.34){\rule{6.600pt}{0.800pt}}
\multiput(1093.00,488.34)(18.301,-4.000){2}{\rule{3.300pt}{0.800pt}}
\put(1125,482.34){\rule{6.400pt}{0.800pt}}
\multiput(1125.00,484.34)(17.716,-4.000){2}{\rule{3.200pt}{0.800pt}}
\put(1156,478.84){\rule{7.468pt}{0.800pt}}
\multiput(1156.00,480.34)(15.500,-3.000){2}{\rule{3.734pt}{0.800pt}}
\put(1187,475.34){\rule{6.600pt}{0.800pt}}
\multiput(1187.00,477.34)(18.301,-4.000){2}{\rule{3.300pt}{0.800pt}}
\put(1219,471.34){\rule{6.400pt}{0.800pt}}
\multiput(1219.00,473.34)(17.716,-4.000){2}{\rule{3.200pt}{0.800pt}}
\put(1250,467.84){\rule{7.709pt}{0.800pt}}
\multiput(1250.00,469.34)(16.000,-3.000){2}{\rule{3.854pt}{0.800pt}}
\put(1282,464.34){\rule{6.400pt}{0.800pt}}
\multiput(1282.00,466.34)(17.716,-4.000){2}{\rule{3.200pt}{0.800pt}}
\put(1313,460.34){\rule{6.600pt}{0.800pt}}
\multiput(1313.00,462.34)(18.301,-4.000){2}{\rule{3.300pt}{0.800pt}}
\put(1345,456.34){\rule{6.400pt}{0.800pt}}
\multiput(1345.00,458.34)(17.716,-4.000){2}{\rule{3.200pt}{0.800pt}}
\put(1376,452.84){\rule{7.709pt}{0.800pt}}
\multiput(1376.00,454.34)(16.000,-3.000){2}{\rule{3.854pt}{0.800pt}}
\put(1408,449.34){\rule{6.400pt}{0.800pt}}
\multiput(1408.00,451.34)(17.716,-4.000){2}{\rule{3.200pt}{0.800pt}}
\put(1439,449){\usebox{\plotpoint}}
\put(181,172){\usebox{\plotpoint}}
\multiput(181.00,173.41)(0.873,0.506){29}{\rule{1.578pt}{0.122pt}}
\multiput(181.00,170.34)(27.725,18.000){2}{\rule{0.789pt}{0.800pt}}
\multiput(212.00,191.41)(0.853,0.506){31}{\rule{1.547pt}{0.122pt}}
\multiput(212.00,188.34)(28.788,19.000){2}{\rule{0.774pt}{0.800pt}}
\multiput(244.00,210.41)(0.825,0.506){31}{\rule{1.505pt}{0.122pt}}
\multiput(244.00,207.34)(27.876,19.000){2}{\rule{0.753pt}{0.800pt}}
\multiput(275.00,229.41)(0.902,0.506){29}{\rule{1.622pt}{0.122pt}}
\multiput(275.00,226.34)(28.633,18.000){2}{\rule{0.811pt}{0.800pt}}
\multiput(307.00,247.41)(0.873,0.506){29}{\rule{1.578pt}{0.122pt}}
\multiput(307.00,244.34)(27.725,18.000){2}{\rule{0.789pt}{0.800pt}}
\multiput(338.00,265.41)(0.902,0.506){29}{\rule{1.622pt}{0.122pt}}
\multiput(338.00,262.34)(28.633,18.000){2}{\rule{0.811pt}{0.800pt}}
\multiput(370.00,283.41)(0.873,0.506){29}{\rule{1.578pt}{0.122pt}}
\multiput(370.00,280.34)(27.725,18.000){2}{\rule{0.789pt}{0.800pt}}
\multiput(401.00,301.41)(0.902,0.506){29}{\rule{1.622pt}{0.122pt}}
\multiput(401.00,298.34)(28.633,18.000){2}{\rule{0.811pt}{0.800pt}}
\multiput(433.00,319.41)(0.873,0.506){29}{\rule{1.578pt}{0.122pt}}
\multiput(433.00,316.34)(27.725,18.000){2}{\rule{0.789pt}{0.800pt}}
\multiput(464.00,337.41)(0.958,0.507){27}{\rule{1.706pt}{0.122pt}}
\multiput(464.00,334.34)(28.459,17.000){2}{\rule{0.853pt}{0.800pt}}
\multiput(496.00,354.41)(0.873,0.506){29}{\rule{1.578pt}{0.122pt}}
\multiput(496.00,351.34)(27.725,18.000){2}{\rule{0.789pt}{0.800pt}}
\multiput(527.00,372.41)(0.927,0.507){27}{\rule{1.659pt}{0.122pt}}
\multiput(527.00,369.34)(27.557,17.000){2}{\rule{0.829pt}{0.800pt}}
\multiput(558.00,389.41)(0.958,0.507){27}{\rule{1.706pt}{0.122pt}}
\multiput(558.00,386.34)(28.459,17.000){2}{\rule{0.853pt}{0.800pt}}
\multiput(590.00,406.41)(0.927,0.507){27}{\rule{1.659pt}{0.122pt}}
\multiput(590.00,403.34)(27.557,17.000){2}{\rule{0.829pt}{0.800pt}}
\multiput(621.00,423.41)(0.958,0.507){27}{\rule{1.706pt}{0.122pt}}
\multiput(621.00,420.34)(28.459,17.000){2}{\rule{0.853pt}{0.800pt}}
\multiput(653.00,440.41)(0.927,0.507){27}{\rule{1.659pt}{0.122pt}}
\multiput(653.00,437.34)(27.557,17.000){2}{\rule{0.829pt}{0.800pt}}
\multiput(684.00,457.41)(0.958,0.507){27}{\rule{1.706pt}{0.122pt}}
\multiput(684.00,454.34)(28.459,17.000){2}{\rule{0.853pt}{0.800pt}}
\multiput(716.00,474.41)(0.927,0.507){27}{\rule{1.659pt}{0.122pt}}
\multiput(716.00,471.34)(27.557,17.000){2}{\rule{0.829pt}{0.800pt}}
\multiput(747.00,491.41)(1.022,0.507){25}{\rule{1.800pt}{0.122pt}}
\multiput(747.00,488.34)(28.264,16.000){2}{\rule{0.900pt}{0.800pt}}
\multiput(779.00,507.41)(0.989,0.507){25}{\rule{1.750pt}{0.122pt}}
\multiput(779.00,504.34)(27.368,16.000){2}{\rule{0.875pt}{0.800pt}}
\multiput(810.00,523.41)(0.927,0.507){27}{\rule{1.659pt}{0.122pt}}
\multiput(810.00,520.34)(27.557,17.000){2}{\rule{0.829pt}{0.800pt}}
\multiput(841.00,540.41)(1.022,0.507){25}{\rule{1.800pt}{0.122pt}}
\multiput(841.00,537.34)(28.264,16.000){2}{\rule{0.900pt}{0.800pt}}
\multiput(873.00,556.41)(0.989,0.507){25}{\rule{1.750pt}{0.122pt}}
\multiput(873.00,553.34)(27.368,16.000){2}{\rule{0.875pt}{0.800pt}}
\multiput(904.00,572.41)(1.022,0.507){25}{\rule{1.800pt}{0.122pt}}
\multiput(904.00,569.34)(28.264,16.000){2}{\rule{0.900pt}{0.800pt}}
\multiput(936.00,588.41)(0.989,0.507){25}{\rule{1.750pt}{0.122pt}}
\multiput(936.00,585.34)(27.368,16.000){2}{\rule{0.875pt}{0.800pt}}
\multiput(967.00,604.41)(1.022,0.507){25}{\rule{1.800pt}{0.122pt}}
\multiput(967.00,601.34)(28.264,16.000){2}{\rule{0.900pt}{0.800pt}}
\multiput(999.00,620.41)(0.989,0.507){25}{\rule{1.750pt}{0.122pt}}
\multiput(999.00,617.34)(27.368,16.000){2}{\rule{0.875pt}{0.800pt}}
\multiput(1030.00,636.41)(1.095,0.508){23}{\rule{1.907pt}{0.122pt}}
\multiput(1030.00,633.34)(28.043,15.000){2}{\rule{0.953pt}{0.800pt}}
\multiput(1062.00,651.41)(0.989,0.507){25}{\rule{1.750pt}{0.122pt}}
\multiput(1062.00,648.34)(27.368,16.000){2}{\rule{0.875pt}{0.800pt}}
\multiput(1093.00,667.41)(1.095,0.508){23}{\rule{1.907pt}{0.122pt}}
\multiput(1093.00,664.34)(28.043,15.000){2}{\rule{0.953pt}{0.800pt}}
\multiput(1125.00,682.41)(0.989,0.507){25}{\rule{1.750pt}{0.122pt}}
\multiput(1125.00,679.34)(27.368,16.000){2}{\rule{0.875pt}{0.800pt}}
\multiput(1156.00,698.41)(1.059,0.508){23}{\rule{1.853pt}{0.122pt}}
\multiput(1156.00,695.34)(27.153,15.000){2}{\rule{0.927pt}{0.800pt}}
\multiput(1187.00,713.41)(1.022,0.507){25}{\rule{1.800pt}{0.122pt}}
\multiput(1187.00,710.34)(28.264,16.000){2}{\rule{0.900pt}{0.800pt}}
\multiput(1219.00,729.41)(1.059,0.508){23}{\rule{1.853pt}{0.122pt}}
\multiput(1219.00,726.34)(27.153,15.000){2}{\rule{0.927pt}{0.800pt}}
\multiput(1250.00,744.41)(1.095,0.508){23}{\rule{1.907pt}{0.122pt}}
\multiput(1250.00,741.34)(28.043,15.000){2}{\rule{0.953pt}{0.800pt}}
\multiput(1282.00,759.41)(1.059,0.508){23}{\rule{1.853pt}{0.122pt}}
\multiput(1282.00,756.34)(27.153,15.000){2}{\rule{0.927pt}{0.800pt}}
\multiput(1313.00,774.41)(1.095,0.508){23}{\rule{1.907pt}{0.122pt}}
\multiput(1313.00,771.34)(28.043,15.000){2}{\rule{0.953pt}{0.800pt}}
\multiput(1345.00,789.41)(0.989,0.507){25}{\rule{1.750pt}{0.122pt}}
\multiput(1345.00,786.34)(27.368,16.000){2}{\rule{0.875pt}{0.800pt}}
\multiput(1376.00,805.41)(1.095,0.508){23}{\rule{1.907pt}{0.122pt}}
\multiput(1376.00,802.34)(28.043,15.000){2}{\rule{0.953pt}{0.800pt}}
\multiput(1408.00,820.41)(1.059,0.508){23}{\rule{1.853pt}{0.122pt}}
\multiput(1408.00,817.34)(27.153,15.000){2}{\rule{0.927pt}{0.800pt}}
\put(181,732){\usebox{\plotpoint}}
\multiput(181.00,730.08)(1.656,-0.514){13}{\rule{2.680pt}{0.124pt}}
\multiput(181.00,730.34)(25.438,-10.000){2}{\rule{1.340pt}{0.800pt}}
\multiput(212.00,720.08)(1.536,-0.512){15}{\rule{2.527pt}{0.123pt}}
\multiput(212.00,720.34)(26.755,-11.000){2}{\rule{1.264pt}{0.800pt}}
\multiput(244.00,709.08)(1.656,-0.514){13}{\rule{2.680pt}{0.124pt}}
\multiput(244.00,709.34)(25.438,-10.000){2}{\rule{1.340pt}{0.800pt}}
\multiput(275.00,699.08)(1.536,-0.512){15}{\rule{2.527pt}{0.123pt}}
\multiput(275.00,699.34)(26.755,-11.000){2}{\rule{1.264pt}{0.800pt}}
\multiput(307.00,688.08)(1.486,-0.512){15}{\rule{2.455pt}{0.123pt}}
\multiput(307.00,688.34)(25.905,-11.000){2}{\rule{1.227pt}{0.800pt}}
\multiput(338.00,677.08)(1.712,-0.514){13}{\rule{2.760pt}{0.124pt}}
\multiput(338.00,677.34)(26.271,-10.000){2}{\rule{1.380pt}{0.800pt}}
\multiput(370.00,667.08)(1.486,-0.512){15}{\rule{2.455pt}{0.123pt}}
\multiput(370.00,667.34)(25.905,-11.000){2}{\rule{1.227pt}{0.800pt}}
\multiput(401.00,656.08)(1.712,-0.514){13}{\rule{2.760pt}{0.124pt}}
\multiput(401.00,656.34)(26.271,-10.000){2}{\rule{1.380pt}{0.800pt}}
\multiput(433.00,646.08)(1.486,-0.512){15}{\rule{2.455pt}{0.123pt}}
\multiput(433.00,646.34)(25.905,-11.000){2}{\rule{1.227pt}{0.800pt}}
\multiput(464.00,635.08)(1.712,-0.514){13}{\rule{2.760pt}{0.124pt}}
\multiput(464.00,635.34)(26.271,-10.000){2}{\rule{1.380pt}{0.800pt}}
\multiput(496.00,625.08)(1.486,-0.512){15}{\rule{2.455pt}{0.123pt}}
\multiput(496.00,625.34)(25.905,-11.000){2}{\rule{1.227pt}{0.800pt}}
\multiput(527.00,614.08)(1.656,-0.514){13}{\rule{2.680pt}{0.124pt}}
\multiput(527.00,614.34)(25.438,-10.000){2}{\rule{1.340pt}{0.800pt}}
\multiput(558.00,604.08)(1.712,-0.514){13}{\rule{2.760pt}{0.124pt}}
\multiput(558.00,604.34)(26.271,-10.000){2}{\rule{1.380pt}{0.800pt}}
\multiput(590.00,594.08)(1.486,-0.512){15}{\rule{2.455pt}{0.123pt}}
\multiput(590.00,594.34)(25.905,-11.000){2}{\rule{1.227pt}{0.800pt}}
\multiput(621.00,583.08)(1.712,-0.514){13}{\rule{2.760pt}{0.124pt}}
\multiput(621.00,583.34)(26.271,-10.000){2}{\rule{1.380pt}{0.800pt}}
\multiput(653.00,573.08)(1.486,-0.512){15}{\rule{2.455pt}{0.123pt}}
\multiput(653.00,573.34)(25.905,-11.000){2}{\rule{1.227pt}{0.800pt}}
\multiput(684.00,562.08)(1.712,-0.514){13}{\rule{2.760pt}{0.124pt}}
\multiput(684.00,562.34)(26.271,-10.000){2}{\rule{1.380pt}{0.800pt}}
\multiput(716.00,552.08)(1.486,-0.512){15}{\rule{2.455pt}{0.123pt}}
\multiput(716.00,552.34)(25.905,-11.000){2}{\rule{1.227pt}{0.800pt}}
\multiput(747.00,541.08)(1.712,-0.514){13}{\rule{2.760pt}{0.124pt}}
\multiput(747.00,541.34)(26.271,-10.000){2}{\rule{1.380pt}{0.800pt}}
\multiput(779.00,531.08)(1.486,-0.512){15}{\rule{2.455pt}{0.123pt}}
\multiput(779.00,531.34)(25.905,-11.000){2}{\rule{1.227pt}{0.800pt}}
\multiput(810.00,520.08)(1.656,-0.514){13}{\rule{2.680pt}{0.124pt}}
\multiput(810.00,520.34)(25.438,-10.000){2}{\rule{1.340pt}{0.800pt}}
\multiput(841.00,510.08)(1.712,-0.514){13}{\rule{2.760pt}{0.124pt}}
\multiput(841.00,510.34)(26.271,-10.000){2}{\rule{1.380pt}{0.800pt}}
\multiput(873.00,500.08)(1.486,-0.512){15}{\rule{2.455pt}{0.123pt}}
\multiput(873.00,500.34)(25.905,-11.000){2}{\rule{1.227pt}{0.800pt}}
\multiput(904.00,489.08)(1.712,-0.514){13}{\rule{2.760pt}{0.124pt}}
\multiput(904.00,489.34)(26.271,-10.000){2}{\rule{1.380pt}{0.800pt}}
\multiput(936.00,479.08)(1.486,-0.512){15}{\rule{2.455pt}{0.123pt}}
\multiput(936.00,479.34)(25.905,-11.000){2}{\rule{1.227pt}{0.800pt}}
\multiput(967.00,468.08)(1.712,-0.514){13}{\rule{2.760pt}{0.124pt}}
\multiput(967.00,468.34)(26.271,-10.000){2}{\rule{1.380pt}{0.800pt}}
\multiput(999.00,458.08)(1.656,-0.514){13}{\rule{2.680pt}{0.124pt}}
\multiput(999.00,458.34)(25.438,-10.000){2}{\rule{1.340pt}{0.800pt}}
\multiput(1030.00,448.08)(1.536,-0.512){15}{\rule{2.527pt}{0.123pt}}
\multiput(1030.00,448.34)(26.755,-11.000){2}{\rule{1.264pt}{0.800pt}}
\multiput(1062.00,437.08)(1.656,-0.514){13}{\rule{2.680pt}{0.124pt}}
\multiput(1062.00,437.34)(25.438,-10.000){2}{\rule{1.340pt}{0.800pt}}
\multiput(1093.00,427.08)(1.712,-0.514){13}{\rule{2.760pt}{0.124pt}}
\multiput(1093.00,427.34)(26.271,-10.000){2}{\rule{1.380pt}{0.800pt}}
\multiput(1125.00,417.08)(1.486,-0.512){15}{\rule{2.455pt}{0.123pt}}
\multiput(1125.00,417.34)(25.905,-11.000){2}{\rule{1.227pt}{0.800pt}}
\multiput(1156.00,406.08)(1.656,-0.514){13}{\rule{2.680pt}{0.124pt}}
\multiput(1156.00,406.34)(25.438,-10.000){2}{\rule{1.340pt}{0.800pt}}
\multiput(1187.00,396.08)(1.712,-0.514){13}{\rule{2.760pt}{0.124pt}}
\multiput(1187.00,396.34)(26.271,-10.000){2}{\rule{1.380pt}{0.800pt}}
\multiput(1219.00,386.08)(1.486,-0.512){15}{\rule{2.455pt}{0.123pt}}
\multiput(1219.00,386.34)(25.905,-11.000){2}{\rule{1.227pt}{0.800pt}}
\multiput(1250.00,375.08)(1.712,-0.514){13}{\rule{2.760pt}{0.124pt}}
\multiput(1250.00,375.34)(26.271,-10.000){2}{\rule{1.380pt}{0.800pt}}
\multiput(1282.00,365.08)(1.656,-0.514){13}{\rule{2.680pt}{0.124pt}}
\multiput(1282.00,365.34)(25.438,-10.000){2}{\rule{1.340pt}{0.800pt}}
\multiput(1313.00,355.08)(1.536,-0.512){15}{\rule{2.527pt}{0.123pt}}
\multiput(1313.00,355.34)(26.755,-11.000){2}{\rule{1.264pt}{0.800pt}}
\multiput(1345.00,344.08)(1.656,-0.514){13}{\rule{2.680pt}{0.124pt}}
\multiput(1345.00,344.34)(25.438,-10.000){2}{\rule{1.340pt}{0.800pt}}
\multiput(1376.00,334.08)(1.712,-0.514){13}{\rule{2.760pt}{0.124pt}}
\multiput(1376.00,334.34)(26.271,-10.000){2}{\rule{1.380pt}{0.800pt}}
\multiput(1408.00,324.08)(1.486,-0.512){15}{\rule{2.455pt}{0.123pt}}
\multiput(1408.00,324.34)(25.905,-11.000){2}{\rule{1.227pt}{0.800pt}}
\put(1439,315){\usebox{\plotpoint}}
\end{picture}

%% file: newfig3.tex
\setlength{\unitlength}{0.240900pt}
\ifx\plotpoint\undefined\newsavebox{\plotpoint}\fi
\sbox{\plotpoint}{\rule[-0.200pt]{0.400pt}{0.400pt}}%
\begin{picture}(1500,900)(0,0)
\font\gnuplot=cmr10 at 10pt
\gnuplot
\sbox{\plotpoint}{\rule[-0.200pt]{0.400pt}{0.400pt}}%
\put(181.0,123.0){\rule[-0.200pt]{4.818pt}{0.400pt}}
\put(161,123){\makebox(0,0)[r]{0.09}}
\put(1419.0,123.0){\rule[-0.200pt]{4.818pt}{0.400pt}}
\put(181.0,215.0){\rule[-0.200pt]{4.818pt}{0.400pt}}
\put(161,215){\makebox(0,0)[r]{0.1}}
\put(1419.0,215.0){\rule[-0.200pt]{4.818pt}{0.400pt}}
\put(181.0,307.0){\rule[-0.200pt]{4.818pt}{0.400pt}}
\put(161,307){\makebox(0,0)[r]{0.11}}
\put(1419.0,307.0){\rule[-0.200pt]{4.818pt}{0.400pt}}
\put(181.0,399.0){\rule[-0.200pt]{4.818pt}{0.400pt}}
\put(161,399){\makebox(0,0)[r]{0.12}}
\put(1419.0,399.0){\rule[-0.200pt]{4.818pt}{0.400pt}}
\put(181.0,491.0){\rule[-0.200pt]{4.818pt}{0.400pt}}
\put(161,491){\makebox(0,0)[r]{0.13}}
\put(1419.0,491.0){\rule[-0.200pt]{4.818pt}{0.400pt}}
\put(181.0,584.0){\rule[-0.200pt]{4.818pt}{0.400pt}}
\put(161,584){\makebox(0,0)[r]{0.14}}
\put(1419.0,584.0){\rule[-0.200pt]{4.818pt}{0.400pt}}
\put(181.0,676.0){\rule[-0.200pt]{4.818pt}{0.400pt}}
\put(161,676){\makebox(0,0)[r]{0.15}}
\put(1419.0,676.0){\rule[-0.200pt]{4.818pt}{0.400pt}}
\put(181.0,768.0){\rule[-0.200pt]{4.818pt}{0.400pt}}
\put(161,768){\makebox(0,0)[r]{0.16}}
\put(1419.0,768.0){\rule[-0.200pt]{4.818pt}{0.400pt}}
\put(181.0,860.0){\rule[-0.200pt]{4.818pt}{0.400pt}}
\put(161,860){\makebox(0,0)[r]{0.17}}
\put(1419.0,860.0){\rule[-0.200pt]{4.818pt}{0.400pt}}
\put(181.0,123.0){\rule[-0.200pt]{0.400pt}{4.818pt}}
\put(181,82){\makebox(0,0){0}}
\put(181.0,840.0){\rule[-0.200pt]{0.400pt}{4.818pt}}
\put(496.0,123.0){\rule[-0.200pt]{0.400pt}{4.818pt}}
\put(496,82){\makebox(0,0){0.5}}
\put(496.0,840.0){\rule[-0.200pt]{0.400pt}{4.818pt}}
\put(810.0,123.0){\rule[-0.200pt]{0.400pt}{4.818pt}}
\put(810,82){\makebox(0,0){1}}
\put(810.0,840.0){\rule[-0.200pt]{0.400pt}{4.818pt}}
\put(1125.0,123.0){\rule[-0.200pt]{0.400pt}{4.818pt}}
\put(1125,82){\makebox(0,0){1.5}}
\put(1125.0,840.0){\rule[-0.200pt]{0.400pt}{4.818pt}}
\put(1439.0,123.0){\rule[-0.200pt]{0.400pt}{4.818pt}}
\put(1439,82){\makebox(0,0){2}}
\put(1439.0,840.0){\rule[-0.200pt]{0.400pt}{4.818pt}}
\put(181.0,123.0){\rule[-0.200pt]{303.052pt}{0.400pt}}
\put(1439.0,123.0){\rule[-0.200pt]{0.400pt}{177.543pt}}
\put(181.0,860.0){\rule[-0.200pt]{303.052pt}{0.400pt}}
\put(30,491){\makebox(0,0){$\alpha_3(M_Z)$}}
\put(810,21){\makebox(0,0){$\theta/\pi$}}
\put(1376,390){\makebox(0,0)[r]{experiment}}
\put(1376,492){\makebox(0,0)[r]{$x=0.01$}}
\put(1376,630){\makebox(0,0)[r]{$x=0.05$}}
\put(1062,768){\makebox(0,0)[r]{$x=0.1$}}
\put(181.0,123.0){\rule[-0.200pt]{0.400pt}{177.543pt}}
\sbox{\plotpoint}{\rule[-0.400pt]{0.800pt}{0.800pt}}%
\put(181,451){\usebox{\plotpoint}}
\put(181,447.84){\rule{3.132pt}{0.800pt}}
\multiput(181.00,449.34)(6.500,-3.000){2}{\rule{1.566pt}{0.800pt}}
\put(194,445.34){\rule{2.891pt}{0.800pt}}
\multiput(194.00,446.34)(6.000,-2.000){2}{\rule{1.445pt}{0.800pt}}
\put(206,442.84){\rule{3.132pt}{0.800pt}}
\multiput(206.00,444.34)(6.500,-3.000){2}{\rule{1.566pt}{0.800pt}}
\put(219,440.34){\rule{2.891pt}{0.800pt}}
\multiput(219.00,441.34)(6.000,-2.000){2}{\rule{1.445pt}{0.800pt}}
\put(231,437.34){\rule{2.800pt}{0.800pt}}
\multiput(231.00,439.34)(7.188,-4.000){2}{\rule{1.400pt}{0.800pt}}
\put(244,433.84){\rule{2.891pt}{0.800pt}}
\multiput(244.00,435.34)(6.000,-3.000){2}{\rule{1.445pt}{0.800pt}}
\put(256,430.84){\rule{3.132pt}{0.800pt}}
\multiput(256.00,432.34)(6.500,-3.000){2}{\rule{1.566pt}{0.800pt}}
\put(269,427.34){\rule{2.800pt}{0.800pt}}
\multiput(269.00,429.34)(7.188,-4.000){2}{\rule{1.400pt}{0.800pt}}
\put(282,423.84){\rule{2.891pt}{0.800pt}}
\multiput(282.00,425.34)(6.000,-3.000){2}{\rule{1.445pt}{0.800pt}}
\put(294,420.34){\rule{2.800pt}{0.800pt}}
\multiput(294.00,422.34)(7.188,-4.000){2}{\rule{1.400pt}{0.800pt}}
\put(307,416.34){\rule{2.600pt}{0.800pt}}
\multiput(307.00,418.34)(6.604,-4.000){2}{\rule{1.300pt}{0.800pt}}
\put(319,412.84){\rule{3.132pt}{0.800pt}}
\multiput(319.00,414.34)(6.500,-3.000){2}{\rule{1.566pt}{0.800pt}}
\put(332,409.34){\rule{2.800pt}{0.800pt}}
\multiput(332.00,411.34)(7.188,-4.000){2}{\rule{1.400pt}{0.800pt}}
\put(345,405.34){\rule{2.600pt}{0.800pt}}
\multiput(345.00,407.34)(6.604,-4.000){2}{\rule{1.300pt}{0.800pt}}
\put(357,401.34){\rule{2.800pt}{0.800pt}}
\multiput(357.00,403.34)(7.188,-4.000){2}{\rule{1.400pt}{0.800pt}}
\put(370,397.34){\rule{2.600pt}{0.800pt}}
\multiput(370.00,399.34)(6.604,-4.000){2}{\rule{1.300pt}{0.800pt}}
\put(382,393.34){\rule{2.800pt}{0.800pt}}
\multiput(382.00,395.34)(7.188,-4.000){2}{\rule{1.400pt}{0.800pt}}
\put(395,389.34){\rule{2.600pt}{0.800pt}}
\multiput(395.00,391.34)(6.604,-4.000){2}{\rule{1.300pt}{0.800pt}}
\put(407,385.34){\rule{2.800pt}{0.800pt}}
\multiput(407.00,387.34)(7.188,-4.000){2}{\rule{1.400pt}{0.800pt}}
\put(420,381.34){\rule{2.800pt}{0.800pt}}
\multiput(420.00,383.34)(7.188,-4.000){2}{\rule{1.400pt}{0.800pt}}
\put(433,377.84){\rule{2.891pt}{0.800pt}}
\multiput(433.00,379.34)(6.000,-3.000){2}{\rule{1.445pt}{0.800pt}}
\put(445,374.34){\rule{2.800pt}{0.800pt}}
\multiput(445.00,376.34)(7.188,-4.000){2}{\rule{1.400pt}{0.800pt}}
\put(458,370.84){\rule{2.891pt}{0.800pt}}
\multiput(458.00,372.34)(6.000,-3.000){2}{\rule{1.445pt}{0.800pt}}
\put(470,367.34){\rule{2.800pt}{0.800pt}}
\multiput(470.00,369.34)(7.188,-4.000){2}{\rule{1.400pt}{0.800pt}}
\put(483,363.84){\rule{3.132pt}{0.800pt}}
\multiput(483.00,365.34)(6.500,-3.000){2}{\rule{1.566pt}{0.800pt}}
\put(496,360.84){\rule{2.891pt}{0.800pt}}
\multiput(496.00,362.34)(6.000,-3.000){2}{\rule{1.445pt}{0.800pt}}
\put(508,357.84){\rule{3.132pt}{0.800pt}}
\multiput(508.00,359.34)(6.500,-3.000){2}{\rule{1.566pt}{0.800pt}}
\put(521,354.84){\rule{2.891pt}{0.800pt}}
\multiput(521.00,356.34)(6.000,-3.000){2}{\rule{1.445pt}{0.800pt}}
\put(533,351.84){\rule{3.132pt}{0.800pt}}
\multiput(533.00,353.34)(6.500,-3.000){2}{\rule{1.566pt}{0.800pt}}
\put(546,349.34){\rule{2.891pt}{0.800pt}}
\multiput(546.00,350.34)(6.000,-2.000){2}{\rule{1.445pt}{0.800pt}}
\put(558,346.84){\rule{3.132pt}{0.800pt}}
\multiput(558.00,348.34)(6.500,-3.000){2}{\rule{1.566pt}{0.800pt}}
\put(571,344.34){\rule{3.132pt}{0.800pt}}
\multiput(571.00,345.34)(6.500,-2.000){2}{\rule{1.566pt}{0.800pt}}
\put(584,342.34){\rule{2.891pt}{0.800pt}}
\multiput(584.00,343.34)(6.000,-2.000){2}{\rule{1.445pt}{0.800pt}}
\put(596,340.34){\rule{3.132pt}{0.800pt}}
\multiput(596.00,341.34)(6.500,-2.000){2}{\rule{1.566pt}{0.800pt}}
\put(609,338.84){\rule{2.891pt}{0.800pt}}
\multiput(609.00,339.34)(6.000,-1.000){2}{\rule{1.445pt}{0.800pt}}
\put(621,337.84){\rule{3.132pt}{0.800pt}}
\multiput(621.00,338.34)(6.500,-1.000){2}{\rule{1.566pt}{0.800pt}}
\put(634,336.84){\rule{2.891pt}{0.800pt}}
\multiput(634.00,337.34)(6.000,-1.000){2}{\rule{1.445pt}{0.800pt}}
\put(646,335.84){\rule{3.132pt}{0.800pt}}
\multiput(646.00,336.34)(6.500,-1.000){2}{\rule{1.566pt}{0.800pt}}
\put(659,334.84){\rule{3.132pt}{0.800pt}}
\multiput(659.00,335.34)(6.500,-1.000){2}{\rule{1.566pt}{0.800pt}}
\put(672,333.84){\rule{2.891pt}{0.800pt}}
\multiput(672.00,334.34)(6.000,-1.000){2}{\rule{1.445pt}{0.800pt}}
\put(709,333.84){\rule{3.132pt}{0.800pt}}
\multiput(709.00,333.34)(6.500,1.000){2}{\rule{1.566pt}{0.800pt}}
\put(684.0,335.0){\rule[-0.400pt]{6.022pt}{0.800pt}}
\put(735,334.84){\rule{2.891pt}{0.800pt}}
\multiput(735.00,334.34)(6.000,1.000){2}{\rule{1.445pt}{0.800pt}}
\put(747,335.84){\rule{3.132pt}{0.800pt}}
\multiput(747.00,335.34)(6.500,1.000){2}{\rule{1.566pt}{0.800pt}}
\put(760,336.84){\rule{2.891pt}{0.800pt}}
\multiput(760.00,336.34)(6.000,1.000){2}{\rule{1.445pt}{0.800pt}}
\put(772,337.84){\rule{3.132pt}{0.800pt}}
\multiput(772.00,337.34)(6.500,1.000){2}{\rule{1.566pt}{0.800pt}}
\put(785,339.34){\rule{2.891pt}{0.800pt}}
\multiput(785.00,338.34)(6.000,2.000){2}{\rule{1.445pt}{0.800pt}}
\put(797,340.84){\rule{3.132pt}{0.800pt}}
\multiput(797.00,340.34)(6.500,1.000){2}{\rule{1.566pt}{0.800pt}}
\put(810,342.34){\rule{3.132pt}{0.800pt}}
\multiput(810.00,341.34)(6.500,2.000){2}{\rule{1.566pt}{0.800pt}}
\put(823,344.34){\rule{2.891pt}{0.800pt}}
\multiput(823.00,343.34)(6.000,2.000){2}{\rule{1.445pt}{0.800pt}}
\put(835,346.84){\rule{3.132pt}{0.800pt}}
\multiput(835.00,345.34)(6.500,3.000){2}{\rule{1.566pt}{0.800pt}}
\put(848,349.34){\rule{2.891pt}{0.800pt}}
\multiput(848.00,348.34)(6.000,2.000){2}{\rule{1.445pt}{0.800pt}}
\put(860,351.84){\rule{3.132pt}{0.800pt}}
\multiput(860.00,350.34)(6.500,3.000){2}{\rule{1.566pt}{0.800pt}}
\put(873,354.84){\rule{2.891pt}{0.800pt}}
\multiput(873.00,353.34)(6.000,3.000){2}{\rule{1.445pt}{0.800pt}}
\put(885,357.84){\rule{3.132pt}{0.800pt}}
\multiput(885.00,356.34)(6.500,3.000){2}{\rule{1.566pt}{0.800pt}}
\put(898,360.84){\rule{3.132pt}{0.800pt}}
\multiput(898.00,359.34)(6.500,3.000){2}{\rule{1.566pt}{0.800pt}}
\put(911,364.34){\rule{2.600pt}{0.800pt}}
\multiput(911.00,362.34)(6.604,4.000){2}{\rule{1.300pt}{0.800pt}}
\put(923,367.84){\rule{3.132pt}{0.800pt}}
\multiput(923.00,366.34)(6.500,3.000){2}{\rule{1.566pt}{0.800pt}}
\put(936,371.34){\rule{2.600pt}{0.800pt}}
\multiput(936.00,369.34)(6.604,4.000){2}{\rule{1.300pt}{0.800pt}}
\put(948,375.34){\rule{2.800pt}{0.800pt}}
\multiput(948.00,373.34)(7.188,4.000){2}{\rule{1.400pt}{0.800pt}}
\put(961,379.34){\rule{2.800pt}{0.800pt}}
\multiput(961.00,377.34)(7.188,4.000){2}{\rule{1.400pt}{0.800pt}}
\put(974,382.84){\rule{2.891pt}{0.800pt}}
\multiput(974.00,381.34)(6.000,3.000){2}{\rule{1.445pt}{0.800pt}}
\put(986,386.34){\rule{2.800pt}{0.800pt}}
\multiput(986.00,384.34)(7.188,4.000){2}{\rule{1.400pt}{0.800pt}}
\put(999,390.34){\rule{2.600pt}{0.800pt}}
\multiput(999.00,388.34)(6.604,4.000){2}{\rule{1.300pt}{0.800pt}}
\multiput(1011.00,395.38)(1.768,0.560){3}{\rule{2.280pt}{0.135pt}}
\multiput(1011.00,392.34)(8.268,5.000){2}{\rule{1.140pt}{0.800pt}}
\put(1024,399.34){\rule{2.600pt}{0.800pt}}
\multiput(1024.00,397.34)(6.604,4.000){2}{\rule{1.300pt}{0.800pt}}
\put(1036,403.34){\rule{2.800pt}{0.800pt}}
\multiput(1036.00,401.34)(7.188,4.000){2}{\rule{1.400pt}{0.800pt}}
\put(1049,407.34){\rule{2.800pt}{0.800pt}}
\multiput(1049.00,405.34)(7.188,4.000){2}{\rule{1.400pt}{0.800pt}}
\put(1062,411.34){\rule{2.600pt}{0.800pt}}
\multiput(1062.00,409.34)(6.604,4.000){2}{\rule{1.300pt}{0.800pt}}
\put(1074,415.34){\rule{2.800pt}{0.800pt}}
\multiput(1074.00,413.34)(7.188,4.000){2}{\rule{1.400pt}{0.800pt}}
\put(1087,419.34){\rule{2.600pt}{0.800pt}}
\multiput(1087.00,417.34)(6.604,4.000){2}{\rule{1.300pt}{0.800pt}}
\put(1099,423.34){\rule{2.800pt}{0.800pt}}
\multiput(1099.00,421.34)(7.188,4.000){2}{\rule{1.400pt}{0.800pt}}
\put(1112,427.34){\rule{2.800pt}{0.800pt}}
\multiput(1112.00,425.34)(7.188,4.000){2}{\rule{1.400pt}{0.800pt}}
\put(1125,431.34){\rule{2.600pt}{0.800pt}}
\multiput(1125.00,429.34)(6.604,4.000){2}{\rule{1.300pt}{0.800pt}}
\put(1137,434.84){\rule{3.132pt}{0.800pt}}
\multiput(1137.00,433.34)(6.500,3.000){2}{\rule{1.566pt}{0.800pt}}
\put(1150,438.34){\rule{2.600pt}{0.800pt}}
\multiput(1150.00,436.34)(6.604,4.000){2}{\rule{1.300pt}{0.800pt}}
\put(1162,441.84){\rule{3.132pt}{0.800pt}}
\multiput(1162.00,440.34)(6.500,3.000){2}{\rule{1.566pt}{0.800pt}}
\put(1175,444.84){\rule{2.891pt}{0.800pt}}
\multiput(1175.00,443.34)(6.000,3.000){2}{\rule{1.445pt}{0.800pt}}
\put(1187,447.84){\rule{3.132pt}{0.800pt}}
\multiput(1187.00,446.34)(6.500,3.000){2}{\rule{1.566pt}{0.800pt}}
\put(1200,450.84){\rule{3.132pt}{0.800pt}}
\multiput(1200.00,449.34)(6.500,3.000){2}{\rule{1.566pt}{0.800pt}}
\put(1213,453.84){\rule{2.891pt}{0.800pt}}
\multiput(1213.00,452.34)(6.000,3.000){2}{\rule{1.445pt}{0.800pt}}
\put(1225,456.34){\rule{3.132pt}{0.800pt}}
\multiput(1225.00,455.34)(6.500,2.000){2}{\rule{1.566pt}{0.800pt}}
\put(1238,458.34){\rule{2.891pt}{0.800pt}}
\multiput(1238.00,457.34)(6.000,2.000){2}{\rule{1.445pt}{0.800pt}}
\put(1250,460.34){\rule{3.132pt}{0.800pt}}
\multiput(1250.00,459.34)(6.500,2.000){2}{\rule{1.566pt}{0.800pt}}
\put(1263,461.84){\rule{2.891pt}{0.800pt}}
\multiput(1263.00,461.34)(6.000,1.000){2}{\rule{1.445pt}{0.800pt}}
\put(1275,463.34){\rule{3.132pt}{0.800pt}}
\multiput(1275.00,462.34)(6.500,2.000){2}{\rule{1.566pt}{0.800pt}}
\put(1288,464.84){\rule{3.132pt}{0.800pt}}
\multiput(1288.00,464.34)(6.500,1.000){2}{\rule{1.566pt}{0.800pt}}
\put(1301,465.84){\rule{2.891pt}{0.800pt}}
\multiput(1301.00,465.34)(6.000,1.000){2}{\rule{1.445pt}{0.800pt}}
\put(722.0,336.0){\rule[-0.400pt]{3.132pt}{0.800pt}}
\put(1364,465.84){\rule{2.891pt}{0.800pt}}
\multiput(1364.00,466.34)(6.000,-1.000){2}{\rule{1.445pt}{0.800pt}}
\put(1313.0,468.0){\rule[-0.400pt]{12.286pt}{0.800pt}}
\put(1389,464.34){\rule{2.891pt}{0.800pt}}
\multiput(1389.00,465.34)(6.000,-2.000){2}{\rule{1.445pt}{0.800pt}}
\put(1401,462.84){\rule{3.132pt}{0.800pt}}
\multiput(1401.00,463.34)(6.500,-1.000){2}{\rule{1.566pt}{0.800pt}}
\put(1414,461.34){\rule{2.891pt}{0.800pt}}
\multiput(1414.00,462.34)(6.000,-2.000){2}{\rule{1.445pt}{0.800pt}}
\put(1426,459.34){\rule{3.132pt}{0.800pt}}
\multiput(1426.00,460.34)(6.500,-2.000){2}{\rule{1.566pt}{0.800pt}}
\put(1376.0,467.0){\rule[-0.400pt]{3.132pt}{0.800pt}}
\put(181,709){\usebox{\plotpoint}}
\multiput(182.41,703.83)(0.509,-0.657){19}{\rule{0.123pt}{1.246pt}}
\multiput(179.34,706.41)(13.000,-14.414){2}{\rule{0.800pt}{0.623pt}}
\multiput(195.41,686.47)(0.511,-0.717){17}{\rule{0.123pt}{1.333pt}}
\multiput(192.34,689.23)(12.000,-14.233){2}{\rule{0.800pt}{0.667pt}}
\multiput(207.41,669.32)(0.509,-0.740){19}{\rule{0.123pt}{1.369pt}}
\multiput(204.34,672.16)(13.000,-16.158){2}{\rule{0.800pt}{0.685pt}}
\multiput(220.41,649.91)(0.511,-0.807){17}{\rule{0.123pt}{1.467pt}}
\multiput(217.34,652.96)(12.000,-15.956){2}{\rule{0.800pt}{0.733pt}}
\multiput(232.41,631.06)(0.509,-0.781){19}{\rule{0.123pt}{1.431pt}}
\multiput(229.34,634.03)(13.000,-17.030){2}{\rule{0.800pt}{0.715pt}}
\multiput(245.41,610.36)(0.511,-0.897){17}{\rule{0.123pt}{1.600pt}}
\multiput(242.34,613.68)(12.000,-17.679){2}{\rule{0.800pt}{0.800pt}}
\multiput(257.41,589.81)(0.509,-0.823){19}{\rule{0.123pt}{1.492pt}}
\multiput(254.34,592.90)(13.000,-17.903){2}{\rule{0.800pt}{0.746pt}}
\multiput(270.41,568.81)(0.509,-0.823){19}{\rule{0.123pt}{1.492pt}}
\multiput(267.34,571.90)(13.000,-17.903){2}{\rule{0.800pt}{0.746pt}}
\multiput(283.41,547.08)(0.511,-0.943){17}{\rule{0.123pt}{1.667pt}}
\multiput(280.34,550.54)(12.000,-18.541){2}{\rule{0.800pt}{0.833pt}}
\multiput(295.41,525.81)(0.509,-0.823){19}{\rule{0.123pt}{1.492pt}}
\multiput(292.34,528.90)(13.000,-17.903){2}{\rule{0.800pt}{0.746pt}}
\multiput(308.41,504.08)(0.511,-0.943){17}{\rule{0.123pt}{1.667pt}}
\multiput(305.34,507.54)(12.000,-18.541){2}{\rule{0.800pt}{0.833pt}}
\multiput(320.41,482.81)(0.509,-0.823){19}{\rule{0.123pt}{1.492pt}}
\multiput(317.34,485.90)(13.000,-17.903){2}{\rule{0.800pt}{0.746pt}}
\multiput(333.41,461.81)(0.509,-0.823){19}{\rule{0.123pt}{1.492pt}}
\multiput(330.34,464.90)(13.000,-17.903){2}{\rule{0.800pt}{0.746pt}}
\multiput(346.41,440.63)(0.511,-0.852){17}{\rule{0.123pt}{1.533pt}}
\multiput(343.34,443.82)(12.000,-16.817){2}{\rule{0.800pt}{0.767pt}}
\multiput(358.41,421.06)(0.509,-0.781){19}{\rule{0.123pt}{1.431pt}}
\multiput(355.34,424.03)(13.000,-17.030){2}{\rule{0.800pt}{0.715pt}}
\multiput(371.41,400.63)(0.511,-0.852){17}{\rule{0.123pt}{1.533pt}}
\multiput(368.34,403.82)(12.000,-16.817){2}{\rule{0.800pt}{0.767pt}}
\multiput(383.41,381.32)(0.509,-0.740){19}{\rule{0.123pt}{1.369pt}}
\multiput(380.34,384.16)(13.000,-16.158){2}{\rule{0.800pt}{0.685pt}}
\multiput(396.41,362.19)(0.511,-0.762){17}{\rule{0.123pt}{1.400pt}}
\multiput(393.34,365.09)(12.000,-15.094){2}{\rule{0.800pt}{0.700pt}}
\multiput(408.41,344.57)(0.509,-0.698){19}{\rule{0.123pt}{1.308pt}}
\multiput(405.34,347.29)(13.000,-15.286){2}{\rule{0.800pt}{0.654pt}}
\multiput(421.41,326.83)(0.509,-0.657){19}{\rule{0.123pt}{1.246pt}}
\multiput(418.34,329.41)(13.000,-14.414){2}{\rule{0.800pt}{0.623pt}}
\multiput(434.41,309.74)(0.511,-0.671){17}{\rule{0.123pt}{1.267pt}}
\multiput(431.34,312.37)(12.000,-13.371){2}{\rule{0.800pt}{0.633pt}}
\multiput(446.41,294.08)(0.509,-0.616){19}{\rule{0.123pt}{1.185pt}}
\multiput(443.34,296.54)(13.000,-13.541){2}{\rule{0.800pt}{0.592pt}}
\multiput(459.41,278.02)(0.511,-0.626){17}{\rule{0.123pt}{1.200pt}}
\multiput(456.34,280.51)(12.000,-12.509){2}{\rule{0.800pt}{0.600pt}}
\multiput(471.41,263.59)(0.509,-0.533){19}{\rule{0.123pt}{1.062pt}}
\multiput(468.34,265.80)(13.000,-11.797){2}{\rule{0.800pt}{0.531pt}}
\multiput(483.00,252.08)(0.492,-0.509){19}{\rule{1.000pt}{0.123pt}}
\multiput(483.00,252.34)(10.924,-13.000){2}{\rule{0.500pt}{0.800pt}}
\multiput(497.41,236.57)(0.511,-0.536){17}{\rule{0.123pt}{1.067pt}}
\multiput(494.34,238.79)(12.000,-10.786){2}{\rule{0.800pt}{0.533pt}}
\multiput(508.00,226.08)(0.536,-0.511){17}{\rule{1.067pt}{0.123pt}}
\multiput(508.00,226.34)(10.786,-12.000){2}{\rule{0.533pt}{0.800pt}}
\multiput(521.00,214.08)(0.539,-0.512){15}{\rule{1.073pt}{0.123pt}}
\multiput(521.00,214.34)(9.774,-11.000){2}{\rule{0.536pt}{0.800pt}}
\multiput(533.00,203.08)(0.654,-0.514){13}{\rule{1.240pt}{0.124pt}}
\multiput(533.00,203.34)(10.426,-10.000){2}{\rule{0.620pt}{0.800pt}}
\multiput(546.00,193.08)(0.674,-0.516){11}{\rule{1.267pt}{0.124pt}}
\multiput(546.00,193.34)(9.371,-9.000){2}{\rule{0.633pt}{0.800pt}}
\multiput(558.00,184.08)(0.737,-0.516){11}{\rule{1.356pt}{0.124pt}}
\multiput(558.00,184.34)(10.186,-9.000){2}{\rule{0.678pt}{0.800pt}}
\multiput(571.00,175.08)(0.847,-0.520){9}{\rule{1.500pt}{0.125pt}}
\multiput(571.00,175.34)(9.887,-8.000){2}{\rule{0.750pt}{0.800pt}}
\multiput(584.00,167.08)(0.913,-0.526){7}{\rule{1.571pt}{0.127pt}}
\multiput(584.00,167.34)(8.738,-7.000){2}{\rule{0.786pt}{0.800pt}}
\multiput(596.00,160.07)(1.244,-0.536){5}{\rule{1.933pt}{0.129pt}}
\multiput(596.00,160.34)(8.987,-6.000){2}{\rule{0.967pt}{0.800pt}}
\multiput(609.00,154.06)(1.600,-0.560){3}{\rule{2.120pt}{0.135pt}}
\multiput(609.00,154.34)(7.600,-5.000){2}{\rule{1.060pt}{0.800pt}}
\multiput(621.00,149.06)(1.768,-0.560){3}{\rule{2.280pt}{0.135pt}}
\multiput(621.00,149.34)(8.268,-5.000){2}{\rule{1.140pt}{0.800pt}}
\put(634,142.34){\rule{2.600pt}{0.800pt}}
\multiput(634.00,144.34)(6.604,-4.000){2}{\rule{1.300pt}{0.800pt}}
\put(646,138.84){\rule{3.132pt}{0.800pt}}
\multiput(646.00,140.34)(6.500,-3.000){2}{\rule{1.566pt}{0.800pt}}
\put(659,136.34){\rule{3.132pt}{0.800pt}}
\multiput(659.00,137.34)(6.500,-2.000){2}{\rule{1.566pt}{0.800pt}}
\put(672,134.34){\rule{2.891pt}{0.800pt}}
\multiput(672.00,135.34)(6.000,-2.000){2}{\rule{1.445pt}{0.800pt}}
\put(684,132.84){\rule{3.132pt}{0.800pt}}
\multiput(684.00,133.34)(6.500,-1.000){2}{\rule{1.566pt}{0.800pt}}
\put(709,132.84){\rule{3.132pt}{0.800pt}}
\multiput(709.00,132.34)(6.500,1.000){2}{\rule{1.566pt}{0.800pt}}
\put(722,134.34){\rule{3.132pt}{0.800pt}}
\multiput(722.00,133.34)(6.500,2.000){2}{\rule{1.566pt}{0.800pt}}
\put(735,136.34){\rule{2.891pt}{0.800pt}}
\multiput(735.00,135.34)(6.000,2.000){2}{\rule{1.445pt}{0.800pt}}
\put(747,138.84){\rule{3.132pt}{0.800pt}}
\multiput(747.00,137.34)(6.500,3.000){2}{\rule{1.566pt}{0.800pt}}
\put(760,142.34){\rule{2.600pt}{0.800pt}}
\multiput(760.00,140.34)(6.604,4.000){2}{\rule{1.300pt}{0.800pt}}
\multiput(772.00,147.38)(1.768,0.560){3}{\rule{2.280pt}{0.135pt}}
\multiput(772.00,144.34)(8.268,5.000){2}{\rule{1.140pt}{0.800pt}}
\multiput(785.00,152.38)(1.600,0.560){3}{\rule{2.120pt}{0.135pt}}
\multiput(785.00,149.34)(7.600,5.000){2}{\rule{1.060pt}{0.800pt}}
\multiput(797.00,157.40)(1.000,0.526){7}{\rule{1.686pt}{0.127pt}}
\multiput(797.00,154.34)(9.501,7.000){2}{\rule{0.843pt}{0.800pt}}
\multiput(810.00,164.40)(1.000,0.526){7}{\rule{1.686pt}{0.127pt}}
\multiput(810.00,161.34)(9.501,7.000){2}{\rule{0.843pt}{0.800pt}}
\multiput(823.00,171.40)(0.774,0.520){9}{\rule{1.400pt}{0.125pt}}
\multiput(823.00,168.34)(9.094,8.000){2}{\rule{0.700pt}{0.800pt}}
\multiput(835.00,179.40)(0.847,0.520){9}{\rule{1.500pt}{0.125pt}}
\multiput(835.00,176.34)(9.887,8.000){2}{\rule{0.750pt}{0.800pt}}
\multiput(848.00,187.40)(0.599,0.514){13}{\rule{1.160pt}{0.124pt}}
\multiput(848.00,184.34)(9.592,10.000){2}{\rule{0.580pt}{0.800pt}}
\multiput(860.00,197.40)(0.654,0.514){13}{\rule{1.240pt}{0.124pt}}
\multiput(860.00,194.34)(10.426,10.000){2}{\rule{0.620pt}{0.800pt}}
\multiput(873.00,207.41)(0.491,0.511){17}{\rule{1.000pt}{0.123pt}}
\multiput(873.00,204.34)(9.924,12.000){2}{\rule{0.500pt}{0.800pt}}
\multiput(885.00,219.41)(0.536,0.511){17}{\rule{1.067pt}{0.123pt}}
\multiput(885.00,216.34)(10.786,12.000){2}{\rule{0.533pt}{0.800pt}}
\multiput(898.00,231.41)(0.492,0.509){19}{\rule{1.000pt}{0.123pt}}
\multiput(898.00,228.34)(10.924,13.000){2}{\rule{0.500pt}{0.800pt}}
\multiput(912.41,243.00)(0.511,0.536){17}{\rule{0.123pt}{1.067pt}}
\multiput(909.34,243.00)(12.000,10.786){2}{\rule{0.800pt}{0.533pt}}
\multiput(924.41,256.00)(0.509,0.574){19}{\rule{0.123pt}{1.123pt}}
\multiput(921.34,256.00)(13.000,12.669){2}{\rule{0.800pt}{0.562pt}}
\multiput(937.41,271.00)(0.511,0.626){17}{\rule{0.123pt}{1.200pt}}
\multiput(934.34,271.00)(12.000,12.509){2}{\rule{0.800pt}{0.600pt}}
\multiput(949.41,286.00)(0.509,0.616){19}{\rule{0.123pt}{1.185pt}}
\multiput(946.34,286.00)(13.000,13.541){2}{\rule{0.800pt}{0.592pt}}
\multiput(962.41,302.00)(0.509,0.657){19}{\rule{0.123pt}{1.246pt}}
\multiput(959.34,302.00)(13.000,14.414){2}{\rule{0.800pt}{0.623pt}}
\multiput(975.41,319.00)(0.511,0.762){17}{\rule{0.123pt}{1.400pt}}
\multiput(972.34,319.00)(12.000,15.094){2}{\rule{0.800pt}{0.700pt}}
\multiput(987.41,337.00)(0.509,0.740){19}{\rule{0.123pt}{1.369pt}}
\multiput(984.34,337.00)(13.000,16.158){2}{\rule{0.800pt}{0.685pt}}
\multiput(1000.41,356.00)(0.511,0.807){17}{\rule{0.123pt}{1.467pt}}
\multiput(997.34,356.00)(12.000,15.956){2}{\rule{0.800pt}{0.733pt}}
\multiput(1012.41,375.00)(0.509,0.781){19}{\rule{0.123pt}{1.431pt}}
\multiput(1009.34,375.00)(13.000,17.030){2}{\rule{0.800pt}{0.715pt}}
\multiput(1025.41,395.00)(0.511,0.897){17}{\rule{0.123pt}{1.600pt}}
\multiput(1022.34,395.00)(12.000,17.679){2}{\rule{0.800pt}{0.800pt}}
\multiput(1037.41,416.00)(0.509,0.823){19}{\rule{0.123pt}{1.492pt}}
\multiput(1034.34,416.00)(13.000,17.903){2}{\rule{0.800pt}{0.746pt}}
\multiput(1050.41,437.00)(0.509,0.864){19}{\rule{0.123pt}{1.554pt}}
\multiput(1047.34,437.00)(13.000,18.775){2}{\rule{0.800pt}{0.777pt}}
\multiput(1063.41,459.00)(0.511,0.988){17}{\rule{0.123pt}{1.733pt}}
\multiput(1060.34,459.00)(12.000,19.402){2}{\rule{0.800pt}{0.867pt}}
\multiput(1075.41,482.00)(0.509,0.905){19}{\rule{0.123pt}{1.615pt}}
\multiput(1072.34,482.00)(13.000,19.647){2}{\rule{0.800pt}{0.808pt}}
\multiput(1088.41,505.00)(0.511,0.988){17}{\rule{0.123pt}{1.733pt}}
\multiput(1085.34,505.00)(12.000,19.402){2}{\rule{0.800pt}{0.867pt}}
\multiput(1100.41,528.00)(0.509,0.947){19}{\rule{0.123pt}{1.677pt}}
\multiput(1097.34,528.00)(13.000,20.519){2}{\rule{0.800pt}{0.838pt}}
\multiput(1113.41,552.00)(0.509,0.947){19}{\rule{0.123pt}{1.677pt}}
\multiput(1110.34,552.00)(13.000,20.519){2}{\rule{0.800pt}{0.838pt}}
\multiput(1126.41,576.00)(0.511,0.988){17}{\rule{0.123pt}{1.733pt}}
\multiput(1123.34,576.00)(12.000,19.402){2}{\rule{0.800pt}{0.867pt}}
\multiput(1138.41,599.00)(0.509,0.947){19}{\rule{0.123pt}{1.677pt}}
\multiput(1135.34,599.00)(13.000,20.519){2}{\rule{0.800pt}{0.838pt}}
\multiput(1151.41,623.00)(0.511,0.988){17}{\rule{0.123pt}{1.733pt}}
\multiput(1148.34,623.00)(12.000,19.402){2}{\rule{0.800pt}{0.867pt}}
\multiput(1163.41,646.00)(0.509,0.905){19}{\rule{0.123pt}{1.615pt}}
\multiput(1160.34,646.00)(13.000,19.647){2}{\rule{0.800pt}{0.808pt}}
\multiput(1176.41,669.00)(0.511,0.988){17}{\rule{0.123pt}{1.733pt}}
\multiput(1173.34,669.00)(12.000,19.402){2}{\rule{0.800pt}{0.867pt}}
\multiput(1188.41,692.00)(0.509,0.864){19}{\rule{0.123pt}{1.554pt}}
\multiput(1185.34,692.00)(13.000,18.775){2}{\rule{0.800pt}{0.777pt}}
\multiput(1201.41,714.00)(0.509,0.781){19}{\rule{0.123pt}{1.431pt}}
\multiput(1198.34,714.00)(13.000,17.030){2}{\rule{0.800pt}{0.715pt}}
\multiput(1214.41,734.00)(0.511,0.852){17}{\rule{0.123pt}{1.533pt}}
\multiput(1211.34,734.00)(12.000,16.817){2}{\rule{0.800pt}{0.767pt}}
\multiput(1226.41,754.00)(0.509,0.740){19}{\rule{0.123pt}{1.369pt}}
\multiput(1223.34,754.00)(13.000,16.158){2}{\rule{0.800pt}{0.685pt}}
\multiput(1239.41,773.00)(0.511,0.671){17}{\rule{0.123pt}{1.267pt}}
\multiput(1236.34,773.00)(12.000,13.371){2}{\rule{0.800pt}{0.633pt}}
\multiput(1251.41,789.00)(0.509,0.616){19}{\rule{0.123pt}{1.185pt}}
\multiput(1248.34,789.00)(13.000,13.541){2}{\rule{0.800pt}{0.592pt}}
\multiput(1264.41,805.00)(0.511,0.536){17}{\rule{0.123pt}{1.067pt}}
\multiput(1261.34,805.00)(12.000,10.786){2}{\rule{0.800pt}{0.533pt}}
\multiput(1275.00,819.40)(0.589,0.512){15}{\rule{1.145pt}{0.123pt}}
\multiput(1275.00,816.34)(10.623,11.000){2}{\rule{0.573pt}{0.800pt}}
\multiput(1288.00,830.40)(0.737,0.516){11}{\rule{1.356pt}{0.124pt}}
\multiput(1288.00,827.34)(10.186,9.000){2}{\rule{0.678pt}{0.800pt}}
\multiput(1301.00,839.40)(0.913,0.526){7}{\rule{1.571pt}{0.127pt}}
\multiput(1301.00,836.34)(8.738,7.000){2}{\rule{0.786pt}{0.800pt}}
\multiput(1313.00,846.38)(1.768,0.560){3}{\rule{2.280pt}{0.135pt}}
\multiput(1313.00,843.34)(8.268,5.000){2}{\rule{1.140pt}{0.800pt}}
\put(1326,849.34){\rule{2.891pt}{0.800pt}}
\multiput(1326.00,848.34)(6.000,2.000){2}{\rule{1.445pt}{0.800pt}}
\put(1338,849.84){\rule{3.132pt}{0.800pt}}
\multiput(1338.00,850.34)(6.500,-1.000){2}{\rule{1.566pt}{0.800pt}}
\put(1351,847.84){\rule{3.132pt}{0.800pt}}
\multiput(1351.00,849.34)(6.500,-3.000){2}{\rule{1.566pt}{0.800pt}}
\multiput(1364.00,846.06)(1.600,-0.560){3}{\rule{2.120pt}{0.135pt}}
\multiput(1364.00,846.34)(7.600,-5.000){2}{\rule{1.060pt}{0.800pt}}
\multiput(1376.00,841.08)(1.000,-0.526){7}{\rule{1.686pt}{0.127pt}}
\multiput(1376.00,841.34)(9.501,-7.000){2}{\rule{0.843pt}{0.800pt}}
\multiput(1389.00,834.08)(0.599,-0.514){13}{\rule{1.160pt}{0.124pt}}
\multiput(1389.00,834.34)(9.592,-10.000){2}{\rule{0.580pt}{0.800pt}}
\multiput(1401.00,824.08)(0.536,-0.511){17}{\rule{1.067pt}{0.123pt}}
\multiput(1401.00,824.34)(10.786,-12.000){2}{\rule{0.533pt}{0.800pt}}
\multiput(1415.41,809.30)(0.511,-0.581){17}{\rule{0.123pt}{1.133pt}}
\multiput(1412.34,811.65)(12.000,-11.648){2}{\rule{0.800pt}{0.567pt}}
\multiput(1427.41,795.34)(0.509,-0.574){19}{\rule{0.123pt}{1.123pt}}
\multiput(1424.34,797.67)(13.000,-12.669){2}{\rule{0.800pt}{0.562pt}}
\put(697.0,134.0){\rule[-0.400pt]{2.891pt}{0.800pt}}
\multiput(260.40,842.90)(0.514,-2.658){13}{\rule{0.124pt}{4.120pt}}
\multiput(257.34,851.45)(10.000,-40.449){2}{\rule{0.800pt}{2.060pt}}
\multiput(270.41,795.35)(0.509,-2.353){19}{\rule{0.123pt}{3.769pt}}
\multiput(267.34,803.18)(13.000,-50.177){2}{\rule{0.800pt}{1.885pt}}
\multiput(283.41,736.95)(0.511,-2.434){17}{\rule{0.123pt}{3.867pt}}
\multiput(280.34,744.97)(12.000,-46.975){2}{\rule{0.800pt}{1.933pt}}
\multiput(295.41,683.38)(0.509,-2.188){19}{\rule{0.123pt}{3.523pt}}
\multiput(292.34,690.69)(13.000,-46.688){2}{\rule{0.800pt}{1.762pt}}
\multiput(308.41,629.06)(0.511,-2.254){17}{\rule{0.123pt}{3.600pt}}
\multiput(305.34,636.53)(12.000,-43.528){2}{\rule{0.800pt}{1.800pt}}
\multiput(320.41,579.91)(0.509,-1.939){19}{\rule{0.123pt}{3.154pt}}
\multiput(317.34,586.45)(13.000,-41.454){2}{\rule{0.800pt}{1.577pt}}
\multiput(333.41,532.42)(0.509,-1.857){19}{\rule{0.123pt}{3.031pt}}
\multiput(330.34,538.71)(13.000,-39.709){2}{\rule{0.800pt}{1.515pt}}
\multiput(346.41,485.99)(0.511,-1.937){17}{\rule{0.123pt}{3.133pt}}
\multiput(343.34,492.50)(12.000,-37.497){2}{\rule{0.800pt}{1.567pt}}
\multiput(358.41,443.70)(0.509,-1.650){19}{\rule{0.123pt}{2.723pt}}
\multiput(355.34,449.35)(13.000,-35.348){2}{\rule{0.800pt}{1.362pt}}
\multiput(371.41,402.38)(0.511,-1.711){17}{\rule{0.123pt}{2.800pt}}
\multiput(368.34,408.19)(12.000,-33.188){2}{\rule{0.800pt}{1.400pt}}
\multiput(383.41,364.97)(0.509,-1.443){19}{\rule{0.123pt}{2.415pt}}
\multiput(380.34,369.99)(13.000,-30.987){2}{\rule{0.800pt}{1.208pt}}
\multiput(396.41,328.48)(0.511,-1.530){17}{\rule{0.123pt}{2.533pt}}
\multiput(393.34,333.74)(12.000,-29.742){2}{\rule{0.800pt}{1.267pt}}
\multiput(408.41,295.00)(0.509,-1.278){19}{\rule{0.123pt}{2.169pt}}
\multiput(405.34,299.50)(13.000,-27.498){2}{\rule{0.800pt}{1.085pt}}
\multiput(421.41,263.51)(0.509,-1.195){19}{\rule{0.123pt}{2.046pt}}
\multiput(418.34,267.75)(13.000,-25.753){2}{\rule{0.800pt}{1.023pt}}
\multiput(434.41,233.42)(0.511,-1.214){17}{\rule{0.123pt}{2.067pt}}
\multiput(431.34,237.71)(12.000,-23.711){2}{\rule{0.800pt}{1.033pt}}
\multiput(446.41,206.27)(0.509,-1.071){19}{\rule{0.123pt}{1.862pt}}
\multiput(443.34,210.14)(13.000,-23.136){2}{\rule{0.800pt}{0.931pt}}
\multiput(459.41,179.53)(0.511,-1.033){17}{\rule{0.123pt}{1.800pt}}
\multiput(456.34,183.26)(12.000,-20.264){2}{\rule{0.800pt}{0.900pt}}
\multiput(471.41,156.29)(0.509,-0.905){19}{\rule{0.123pt}{1.615pt}}
\multiput(468.34,159.65)(13.000,-19.647){2}{\rule{0.800pt}{0.808pt}}
\multiput(484.40,133.52)(0.514,-0.877){13}{\rule{0.124pt}{1.560pt}}
\multiput(481.34,136.76)(10.000,-13.762){2}{\rule{0.800pt}{0.780pt}}
\multiput(913.40,123.00)(0.512,0.938){15}{\rule{0.123pt}{1.655pt}}
\multiput(910.34,123.00)(11.000,16.566){2}{\rule{0.800pt}{0.827pt}}
\multiput(924.41,143.00)(0.509,0.947){19}{\rule{0.123pt}{1.677pt}}
\multiput(921.34,143.00)(13.000,20.519){2}{\rule{0.800pt}{0.838pt}}
\multiput(937.41,167.00)(0.511,1.078){17}{\rule{0.123pt}{1.867pt}}
\multiput(934.34,167.00)(12.000,21.126){2}{\rule{0.800pt}{0.933pt}}
\multiput(949.41,192.00)(0.509,1.112){19}{\rule{0.123pt}{1.923pt}}
\multiput(946.34,192.00)(13.000,24.009){2}{\rule{0.800pt}{0.962pt}}
\multiput(962.41,220.00)(0.509,1.195){19}{\rule{0.123pt}{2.046pt}}
\multiput(959.34,220.00)(13.000,25.753){2}{\rule{0.800pt}{1.023pt}}
\multiput(975.41,250.00)(0.511,1.349){17}{\rule{0.123pt}{2.267pt}}
\multiput(972.34,250.00)(12.000,26.295){2}{\rule{0.800pt}{1.133pt}}
\multiput(987.41,281.00)(0.509,1.402){19}{\rule{0.123pt}{2.354pt}}
\multiput(984.34,281.00)(13.000,30.114){2}{\rule{0.800pt}{1.177pt}}
\multiput(1000.41,316.00)(0.511,1.575){17}{\rule{0.123pt}{2.600pt}}
\multiput(997.34,316.00)(12.000,30.604){2}{\rule{0.800pt}{1.300pt}}
\multiput(1012.41,352.00)(0.509,1.567){19}{\rule{0.123pt}{2.600pt}}
\multiput(1009.34,352.00)(13.000,33.604){2}{\rule{0.800pt}{1.300pt}}
\multiput(1025.41,391.00)(0.511,1.847){17}{\rule{0.123pt}{3.000pt}}
\multiput(1022.34,391.00)(12.000,35.773){2}{\rule{0.800pt}{1.500pt}}
\multiput(1037.41,433.00)(0.509,1.815){19}{\rule{0.123pt}{2.969pt}}
\multiput(1034.34,433.00)(13.000,38.837){2}{\rule{0.800pt}{1.485pt}}
\multiput(1050.41,478.00)(0.509,1.939){19}{\rule{0.123pt}{3.154pt}}
\multiput(1047.34,478.00)(13.000,41.454){2}{\rule{0.800pt}{1.577pt}}
\multiput(1063.41,526.00)(0.511,2.254){17}{\rule{0.123pt}{3.600pt}}
\multiput(1060.34,526.00)(12.000,43.528){2}{\rule{0.800pt}{1.800pt}}
\multiput(1075.41,577.00)(0.509,2.188){19}{\rule{0.123pt}{3.523pt}}
\multiput(1072.34,577.00)(13.000,46.688){2}{\rule{0.800pt}{1.762pt}}
\multiput(1088.41,631.00)(0.511,2.525){17}{\rule{0.123pt}{4.000pt}}
\multiput(1085.34,631.00)(12.000,48.698){2}{\rule{0.800pt}{2.000pt}}
\multiput(1100.41,688.00)(0.509,2.518){19}{\rule{0.123pt}{4.015pt}}
\multiput(1097.34,688.00)(13.000,53.666){2}{\rule{0.800pt}{2.008pt}}
\multiput(1113.41,750.00)(0.509,2.601){19}{\rule{0.123pt}{4.138pt}}
\multiput(1110.34,750.00)(13.000,55.410){2}{\rule{0.800pt}{2.069pt}}
\multiput(1126.40,814.00)(0.520,3.259){9}{\rule{0.125pt}{4.800pt}}
\multiput(1123.34,814.00)(8.000,36.037){2}{\rule{0.800pt}{2.400pt}}
\sbox{\plotpoint}{\rule[-0.200pt]{0.400pt}{0.400pt}}%
\put(181,372){\usebox{\plotpoint}}
\put(181.00,372.00){\usebox{\plotpoint}}
\put(201.76,372.00){\usebox{\plotpoint}}
\put(222.51,372.00){\usebox{\plotpoint}}
\put(243.27,372.00){\usebox{\plotpoint}}
\put(264.02,372.00){\usebox{\plotpoint}}
\put(284.78,372.00){\usebox{\plotpoint}}
\put(305.53,372.00){\usebox{\plotpoint}}
\put(326.29,372.00){\usebox{\plotpoint}}
\put(347.04,372.00){\usebox{\plotpoint}}
\put(367.80,372.00){\usebox{\plotpoint}}
\put(388.55,372.00){\usebox{\plotpoint}}
\put(409.31,372.00){\usebox{\plotpoint}}
\put(430.07,372.00){\usebox{\plotpoint}}
\put(450.82,372.00){\usebox{\plotpoint}}
\put(471.58,372.00){\usebox{\plotpoint}}
\put(492.33,372.00){\usebox{\plotpoint}}
\put(513.09,372.00){\usebox{\plotpoint}}
\put(533.84,372.00){\usebox{\plotpoint}}
\put(554.60,372.00){\usebox{\plotpoint}}
\put(575.35,372.00){\usebox{\plotpoint}}
\put(596.11,372.00){\usebox{\plotpoint}}
\put(616.87,372.00){\usebox{\plotpoint}}
\put(637.62,372.00){\usebox{\plotpoint}}
\put(658.38,372.00){\usebox{\plotpoint}}
\put(679.13,372.00){\usebox{\plotpoint}}
\put(699.89,372.00){\usebox{\plotpoint}}
\put(720.64,372.00){\usebox{\plotpoint}}
\put(741.40,372.00){\usebox{\plotpoint}}
\put(762.15,372.00){\usebox{\plotpoint}}
\put(782.91,372.00){\usebox{\plotpoint}}
\put(803.66,372.00){\usebox{\plotpoint}}
\put(824.42,372.00){\usebox{\plotpoint}}
\put(845.18,372.00){\usebox{\plotpoint}}
\put(865.93,372.00){\usebox{\plotpoint}}
\put(886.69,372.00){\usebox{\plotpoint}}
\put(907.44,372.00){\usebox{\plotpoint}}
\put(928.20,372.00){\usebox{\plotpoint}}
\put(948.95,372.00){\usebox{\plotpoint}}
\put(969.71,372.00){\usebox{\plotpoint}}
\put(990.46,372.00){\usebox{\plotpoint}}
\put(1011.22,372.00){\usebox{\plotpoint}}
\put(1031.98,372.00){\usebox{\plotpoint}}
\put(1052.73,372.00){\usebox{\plotpoint}}
\put(1073.49,372.00){\usebox{\plotpoint}}
\put(1094.24,372.00){\usebox{\plotpoint}}
\put(1115.00,372.00){\usebox{\plotpoint}}
\put(1135.75,372.00){\usebox{\plotpoint}}
\put(1156.51,372.00){\usebox{\plotpoint}}
\put(1177.26,372.00){\usebox{\plotpoint}}
\put(1198.02,372.00){\usebox{\plotpoint}}
\put(1218.77,372.00){\usebox{\plotpoint}}
\put(1239.53,372.00){\usebox{\plotpoint}}
\put(1260.29,372.00){\usebox{\plotpoint}}
\put(1281.04,372.00){\usebox{\plotpoint}}
\put(1301.80,372.00){\usebox{\plotpoint}}
\put(1322.55,372.00){\usebox{\plotpoint}}
\put(1343.31,372.00){\usebox{\plotpoint}}
\put(1364.06,372.00){\usebox{\plotpoint}}
\put(1384.82,372.00){\usebox{\plotpoint}}
\put(1405.57,372.00){\usebox{\plotpoint}}
\put(1426.33,372.00){\usebox{\plotpoint}}
\put(1439,372){\usebox{\plotpoint}}
\put(181,409){\usebox{\plotpoint}}
\put(181.00,409.00){\usebox{\plotpoint}}
\put(201.76,409.00){\usebox{\plotpoint}}
\put(222.51,409.00){\usebox{\plotpoint}}
\put(243.27,409.00){\usebox{\plotpoint}}
\put(264.02,409.00){\usebox{\plotpoint}}
\put(284.78,409.00){\usebox{\plotpoint}}
\put(305.53,409.00){\usebox{\plotpoint}}
\put(326.29,409.00){\usebox{\plotpoint}}
\put(347.04,409.00){\usebox{\plotpoint}}
\put(367.80,409.00){\usebox{\plotpoint}}
\put(388.55,409.00){\usebox{\plotpoint}}
\put(409.31,409.00){\usebox{\plotpoint}}
\put(430.07,409.00){\usebox{\plotpoint}}
\put(450.82,409.00){\usebox{\plotpoint}}
\put(471.58,409.00){\usebox{\plotpoint}}
\put(492.33,409.00){\usebox{\plotpoint}}
\put(513.09,409.00){\usebox{\plotpoint}}
\put(533.84,409.00){\usebox{\plotpoint}}
\put(554.60,409.00){\usebox{\plotpoint}}
\put(575.35,409.00){\usebox{\plotpoint}}
\put(596.11,409.00){\usebox{\plotpoint}}
\put(616.87,409.00){\usebox{\plotpoint}}
\put(637.62,409.00){\usebox{\plotpoint}}
\put(658.38,409.00){\usebox{\plotpoint}}
\put(679.13,409.00){\usebox{\plotpoint}}
\put(699.89,409.00){\usebox{\plotpoint}}
\put(720.64,409.00){\usebox{\plotpoint}}
\put(741.40,409.00){\usebox{\plotpoint}}
\put(762.15,409.00){\usebox{\plotpoint}}
\put(782.91,409.00){\usebox{\plotpoint}}
\put(803.66,409.00){\usebox{\plotpoint}}
\put(824.42,409.00){\usebox{\plotpoint}}
\put(845.18,409.00){\usebox{\plotpoint}}
\put(865.93,409.00){\usebox{\plotpoint}}
\put(886.69,409.00){\usebox{\plotpoint}}
\put(907.44,409.00){\usebox{\plotpoint}}
\put(928.20,409.00){\usebox{\plotpoint}}
\put(948.95,409.00){\usebox{\plotpoint}}
\put(969.71,409.00){\usebox{\plotpoint}}
\put(990.46,409.00){\usebox{\plotpoint}}
\put(1011.22,409.00){\usebox{\plotpoint}}
\put(1031.98,409.00){\usebox{\plotpoint}}
\put(1052.73,409.00){\usebox{\plotpoint}}
\put(1073.49,409.00){\usebox{\plotpoint}}
\put(1094.24,409.00){\usebox{\plotpoint}}
\put(1115.00,409.00){\usebox{\plotpoint}}
\put(1135.75,409.00){\usebox{\plotpoint}}
\put(1156.51,409.00){\usebox{\plotpoint}}
\put(1177.26,409.00){\usebox{\plotpoint}}
\put(1198.02,409.00){\usebox{\plotpoint}}
\put(1218.77,409.00){\usebox{\plotpoint}}
\put(1239.53,409.00){\usebox{\plotpoint}}
\put(1260.29,409.00){\usebox{\plotpoint}}
\put(1281.04,409.00){\usebox{\plotpoint}}
\put(1301.80,409.00){\usebox{\plotpoint}}
\put(1322.55,409.00){\usebox{\plotpoint}}
\put(1343.31,409.00){\usebox{\plotpoint}}
\put(1364.06,409.00){\usebox{\plotpoint}}
\put(1384.82,409.00){\usebox{\plotpoint}}
\put(1405.57,409.00){\usebox{\plotpoint}}
\put(1426.33,409.00){\usebox{\plotpoint}}
\put(1439,409){\usebox{\plotpoint}}
\end{picture}

%% file: newfig4.tex
\setlength{\unitlength}{0.240900pt}
\ifx\plotpoint\undefined\newsavebox{\plotpoint}\fi
\sbox{\plotpoint}{\rule[-0.200pt]{0.400pt}{0.400pt}}%
\begin{picture}(1500,900)(0,0)
\font\gnuplot=cmr10 at 10pt
\gnuplot
\sbox{\plotpoint}{\rule[-0.200pt]{0.400pt}{0.400pt}}%
\put(181.0,123.0){\rule[-0.200pt]{4.818pt}{0.400pt}}
\put(161,123){\makebox(0,0)[r]{14}}
\put(1419.0,123.0){\rule[-0.200pt]{4.818pt}{0.400pt}}
\put(181.0,228.0){\rule[-0.200pt]{4.818pt}{0.400pt}}
\put(161,228){\makebox(0,0)[r]{14.5}}
\put(1419.0,228.0){\rule[-0.200pt]{4.818pt}{0.400pt}}
\put(181.0,334.0){\rule[-0.200pt]{4.818pt}{0.400pt}}
\put(161,334){\makebox(0,0)[r]{15}}
\put(1419.0,334.0){\rule[-0.200pt]{4.818pt}{0.400pt}}
\put(181.0,439.0){\rule[-0.200pt]{4.818pt}{0.400pt}}
\put(161,439){\makebox(0,0)[r]{15.5}}
\put(1419.0,439.0){\rule[-0.200pt]{4.818pt}{0.400pt}}
\put(181.0,544.0){\rule[-0.200pt]{4.818pt}{0.400pt}}
\put(161,544){\makebox(0,0)[r]{16}}
\put(1419.0,544.0){\rule[-0.200pt]{4.818pt}{0.400pt}}
\put(181.0,649.0){\rule[-0.200pt]{4.818pt}{0.400pt}}
\put(161,649){\makebox(0,0)[r]{16.5}}
\put(1419.0,649.0){\rule[-0.200pt]{4.818pt}{0.400pt}}
\put(181.0,755.0){\rule[-0.200pt]{4.818pt}{0.400pt}}
\put(161,755){\makebox(0,0)[r]{17}}
\put(1419.0,755.0){\rule[-0.200pt]{4.818pt}{0.400pt}}
\put(181.0,860.0){\rule[-0.200pt]{4.818pt}{0.400pt}}
\put(161,860){\makebox(0,0)[r]{17.5}}
\put(1419.0,860.0){\rule[-0.200pt]{4.818pt}{0.400pt}}
\put(181.0,123.0){\rule[-0.200pt]{0.400pt}{4.818pt}}
\put(181,82){\makebox(0,0){0}}
\put(181.0,840.0){\rule[-0.200pt]{0.400pt}{4.818pt}}
\put(496.0,123.0){\rule[-0.200pt]{0.400pt}{4.818pt}}
\put(496,82){\makebox(0,0){0.5}}
\put(496.0,840.0){\rule[-0.200pt]{0.400pt}{4.818pt}}
\put(810.0,123.0){\rule[-0.200pt]{0.400pt}{4.818pt}}
\put(810,82){\makebox(0,0){1}}
\put(810.0,840.0){\rule[-0.200pt]{0.400pt}{4.818pt}}
\put(1125.0,123.0){\rule[-0.200pt]{0.400pt}{4.818pt}}
\put(1125,82){\makebox(0,0){1.5}}
\put(1125.0,840.0){\rule[-0.200pt]{0.400pt}{4.818pt}}
\put(1439.0,123.0){\rule[-0.200pt]{0.400pt}{4.818pt}}
\put(1439,82){\makebox(0,0){2}}
\put(1439.0,840.0){\rule[-0.200pt]{0.400pt}{4.818pt}}
\put(181.0,123.0){\rule[-0.200pt]{303.052pt}{0.400pt}}
\put(1439.0,123.0){\rule[-0.200pt]{0.400pt}{177.543pt}}
\put(181.0,860.0){\rule[-0.200pt]{303.052pt}{0.400pt}}
\put(40,491){\makebox(0,0){$T$}}
\put(810,21){\makebox(0,0){$\theta/\pi$}}
\put(1376,586){\makebox(0,0)[r]{$x=0.01$}}
\put(1426,713){\makebox(0,0)[r]{$x=0.05$}}
\put(1250,839){\makebox(0,0)[r]{$x=0.1$}}
\put(181.0,123.0){\rule[-0.200pt]{0.400pt}{177.543pt}}
\sbox{\plotpoint}{\rule[-0.400pt]{0.800pt}{0.800pt}}%
\put(181,617){\usebox{\plotpoint}}
\put(181,614.84){\rule{3.132pt}{0.800pt}}
\multiput(181.00,615.34)(6.500,-1.000){2}{\rule{1.566pt}{0.800pt}}
\put(194,613.84){\rule{2.891pt}{0.800pt}}
\multiput(194.00,614.34)(6.000,-1.000){2}{\rule{1.445pt}{0.800pt}}
\put(206,612.84){\rule{3.132pt}{0.800pt}}
\multiput(206.00,613.34)(6.500,-1.000){2}{\rule{1.566pt}{0.800pt}}
\put(219,611.84){\rule{2.891pt}{0.800pt}}
\multiput(219.00,612.34)(6.000,-1.000){2}{\rule{1.445pt}{0.800pt}}
\put(231,610.84){\rule{3.132pt}{0.800pt}}
\multiput(231.00,611.34)(6.500,-1.000){2}{\rule{1.566pt}{0.800pt}}
\put(244,609.84){\rule{2.891pt}{0.800pt}}
\multiput(244.00,610.34)(6.000,-1.000){2}{\rule{1.445pt}{0.800pt}}
\put(256,608.84){\rule{3.132pt}{0.800pt}}
\multiput(256.00,609.34)(6.500,-1.000){2}{\rule{1.566pt}{0.800pt}}
\put(269,607.34){\rule{3.132pt}{0.800pt}}
\multiput(269.00,608.34)(6.500,-2.000){2}{\rule{1.566pt}{0.800pt}}
\put(282,605.34){\rule{2.891pt}{0.800pt}}
\multiput(282.00,606.34)(6.000,-2.000){2}{\rule{1.445pt}{0.800pt}}
\put(294,603.84){\rule{3.132pt}{0.800pt}}
\multiput(294.00,604.34)(6.500,-1.000){2}{\rule{1.566pt}{0.800pt}}
\put(307,602.34){\rule{2.891pt}{0.800pt}}
\multiput(307.00,603.34)(6.000,-2.000){2}{\rule{1.445pt}{0.800pt}}
\put(319,600.34){\rule{3.132pt}{0.800pt}}
\multiput(319.00,601.34)(6.500,-2.000){2}{\rule{1.566pt}{0.800pt}}
\put(332,598.34){\rule{3.132pt}{0.800pt}}
\multiput(332.00,599.34)(6.500,-2.000){2}{\rule{1.566pt}{0.800pt}}
\put(345,596.34){\rule{2.891pt}{0.800pt}}
\multiput(345.00,597.34)(6.000,-2.000){2}{\rule{1.445pt}{0.800pt}}
\put(357,594.34){\rule{3.132pt}{0.800pt}}
\multiput(357.00,595.34)(6.500,-2.000){2}{\rule{1.566pt}{0.800pt}}
\put(370,591.84){\rule{2.891pt}{0.800pt}}
\multiput(370.00,593.34)(6.000,-3.000){2}{\rule{1.445pt}{0.800pt}}
\put(382,589.34){\rule{3.132pt}{0.800pt}}
\multiput(382.00,590.34)(6.500,-2.000){2}{\rule{1.566pt}{0.800pt}}
\put(395,587.34){\rule{2.891pt}{0.800pt}}
\multiput(395.00,588.34)(6.000,-2.000){2}{\rule{1.445pt}{0.800pt}}
\put(407,584.84){\rule{3.132pt}{0.800pt}}
\multiput(407.00,586.34)(6.500,-3.000){2}{\rule{1.566pt}{0.800pt}}
\put(420,582.34){\rule{3.132pt}{0.800pt}}
\multiput(420.00,583.34)(6.500,-2.000){2}{\rule{1.566pt}{0.800pt}}
\put(433,580.34){\rule{2.891pt}{0.800pt}}
\multiput(433.00,581.34)(6.000,-2.000){2}{\rule{1.445pt}{0.800pt}}
\put(445,577.84){\rule{3.132pt}{0.800pt}}
\multiput(445.00,579.34)(6.500,-3.000){2}{\rule{1.566pt}{0.800pt}}
\put(458,575.34){\rule{2.891pt}{0.800pt}}
\multiput(458.00,576.34)(6.000,-2.000){2}{\rule{1.445pt}{0.800pt}}
\put(470,572.84){\rule{3.132pt}{0.800pt}}
\multiput(470.00,574.34)(6.500,-3.000){2}{\rule{1.566pt}{0.800pt}}
\put(483,570.34){\rule{3.132pt}{0.800pt}}
\multiput(483.00,571.34)(6.500,-2.000){2}{\rule{1.566pt}{0.800pt}}
\put(496,568.34){\rule{2.891pt}{0.800pt}}
\multiput(496.00,569.34)(6.000,-2.000){2}{\rule{1.445pt}{0.800pt}}
\put(508,565.84){\rule{3.132pt}{0.800pt}}
\multiput(508.00,567.34)(6.500,-3.000){2}{\rule{1.566pt}{0.800pt}}
\put(521,563.34){\rule{2.891pt}{0.800pt}}
\multiput(521.00,564.34)(6.000,-2.000){2}{\rule{1.445pt}{0.800pt}}
\put(533,561.34){\rule{3.132pt}{0.800pt}}
\multiput(533.00,562.34)(6.500,-2.000){2}{\rule{1.566pt}{0.800pt}}
\put(546,559.34){\rule{2.891pt}{0.800pt}}
\multiput(546.00,560.34)(6.000,-2.000){2}{\rule{1.445pt}{0.800pt}}
\put(558,557.34){\rule{3.132pt}{0.800pt}}
\multiput(558.00,558.34)(6.500,-2.000){2}{\rule{1.566pt}{0.800pt}}
\put(571,555.34){\rule{3.132pt}{0.800pt}}
\multiput(571.00,556.34)(6.500,-2.000){2}{\rule{1.566pt}{0.800pt}}
\put(584,553.34){\rule{2.891pt}{0.800pt}}
\multiput(584.00,554.34)(6.000,-2.000){2}{\rule{1.445pt}{0.800pt}}
\put(596,551.34){\rule{3.132pt}{0.800pt}}
\multiput(596.00,552.34)(6.500,-2.000){2}{\rule{1.566pt}{0.800pt}}
\put(609,549.34){\rule{2.891pt}{0.800pt}}
\multiput(609.00,550.34)(6.000,-2.000){2}{\rule{1.445pt}{0.800pt}}
\put(621,547.84){\rule{3.132pt}{0.800pt}}
\multiput(621.00,548.34)(6.500,-1.000){2}{\rule{1.566pt}{0.800pt}}
\put(634,546.34){\rule{2.891pt}{0.800pt}}
\multiput(634.00,547.34)(6.000,-2.000){2}{\rule{1.445pt}{0.800pt}}
\put(646,544.84){\rule{3.132pt}{0.800pt}}
\multiput(646.00,545.34)(6.500,-1.000){2}{\rule{1.566pt}{0.800pt}}
\put(659,543.84){\rule{3.132pt}{0.800pt}}
\multiput(659.00,544.34)(6.500,-1.000){2}{\rule{1.566pt}{0.800pt}}
\put(672,542.84){\rule{2.891pt}{0.800pt}}
\multiput(672.00,543.34)(6.000,-1.000){2}{\rule{1.445pt}{0.800pt}}
\put(684,541.84){\rule{3.132pt}{0.800pt}}
\multiput(684.00,542.34)(6.500,-1.000){2}{\rule{1.566pt}{0.800pt}}
\put(697,540.84){\rule{2.891pt}{0.800pt}}
\multiput(697.00,541.34)(6.000,-1.000){2}{\rule{1.445pt}{0.800pt}}
\put(709,539.84){\rule{3.132pt}{0.800pt}}
\multiput(709.00,540.34)(6.500,-1.000){2}{\rule{1.566pt}{0.800pt}}
\put(810,539.84){\rule{3.132pt}{0.800pt}}
\multiput(810.00,539.34)(6.500,1.000){2}{\rule{1.566pt}{0.800pt}}
\put(823,540.84){\rule{2.891pt}{0.800pt}}
\multiput(823.00,540.34)(6.000,1.000){2}{\rule{1.445pt}{0.800pt}}
\put(835,541.84){\rule{3.132pt}{0.800pt}}
\multiput(835.00,541.34)(6.500,1.000){2}{\rule{1.566pt}{0.800pt}}
\put(848,542.84){\rule{2.891pt}{0.800pt}}
\multiput(848.00,542.34)(6.000,1.000){2}{\rule{1.445pt}{0.800pt}}
\put(860,543.84){\rule{3.132pt}{0.800pt}}
\multiput(860.00,543.34)(6.500,1.000){2}{\rule{1.566pt}{0.800pt}}
\put(873,544.84){\rule{2.891pt}{0.800pt}}
\multiput(873.00,544.34)(6.000,1.000){2}{\rule{1.445pt}{0.800pt}}
\put(885,546.34){\rule{3.132pt}{0.800pt}}
\multiput(885.00,545.34)(6.500,2.000){2}{\rule{1.566pt}{0.800pt}}
\put(898,547.84){\rule{3.132pt}{0.800pt}}
\multiput(898.00,547.34)(6.500,1.000){2}{\rule{1.566pt}{0.800pt}}
\put(911,549.34){\rule{2.891pt}{0.800pt}}
\multiput(911.00,548.34)(6.000,2.000){2}{\rule{1.445pt}{0.800pt}}
\put(923,551.34){\rule{3.132pt}{0.800pt}}
\multiput(923.00,550.34)(6.500,2.000){2}{\rule{1.566pt}{0.800pt}}
\put(936,553.34){\rule{2.891pt}{0.800pt}}
\multiput(936.00,552.34)(6.000,2.000){2}{\rule{1.445pt}{0.800pt}}
\put(948,555.34){\rule{3.132pt}{0.800pt}}
\multiput(948.00,554.34)(6.500,2.000){2}{\rule{1.566pt}{0.800pt}}
\put(961,557.34){\rule{3.132pt}{0.800pt}}
\multiput(961.00,556.34)(6.500,2.000){2}{\rule{1.566pt}{0.800pt}}
\put(974,559.34){\rule{2.891pt}{0.800pt}}
\multiput(974.00,558.34)(6.000,2.000){2}{\rule{1.445pt}{0.800pt}}
\put(986,561.34){\rule{3.132pt}{0.800pt}}
\multiput(986.00,560.34)(6.500,2.000){2}{\rule{1.566pt}{0.800pt}}
\put(999,563.34){\rule{2.891pt}{0.800pt}}
\multiput(999.00,562.34)(6.000,2.000){2}{\rule{1.445pt}{0.800pt}}
\put(1011,565.84){\rule{3.132pt}{0.800pt}}
\multiput(1011.00,564.34)(6.500,3.000){2}{\rule{1.566pt}{0.800pt}}
\put(1024,568.34){\rule{2.891pt}{0.800pt}}
\multiput(1024.00,567.34)(6.000,2.000){2}{\rule{1.445pt}{0.800pt}}
\put(1036,570.34){\rule{3.132pt}{0.800pt}}
\multiput(1036.00,569.34)(6.500,2.000){2}{\rule{1.566pt}{0.800pt}}
\put(1049,572.84){\rule{3.132pt}{0.800pt}}
\multiput(1049.00,571.34)(6.500,3.000){2}{\rule{1.566pt}{0.800pt}}
\put(1062,575.34){\rule{2.891pt}{0.800pt}}
\multiput(1062.00,574.34)(6.000,2.000){2}{\rule{1.445pt}{0.800pt}}
\put(1074,577.84){\rule{3.132pt}{0.800pt}}
\multiput(1074.00,576.34)(6.500,3.000){2}{\rule{1.566pt}{0.800pt}}
\put(1087,580.34){\rule{2.891pt}{0.800pt}}
\multiput(1087.00,579.34)(6.000,2.000){2}{\rule{1.445pt}{0.800pt}}
\put(1099,582.34){\rule{3.132pt}{0.800pt}}
\multiput(1099.00,581.34)(6.500,2.000){2}{\rule{1.566pt}{0.800pt}}
\put(1112,584.84){\rule{3.132pt}{0.800pt}}
\multiput(1112.00,583.34)(6.500,3.000){2}{\rule{1.566pt}{0.800pt}}
\put(1125,587.34){\rule{2.891pt}{0.800pt}}
\multiput(1125.00,586.34)(6.000,2.000){2}{\rule{1.445pt}{0.800pt}}
\put(1137,589.34){\rule{3.132pt}{0.800pt}}
\multiput(1137.00,588.34)(6.500,2.000){2}{\rule{1.566pt}{0.800pt}}
\put(1150,591.84){\rule{2.891pt}{0.800pt}}
\multiput(1150.00,590.34)(6.000,3.000){2}{\rule{1.445pt}{0.800pt}}
\put(1162,594.34){\rule{3.132pt}{0.800pt}}
\multiput(1162.00,593.34)(6.500,2.000){2}{\rule{1.566pt}{0.800pt}}
\put(1175,596.34){\rule{2.891pt}{0.800pt}}
\multiput(1175.00,595.34)(6.000,2.000){2}{\rule{1.445pt}{0.800pt}}
\put(1187,598.34){\rule{3.132pt}{0.800pt}}
\multiput(1187.00,597.34)(6.500,2.000){2}{\rule{1.566pt}{0.800pt}}
\put(1200,600.34){\rule{3.132pt}{0.800pt}}
\multiput(1200.00,599.34)(6.500,2.000){2}{\rule{1.566pt}{0.800pt}}
\put(1213,602.34){\rule{2.891pt}{0.800pt}}
\multiput(1213.00,601.34)(6.000,2.000){2}{\rule{1.445pt}{0.800pt}}
\put(1225,604.34){\rule{3.132pt}{0.800pt}}
\multiput(1225.00,603.34)(6.500,2.000){2}{\rule{1.566pt}{0.800pt}}
\put(1238,605.84){\rule{2.891pt}{0.800pt}}
\multiput(1238.00,605.34)(6.000,1.000){2}{\rule{1.445pt}{0.800pt}}
\put(1250,607.34){\rule{3.132pt}{0.800pt}}
\multiput(1250.00,606.34)(6.500,2.000){2}{\rule{1.566pt}{0.800pt}}
\put(1263,608.84){\rule{2.891pt}{0.800pt}}
\multiput(1263.00,608.34)(6.000,1.000){2}{\rule{1.445pt}{0.800pt}}
\put(1275,610.34){\rule{3.132pt}{0.800pt}}
\multiput(1275.00,609.34)(6.500,2.000){2}{\rule{1.566pt}{0.800pt}}
\put(1288,611.84){\rule{3.132pt}{0.800pt}}
\multiput(1288.00,611.34)(6.500,1.000){2}{\rule{1.566pt}{0.800pt}}
\put(1301,612.84){\rule{2.891pt}{0.800pt}}
\multiput(1301.00,612.34)(6.000,1.000){2}{\rule{1.445pt}{0.800pt}}
\put(1313,613.84){\rule{3.132pt}{0.800pt}}
\multiput(1313.00,613.34)(6.500,1.000){2}{\rule{1.566pt}{0.800pt}}
\put(722.0,541.0){\rule[-0.400pt]{21.199pt}{0.800pt}}
\put(1338,614.84){\rule{3.132pt}{0.800pt}}
\multiput(1338.00,614.34)(6.500,1.000){2}{\rule{1.566pt}{0.800pt}}
\put(1326.0,616.0){\rule[-0.400pt]{2.891pt}{0.800pt}}
\put(1364,615.84){\rule{2.891pt}{0.800pt}}
\multiput(1364.00,615.34)(6.000,1.000){2}{\rule{1.445pt}{0.800pt}}
\put(1351.0,617.0){\rule[-0.400pt]{3.132pt}{0.800pt}}
\put(1426,615.84){\rule{3.132pt}{0.800pt}}
\multiput(1426.00,616.34)(6.500,-1.000){2}{\rule{1.566pt}{0.800pt}}
\put(1376.0,618.0){\rule[-0.400pt]{12.045pt}{0.800pt}}
\put(181,761){\usebox{\plotpoint}}
\put(181,757.84){\rule{3.132pt}{0.800pt}}
\multiput(181.00,759.34)(6.500,-3.000){2}{\rule{1.566pt}{0.800pt}}
\put(194,754.84){\rule{2.891pt}{0.800pt}}
\multiput(194.00,756.34)(6.000,-3.000){2}{\rule{1.445pt}{0.800pt}}
\put(206,751.34){\rule{2.800pt}{0.800pt}}
\multiput(206.00,753.34)(7.188,-4.000){2}{\rule{1.400pt}{0.800pt}}
\multiput(219.00,749.06)(1.600,-0.560){3}{\rule{2.120pt}{0.135pt}}
\multiput(219.00,749.34)(7.600,-5.000){2}{\rule{1.060pt}{0.800pt}}
\multiput(231.00,744.07)(1.244,-0.536){5}{\rule{1.933pt}{0.129pt}}
\multiput(231.00,744.34)(8.987,-6.000){2}{\rule{0.967pt}{0.800pt}}
\multiput(244.00,738.07)(1.132,-0.536){5}{\rule{1.800pt}{0.129pt}}
\multiput(244.00,738.34)(8.264,-6.000){2}{\rule{0.900pt}{0.800pt}}
\multiput(256.00,732.07)(1.244,-0.536){5}{\rule{1.933pt}{0.129pt}}
\multiput(256.00,732.34)(8.987,-6.000){2}{\rule{0.967pt}{0.800pt}}
\multiput(269.00,726.08)(0.847,-0.520){9}{\rule{1.500pt}{0.125pt}}
\multiput(269.00,726.34)(9.887,-8.000){2}{\rule{0.750pt}{0.800pt}}
\multiput(282.00,718.08)(0.774,-0.520){9}{\rule{1.400pt}{0.125pt}}
\multiput(282.00,718.34)(9.094,-8.000){2}{\rule{0.700pt}{0.800pt}}
\multiput(294.00,710.08)(0.847,-0.520){9}{\rule{1.500pt}{0.125pt}}
\multiput(294.00,710.34)(9.887,-8.000){2}{\rule{0.750pt}{0.800pt}}
\multiput(307.00,702.08)(0.674,-0.516){11}{\rule{1.267pt}{0.124pt}}
\multiput(307.00,702.34)(9.371,-9.000){2}{\rule{0.633pt}{0.800pt}}
\multiput(319.00,693.08)(0.737,-0.516){11}{\rule{1.356pt}{0.124pt}}
\multiput(319.00,693.34)(10.186,-9.000){2}{\rule{0.678pt}{0.800pt}}
\multiput(332.00,684.08)(0.654,-0.514){13}{\rule{1.240pt}{0.124pt}}
\multiput(332.00,684.34)(10.426,-10.000){2}{\rule{0.620pt}{0.800pt}}
\multiput(345.00,674.08)(0.599,-0.514){13}{\rule{1.160pt}{0.124pt}}
\multiput(345.00,674.34)(9.592,-10.000){2}{\rule{0.580pt}{0.800pt}}
\multiput(357.00,664.08)(0.589,-0.512){15}{\rule{1.145pt}{0.123pt}}
\multiput(357.00,664.34)(10.623,-11.000){2}{\rule{0.573pt}{0.800pt}}
\multiput(370.00,653.08)(0.539,-0.512){15}{\rule{1.073pt}{0.123pt}}
\multiput(370.00,653.34)(9.774,-11.000){2}{\rule{0.536pt}{0.800pt}}
\multiput(382.00,642.08)(0.589,-0.512){15}{\rule{1.145pt}{0.123pt}}
\multiput(382.00,642.34)(10.623,-11.000){2}{\rule{0.573pt}{0.800pt}}
\multiput(395.00,631.08)(0.491,-0.511){17}{\rule{1.000pt}{0.123pt}}
\multiput(395.00,631.34)(9.924,-12.000){2}{\rule{0.500pt}{0.800pt}}
\multiput(407.00,619.08)(0.536,-0.511){17}{\rule{1.067pt}{0.123pt}}
\multiput(407.00,619.34)(10.786,-12.000){2}{\rule{0.533pt}{0.800pt}}
\multiput(420.00,607.08)(0.536,-0.511){17}{\rule{1.067pt}{0.123pt}}
\multiput(420.00,607.34)(10.786,-12.000){2}{\rule{0.533pt}{0.800pt}}
\multiput(433.00,595.08)(0.491,-0.511){17}{\rule{1.000pt}{0.123pt}}
\multiput(433.00,595.34)(9.924,-12.000){2}{\rule{0.500pt}{0.800pt}}
\multiput(445.00,583.08)(0.536,-0.511){17}{\rule{1.067pt}{0.123pt}}
\multiput(445.00,583.34)(10.786,-12.000){2}{\rule{0.533pt}{0.800pt}}
\multiput(458.00,571.08)(0.491,-0.511){17}{\rule{1.000pt}{0.123pt}}
\multiput(458.00,571.34)(9.924,-12.000){2}{\rule{0.500pt}{0.800pt}}
\multiput(470.00,559.08)(0.536,-0.511){17}{\rule{1.067pt}{0.123pt}}
\multiput(470.00,559.34)(10.786,-12.000){2}{\rule{0.533pt}{0.800pt}}
\multiput(483.00,547.08)(0.536,-0.511){17}{\rule{1.067pt}{0.123pt}}
\multiput(483.00,547.34)(10.786,-12.000){2}{\rule{0.533pt}{0.800pt}}
\multiput(496.00,535.08)(0.491,-0.511){17}{\rule{1.000pt}{0.123pt}}
\multiput(496.00,535.34)(9.924,-12.000){2}{\rule{0.500pt}{0.800pt}}
\multiput(508.00,523.08)(0.536,-0.511){17}{\rule{1.067pt}{0.123pt}}
\multiput(508.00,523.34)(10.786,-12.000){2}{\rule{0.533pt}{0.800pt}}
\multiput(521.00,511.08)(0.539,-0.512){15}{\rule{1.073pt}{0.123pt}}
\multiput(521.00,511.34)(9.774,-11.000){2}{\rule{0.536pt}{0.800pt}}
\multiput(533.00,500.08)(0.536,-0.511){17}{\rule{1.067pt}{0.123pt}}
\multiput(533.00,500.34)(10.786,-12.000){2}{\rule{0.533pt}{0.800pt}}
\multiput(546.00,488.08)(0.539,-0.512){15}{\rule{1.073pt}{0.123pt}}
\multiput(546.00,488.34)(9.774,-11.000){2}{\rule{0.536pt}{0.800pt}}
\multiput(558.00,477.08)(0.654,-0.514){13}{\rule{1.240pt}{0.124pt}}
\multiput(558.00,477.34)(10.426,-10.000){2}{\rule{0.620pt}{0.800pt}}
\multiput(571.00,467.08)(0.589,-0.512){15}{\rule{1.145pt}{0.123pt}}
\multiput(571.00,467.34)(10.623,-11.000){2}{\rule{0.573pt}{0.800pt}}
\multiput(584.00,456.08)(0.599,-0.514){13}{\rule{1.160pt}{0.124pt}}
\multiput(584.00,456.34)(9.592,-10.000){2}{\rule{0.580pt}{0.800pt}}
\multiput(596.00,446.08)(0.737,-0.516){11}{\rule{1.356pt}{0.124pt}}
\multiput(596.00,446.34)(10.186,-9.000){2}{\rule{0.678pt}{0.800pt}}
\multiput(609.00,437.08)(0.674,-0.516){11}{\rule{1.267pt}{0.124pt}}
\multiput(609.00,437.34)(9.371,-9.000){2}{\rule{0.633pt}{0.800pt}}
\multiput(621.00,428.08)(0.847,-0.520){9}{\rule{1.500pt}{0.125pt}}
\multiput(621.00,428.34)(9.887,-8.000){2}{\rule{0.750pt}{0.800pt}}
\multiput(634.00,420.08)(0.913,-0.526){7}{\rule{1.571pt}{0.127pt}}
\multiput(634.00,420.34)(8.738,-7.000){2}{\rule{0.786pt}{0.800pt}}
\multiput(646.00,413.08)(1.000,-0.526){7}{\rule{1.686pt}{0.127pt}}
\multiput(646.00,413.34)(9.501,-7.000){2}{\rule{0.843pt}{0.800pt}}
\multiput(659.00,406.07)(1.244,-0.536){5}{\rule{1.933pt}{0.129pt}}
\multiput(659.00,406.34)(8.987,-6.000){2}{\rule{0.967pt}{0.800pt}}
\multiput(672.00,400.07)(1.132,-0.536){5}{\rule{1.800pt}{0.129pt}}
\multiput(672.00,400.34)(8.264,-6.000){2}{\rule{0.900pt}{0.800pt}}
\multiput(684.00,394.06)(1.768,-0.560){3}{\rule{2.280pt}{0.135pt}}
\multiput(684.00,394.34)(8.268,-5.000){2}{\rule{1.140pt}{0.800pt}}
\put(697,387.34){\rule{2.600pt}{0.800pt}}
\multiput(697.00,389.34)(6.604,-4.000){2}{\rule{1.300pt}{0.800pt}}
\put(709,383.84){\rule{3.132pt}{0.800pt}}
\multiput(709.00,385.34)(6.500,-3.000){2}{\rule{1.566pt}{0.800pt}}
\put(722,381.34){\rule{3.132pt}{0.800pt}}
\multiput(722.00,382.34)(6.500,-2.000){2}{\rule{1.566pt}{0.800pt}}
\put(735,379.34){\rule{2.891pt}{0.800pt}}
\multiput(735.00,380.34)(6.000,-2.000){2}{\rule{1.445pt}{0.800pt}}
\put(785,379.34){\rule{2.891pt}{0.800pt}}
\multiput(785.00,378.34)(6.000,2.000){2}{\rule{1.445pt}{0.800pt}}
\put(797,381.84){\rule{3.132pt}{0.800pt}}
\multiput(797.00,380.34)(6.500,3.000){2}{\rule{1.566pt}{0.800pt}}
\put(810,384.84){\rule{3.132pt}{0.800pt}}
\multiput(810.00,383.34)(6.500,3.000){2}{\rule{1.566pt}{0.800pt}}
\put(823,388.34){\rule{2.600pt}{0.800pt}}
\multiput(823.00,386.34)(6.604,4.000){2}{\rule{1.300pt}{0.800pt}}
\multiput(835.00,393.38)(1.768,0.560){3}{\rule{2.280pt}{0.135pt}}
\multiput(835.00,390.34)(8.268,5.000){2}{\rule{1.140pt}{0.800pt}}
\multiput(848.00,398.38)(1.600,0.560){3}{\rule{2.120pt}{0.135pt}}
\multiput(848.00,395.34)(7.600,5.000){2}{\rule{1.060pt}{0.800pt}}
\multiput(860.00,403.39)(1.244,0.536){5}{\rule{1.933pt}{0.129pt}}
\multiput(860.00,400.34)(8.987,6.000){2}{\rule{0.967pt}{0.800pt}}
\multiput(873.00,409.40)(0.913,0.526){7}{\rule{1.571pt}{0.127pt}}
\multiput(873.00,406.34)(8.738,7.000){2}{\rule{0.786pt}{0.800pt}}
\multiput(885.00,416.40)(0.847,0.520){9}{\rule{1.500pt}{0.125pt}}
\multiput(885.00,413.34)(9.887,8.000){2}{\rule{0.750pt}{0.800pt}}
\multiput(898.00,424.40)(0.847,0.520){9}{\rule{1.500pt}{0.125pt}}
\multiput(898.00,421.34)(9.887,8.000){2}{\rule{0.750pt}{0.800pt}}
\multiput(911.00,432.40)(0.674,0.516){11}{\rule{1.267pt}{0.124pt}}
\multiput(911.00,429.34)(9.371,9.000){2}{\rule{0.633pt}{0.800pt}}
\multiput(923.00,441.40)(0.737,0.516){11}{\rule{1.356pt}{0.124pt}}
\multiput(923.00,438.34)(10.186,9.000){2}{\rule{0.678pt}{0.800pt}}
\multiput(936.00,450.40)(0.599,0.514){13}{\rule{1.160pt}{0.124pt}}
\multiput(936.00,447.34)(9.592,10.000){2}{\rule{0.580pt}{0.800pt}}
\multiput(948.00,460.40)(0.654,0.514){13}{\rule{1.240pt}{0.124pt}}
\multiput(948.00,457.34)(10.426,10.000){2}{\rule{0.620pt}{0.800pt}}
\multiput(961.00,470.40)(0.654,0.514){13}{\rule{1.240pt}{0.124pt}}
\multiput(961.00,467.34)(10.426,10.000){2}{\rule{0.620pt}{0.800pt}}
\multiput(974.00,480.40)(0.539,0.512){15}{\rule{1.073pt}{0.123pt}}
\multiput(974.00,477.34)(9.774,11.000){2}{\rule{0.536pt}{0.800pt}}
\multiput(986.00,491.41)(0.536,0.511){17}{\rule{1.067pt}{0.123pt}}
\multiput(986.00,488.34)(10.786,12.000){2}{\rule{0.533pt}{0.800pt}}
\multiput(999.00,503.40)(0.539,0.512){15}{\rule{1.073pt}{0.123pt}}
\multiput(999.00,500.34)(9.774,11.000){2}{\rule{0.536pt}{0.800pt}}
\multiput(1011.00,514.41)(0.536,0.511){17}{\rule{1.067pt}{0.123pt}}
\multiput(1011.00,511.34)(10.786,12.000){2}{\rule{0.533pt}{0.800pt}}
\multiput(1024.00,526.41)(0.491,0.511){17}{\rule{1.000pt}{0.123pt}}
\multiput(1024.00,523.34)(9.924,12.000){2}{\rule{0.500pt}{0.800pt}}
\multiput(1036.00,538.41)(0.536,0.511){17}{\rule{1.067pt}{0.123pt}}
\multiput(1036.00,535.34)(10.786,12.000){2}{\rule{0.533pt}{0.800pt}}
\multiput(1049.00,550.41)(0.536,0.511){17}{\rule{1.067pt}{0.123pt}}
\multiput(1049.00,547.34)(10.786,12.000){2}{\rule{0.533pt}{0.800pt}}
\multiput(1062.00,562.41)(0.491,0.511){17}{\rule{1.000pt}{0.123pt}}
\multiput(1062.00,559.34)(9.924,12.000){2}{\rule{0.500pt}{0.800pt}}
\multiput(1074.00,574.41)(0.536,0.511){17}{\rule{1.067pt}{0.123pt}}
\multiput(1074.00,571.34)(10.786,12.000){2}{\rule{0.533pt}{0.800pt}}
\multiput(1087.00,586.41)(0.491,0.511){17}{\rule{1.000pt}{0.123pt}}
\multiput(1087.00,583.34)(9.924,12.000){2}{\rule{0.500pt}{0.800pt}}
\multiput(1099.00,598.41)(0.536,0.511){17}{\rule{1.067pt}{0.123pt}}
\multiput(1099.00,595.34)(10.786,12.000){2}{\rule{0.533pt}{0.800pt}}
\multiput(1112.00,610.41)(0.536,0.511){17}{\rule{1.067pt}{0.123pt}}
\multiput(1112.00,607.34)(10.786,12.000){2}{\rule{0.533pt}{0.800pt}}
\multiput(1125.00,622.40)(0.539,0.512){15}{\rule{1.073pt}{0.123pt}}
\multiput(1125.00,619.34)(9.774,11.000){2}{\rule{0.536pt}{0.800pt}}
\multiput(1137.00,633.41)(0.536,0.511){17}{\rule{1.067pt}{0.123pt}}
\multiput(1137.00,630.34)(10.786,12.000){2}{\rule{0.533pt}{0.800pt}}
\multiput(1150.00,645.40)(0.539,0.512){15}{\rule{1.073pt}{0.123pt}}
\multiput(1150.00,642.34)(9.774,11.000){2}{\rule{0.536pt}{0.800pt}}
\multiput(1162.00,656.40)(0.589,0.512){15}{\rule{1.145pt}{0.123pt}}
\multiput(1162.00,653.34)(10.623,11.000){2}{\rule{0.573pt}{0.800pt}}
\multiput(1175.00,667.40)(0.599,0.514){13}{\rule{1.160pt}{0.124pt}}
\multiput(1175.00,664.34)(9.592,10.000){2}{\rule{0.580pt}{0.800pt}}
\multiput(1187.00,677.40)(0.654,0.514){13}{\rule{1.240pt}{0.124pt}}
\multiput(1187.00,674.34)(10.426,10.000){2}{\rule{0.620pt}{0.800pt}}
\multiput(1200.00,687.40)(0.654,0.514){13}{\rule{1.240pt}{0.124pt}}
\multiput(1200.00,684.34)(10.426,10.000){2}{\rule{0.620pt}{0.800pt}}
\multiput(1213.00,697.40)(0.674,0.516){11}{\rule{1.267pt}{0.124pt}}
\multiput(1213.00,694.34)(9.371,9.000){2}{\rule{0.633pt}{0.800pt}}
\multiput(1225.00,706.40)(0.847,0.520){9}{\rule{1.500pt}{0.125pt}}
\multiput(1225.00,703.34)(9.887,8.000){2}{\rule{0.750pt}{0.800pt}}
\multiput(1238.00,714.40)(0.774,0.520){9}{\rule{1.400pt}{0.125pt}}
\multiput(1238.00,711.34)(9.094,8.000){2}{\rule{0.700pt}{0.800pt}}
\multiput(1250.00,722.40)(0.847,0.520){9}{\rule{1.500pt}{0.125pt}}
\multiput(1250.00,719.34)(9.887,8.000){2}{\rule{0.750pt}{0.800pt}}
\multiput(1263.00,730.40)(0.913,0.526){7}{\rule{1.571pt}{0.127pt}}
\multiput(1263.00,727.34)(8.738,7.000){2}{\rule{0.786pt}{0.800pt}}
\multiput(1275.00,737.39)(1.244,0.536){5}{\rule{1.933pt}{0.129pt}}
\multiput(1275.00,734.34)(8.987,6.000){2}{\rule{0.967pt}{0.800pt}}
\multiput(1288.00,743.39)(1.244,0.536){5}{\rule{1.933pt}{0.129pt}}
\multiput(1288.00,740.34)(8.987,6.000){2}{\rule{0.967pt}{0.800pt}}
\multiput(1301.00,749.38)(1.600,0.560){3}{\rule{2.120pt}{0.135pt}}
\multiput(1301.00,746.34)(7.600,5.000){2}{\rule{1.060pt}{0.800pt}}
\put(1313,753.34){\rule{2.800pt}{0.800pt}}
\multiput(1313.00,751.34)(7.188,4.000){2}{\rule{1.400pt}{0.800pt}}
\put(1326,757.34){\rule{2.600pt}{0.800pt}}
\multiput(1326.00,755.34)(6.604,4.000){2}{\rule{1.300pt}{0.800pt}}
\put(1338,760.84){\rule{3.132pt}{0.800pt}}
\multiput(1338.00,759.34)(6.500,3.000){2}{\rule{1.566pt}{0.800pt}}
\put(1351,763.34){\rule{3.132pt}{0.800pt}}
\multiput(1351.00,762.34)(6.500,2.000){2}{\rule{1.566pt}{0.800pt}}
\put(1364,765.34){\rule{2.891pt}{0.800pt}}
\multiput(1364.00,764.34)(6.000,2.000){2}{\rule{1.445pt}{0.800pt}}
\put(747.0,380.0){\rule[-0.400pt]{9.154pt}{0.800pt}}
\put(1414,765.34){\rule{2.891pt}{0.800pt}}
\multiput(1414.00,766.34)(6.000,-2.000){2}{\rule{1.445pt}{0.800pt}}
\put(1426,763.34){\rule{3.132pt}{0.800pt}}
\multiput(1426.00,764.34)(6.500,-2.000){2}{\rule{1.566pt}{0.800pt}}
\put(1376.0,768.0){\rule[-0.400pt]{9.154pt}{0.800pt}}
\put(278,856.34){\rule{0.964pt}{0.800pt}}
\multiput(278.00,858.34)(2.000,-4.000){2}{\rule{0.482pt}{0.800pt}}
\multiput(283.41,850.74)(0.511,-0.671){17}{\rule{0.123pt}{1.267pt}}
\multiput(280.34,853.37)(12.000,-13.371){2}{\rule{0.800pt}{0.633pt}}
\multiput(295.41,835.08)(0.509,-0.616){19}{\rule{0.123pt}{1.185pt}}
\multiput(292.34,837.54)(13.000,-13.541){2}{\rule{0.800pt}{0.592pt}}
\multiput(308.41,818.47)(0.511,-0.717){17}{\rule{0.123pt}{1.333pt}}
\multiput(305.34,821.23)(12.000,-14.233){2}{\rule{0.800pt}{0.667pt}}
\multiput(320.41,801.57)(0.509,-0.698){19}{\rule{0.123pt}{1.308pt}}
\multiput(317.34,804.29)(13.000,-15.286){2}{\rule{0.800pt}{0.654pt}}
\multiput(333.41,783.32)(0.509,-0.740){19}{\rule{0.123pt}{1.369pt}}
\multiput(330.34,786.16)(13.000,-16.158){2}{\rule{0.800pt}{0.685pt}}
\multiput(346.41,763.64)(0.511,-0.852){17}{\rule{0.123pt}{1.533pt}}
\multiput(343.34,766.82)(12.000,-16.817){2}{\rule{0.800pt}{0.767pt}}
\multiput(358.41,744.06)(0.509,-0.781){19}{\rule{0.123pt}{1.431pt}}
\multiput(355.34,747.03)(13.000,-17.030){2}{\rule{0.800pt}{0.715pt}}
\multiput(371.41,723.08)(0.511,-0.943){17}{\rule{0.123pt}{1.667pt}}
\multiput(368.34,726.54)(12.000,-18.541){2}{\rule{0.800pt}{0.833pt}}
\multiput(383.41,701.55)(0.509,-0.864){19}{\rule{0.123pt}{1.554pt}}
\multiput(380.34,704.77)(13.000,-18.775){2}{\rule{0.800pt}{0.777pt}}
\multiput(396.41,678.80)(0.511,-0.988){17}{\rule{0.123pt}{1.733pt}}
\multiput(393.34,682.40)(12.000,-19.402){2}{\rule{0.800pt}{0.867pt}}
\multiput(408.41,656.29)(0.509,-0.905){19}{\rule{0.123pt}{1.615pt}}
\multiput(405.34,659.65)(13.000,-19.647){2}{\rule{0.800pt}{0.808pt}}
\multiput(421.41,633.04)(0.509,-0.947){19}{\rule{0.123pt}{1.677pt}}
\multiput(418.34,636.52)(13.000,-20.519){2}{\rule{0.800pt}{0.838pt}}
\multiput(434.41,608.53)(0.511,-1.033){17}{\rule{0.123pt}{1.800pt}}
\multiput(431.34,612.26)(12.000,-20.264){2}{\rule{0.800pt}{0.900pt}}
\multiput(446.41,585.04)(0.509,-0.947){19}{\rule{0.123pt}{1.677pt}}
\multiput(443.34,588.52)(13.000,-20.519){2}{\rule{0.800pt}{0.838pt}}
\multiput(459.41,560.25)(0.511,-1.078){17}{\rule{0.123pt}{1.867pt}}
\multiput(456.34,564.13)(12.000,-21.126){2}{\rule{0.800pt}{0.933pt}}
\multiput(471.41,536.04)(0.509,-0.947){19}{\rule{0.123pt}{1.677pt}}
\multiput(468.34,539.52)(13.000,-20.519){2}{\rule{0.800pt}{0.838pt}}
\multiput(484.41,511.78)(0.509,-0.988){19}{\rule{0.123pt}{1.738pt}}
\multiput(481.34,515.39)(13.000,-21.392){2}{\rule{0.800pt}{0.869pt}}
\multiput(497.41,486.25)(0.511,-1.078){17}{\rule{0.123pt}{1.867pt}}
\multiput(494.34,490.13)(12.000,-21.126){2}{\rule{0.800pt}{0.933pt}}
\multiput(509.41,462.04)(0.509,-0.947){19}{\rule{0.123pt}{1.677pt}}
\multiput(506.34,465.52)(13.000,-20.519){2}{\rule{0.800pt}{0.838pt}}
\multiput(522.41,437.53)(0.511,-1.033){17}{\rule{0.123pt}{1.800pt}}
\multiput(519.34,441.26)(12.000,-20.264){2}{\rule{0.800pt}{0.900pt}}
\multiput(534.41,414.29)(0.509,-0.905){19}{\rule{0.123pt}{1.615pt}}
\multiput(531.34,417.65)(13.000,-19.647){2}{\rule{0.800pt}{0.808pt}}
\multiput(547.41,390.80)(0.511,-0.988){17}{\rule{0.123pt}{1.733pt}}
\multiput(544.34,394.40)(12.000,-19.402){2}{\rule{0.800pt}{0.867pt}}
\multiput(559.41,368.29)(0.509,-0.905){19}{\rule{0.123pt}{1.615pt}}
\multiput(556.34,371.65)(13.000,-19.647){2}{\rule{0.800pt}{0.808pt}}
\multiput(572.41,345.81)(0.509,-0.823){19}{\rule{0.123pt}{1.492pt}}
\multiput(569.34,348.90)(13.000,-17.903){2}{\rule{0.800pt}{0.746pt}}
\multiput(585.41,324.36)(0.511,-0.897){17}{\rule{0.123pt}{1.600pt}}
\multiput(582.34,327.68)(12.000,-17.679){2}{\rule{0.800pt}{0.800pt}}
\multiput(597.41,304.06)(0.509,-0.781){19}{\rule{0.123pt}{1.431pt}}
\multiput(594.34,307.03)(13.000,-17.030){2}{\rule{0.800pt}{0.715pt}}
\multiput(610.41,284.19)(0.511,-0.762){17}{\rule{0.123pt}{1.400pt}}
\multiput(607.34,287.09)(12.000,-15.094){2}{\rule{0.800pt}{0.700pt}}
\multiput(622.41,266.57)(0.509,-0.698){19}{\rule{0.123pt}{1.308pt}}
\multiput(619.34,269.29)(13.000,-15.286){2}{\rule{0.800pt}{0.654pt}}
\multiput(635.41,248.74)(0.511,-0.671){17}{\rule{0.123pt}{1.267pt}}
\multiput(632.34,251.37)(12.000,-13.371){2}{\rule{0.800pt}{0.633pt}}
\multiput(647.41,233.59)(0.509,-0.533){19}{\rule{0.123pt}{1.062pt}}
\multiput(644.34,235.80)(13.000,-11.797){2}{\rule{0.800pt}{0.531pt}}
\multiput(660.41,219.59)(0.509,-0.533){19}{\rule{0.123pt}{1.062pt}}
\multiput(657.34,221.80)(13.000,-11.797){2}{\rule{0.800pt}{0.531pt}}
\multiput(672.00,208.08)(0.539,-0.512){15}{\rule{1.073pt}{0.123pt}}
\multiput(672.00,208.34)(9.774,-11.000){2}{\rule{0.536pt}{0.800pt}}
\multiput(684.00,197.08)(0.654,-0.514){13}{\rule{1.240pt}{0.124pt}}
\multiput(684.00,197.34)(10.426,-10.000){2}{\rule{0.620pt}{0.800pt}}
\multiput(697.00,187.08)(0.674,-0.516){11}{\rule{1.267pt}{0.124pt}}
\multiput(697.00,187.34)(9.371,-9.000){2}{\rule{0.633pt}{0.800pt}}
\multiput(709.00,178.07)(1.244,-0.536){5}{\rule{1.933pt}{0.129pt}}
\multiput(709.00,178.34)(8.987,-6.000){2}{\rule{0.967pt}{0.800pt}}
\multiput(722.00,172.06)(1.768,-0.560){3}{\rule{2.280pt}{0.135pt}}
\multiput(722.00,172.34)(8.268,-5.000){2}{\rule{1.140pt}{0.800pt}}
\put(735,165.84){\rule{2.891pt}{0.800pt}}
\multiput(735.00,167.34)(6.000,-3.000){2}{\rule{1.445pt}{0.800pt}}
\put(747,163.34){\rule{3.132pt}{0.800pt}}
\multiput(747.00,164.34)(6.500,-2.000){2}{\rule{1.566pt}{0.800pt}}
\put(772,163.84){\rule{3.132pt}{0.800pt}}
\multiput(772.00,162.34)(6.500,3.000){2}{\rule{1.566pt}{0.800pt}}
\put(785,166.84){\rule{2.891pt}{0.800pt}}
\multiput(785.00,165.34)(6.000,3.000){2}{\rule{1.445pt}{0.800pt}}
\multiput(797.00,171.39)(1.244,0.536){5}{\rule{1.933pt}{0.129pt}}
\multiput(797.00,168.34)(8.987,6.000){2}{\rule{0.967pt}{0.800pt}}
\multiput(810.00,177.40)(1.000,0.526){7}{\rule{1.686pt}{0.127pt}}
\multiput(810.00,174.34)(9.501,7.000){2}{\rule{0.843pt}{0.800pt}}
\multiput(823.00,184.40)(0.674,0.516){11}{\rule{1.267pt}{0.124pt}}
\multiput(823.00,181.34)(9.371,9.000){2}{\rule{0.633pt}{0.800pt}}
\multiput(835.00,193.40)(0.589,0.512){15}{\rule{1.145pt}{0.123pt}}
\multiput(835.00,190.34)(10.623,11.000){2}{\rule{0.573pt}{0.800pt}}
\multiput(848.00,204.41)(0.491,0.511){17}{\rule{1.000pt}{0.123pt}}
\multiput(848.00,201.34)(9.924,12.000){2}{\rule{0.500pt}{0.800pt}}
\multiput(860.00,216.41)(0.492,0.509){19}{\rule{1.000pt}{0.123pt}}
\multiput(860.00,213.34)(10.924,13.000){2}{\rule{0.500pt}{0.800pt}}
\multiput(874.41,228.00)(0.511,0.626){17}{\rule{0.123pt}{1.200pt}}
\multiput(871.34,228.00)(12.000,12.509){2}{\rule{0.800pt}{0.600pt}}
\multiput(886.41,243.00)(0.509,0.616){19}{\rule{0.123pt}{1.185pt}}
\multiput(883.34,243.00)(13.000,13.541){2}{\rule{0.800pt}{0.592pt}}
\multiput(899.41,259.00)(0.509,0.657){19}{\rule{0.123pt}{1.246pt}}
\multiput(896.34,259.00)(13.000,14.414){2}{\rule{0.800pt}{0.623pt}}
\multiput(912.41,276.00)(0.511,0.807){17}{\rule{0.123pt}{1.467pt}}
\multiput(909.34,276.00)(12.000,15.956){2}{\rule{0.800pt}{0.733pt}}
\multiput(924.41,295.00)(0.509,0.740){19}{\rule{0.123pt}{1.369pt}}
\multiput(921.34,295.00)(13.000,16.158){2}{\rule{0.800pt}{0.685pt}}
\multiput(937.41,314.00)(0.511,0.852){17}{\rule{0.123pt}{1.533pt}}
\multiput(934.34,314.00)(12.000,16.817){2}{\rule{0.800pt}{0.767pt}}
\multiput(949.41,334.00)(0.509,0.823){19}{\rule{0.123pt}{1.492pt}}
\multiput(946.34,334.00)(13.000,17.903){2}{\rule{0.800pt}{0.746pt}}
\multiput(962.41,355.00)(0.509,0.864){19}{\rule{0.123pt}{1.554pt}}
\multiput(959.34,355.00)(13.000,18.775){2}{\rule{0.800pt}{0.777pt}}
\multiput(975.41,377.00)(0.511,0.988){17}{\rule{0.123pt}{1.733pt}}
\multiput(972.34,377.00)(12.000,19.402){2}{\rule{0.800pt}{0.867pt}}
\multiput(987.41,400.00)(0.509,0.905){19}{\rule{0.123pt}{1.615pt}}
\multiput(984.34,400.00)(13.000,19.647){2}{\rule{0.800pt}{0.808pt}}
\multiput(1000.41,423.00)(0.511,0.988){17}{\rule{0.123pt}{1.733pt}}
\multiput(997.34,423.00)(12.000,19.402){2}{\rule{0.800pt}{0.867pt}}
\multiput(1012.41,446.00)(0.509,0.947){19}{\rule{0.123pt}{1.677pt}}
\multiput(1009.34,446.00)(13.000,20.519){2}{\rule{0.800pt}{0.838pt}}
\multiput(1025.41,470.00)(0.511,1.033){17}{\rule{0.123pt}{1.800pt}}
\multiput(1022.34,470.00)(12.000,20.264){2}{\rule{0.800pt}{0.900pt}}
\multiput(1037.41,494.00)(0.509,0.947){19}{\rule{0.123pt}{1.677pt}}
\multiput(1034.34,494.00)(13.000,20.519){2}{\rule{0.800pt}{0.838pt}}
\multiput(1050.41,518.00)(0.509,0.988){19}{\rule{0.123pt}{1.738pt}}
\multiput(1047.34,518.00)(13.000,21.392){2}{\rule{0.800pt}{0.869pt}}
\multiput(1063.41,543.00)(0.511,1.033){17}{\rule{0.123pt}{1.800pt}}
\multiput(1060.34,543.00)(12.000,20.264){2}{\rule{0.800pt}{0.900pt}}
\multiput(1075.41,567.00)(0.509,0.947){19}{\rule{0.123pt}{1.677pt}}
\multiput(1072.34,567.00)(13.000,20.519){2}{\rule{0.800pt}{0.838pt}}
\multiput(1088.41,591.00)(0.511,1.033){17}{\rule{0.123pt}{1.800pt}}
\multiput(1085.34,591.00)(12.000,20.264){2}{\rule{0.800pt}{0.900pt}}
\multiput(1100.41,615.00)(0.509,0.947){19}{\rule{0.123pt}{1.677pt}}
\multiput(1097.34,615.00)(13.000,20.519){2}{\rule{0.800pt}{0.838pt}}
\multiput(1113.41,639.00)(0.509,0.947){19}{\rule{0.123pt}{1.677pt}}
\multiput(1110.34,639.00)(13.000,20.519){2}{\rule{0.800pt}{0.838pt}}
\multiput(1126.41,663.00)(0.511,0.988){17}{\rule{0.123pt}{1.733pt}}
\multiput(1123.34,663.00)(12.000,19.402){2}{\rule{0.800pt}{0.867pt}}
\multiput(1138.41,686.00)(0.509,0.905){19}{\rule{0.123pt}{1.615pt}}
\multiput(1135.34,686.00)(13.000,19.647){2}{\rule{0.800pt}{0.808pt}}
\multiput(1151.41,709.00)(0.511,0.943){17}{\rule{0.123pt}{1.667pt}}
\multiput(1148.34,709.00)(12.000,18.541){2}{\rule{0.800pt}{0.833pt}}
\multiput(1163.41,731.00)(0.509,0.823){19}{\rule{0.123pt}{1.492pt}}
\multiput(1160.34,731.00)(13.000,17.903){2}{\rule{0.800pt}{0.746pt}}
\multiput(1176.41,752.00)(0.511,0.897){17}{\rule{0.123pt}{1.600pt}}
\multiput(1173.34,752.00)(12.000,17.679){2}{\rule{0.800pt}{0.800pt}}
\multiput(1188.41,773.00)(0.509,0.781){19}{\rule{0.123pt}{1.431pt}}
\multiput(1185.34,773.00)(13.000,17.030){2}{\rule{0.800pt}{0.715pt}}
\put(760.0,164.0){\rule[-0.400pt]{2.891pt}{0.800pt}}
\sbox{\plotpoint}{\rule[-0.200pt]{0.400pt}{0.400pt}}%
\put(376,123){\usebox{\plotpoint}}
\multiput(376,123)(0.000,20.756){36}{\usebox{\plotpoint}}
\put(376,860){\usebox{\plotpoint}}
\put(1030,123){\usebox{\plotpoint}}
\multiput(1030,123)(0.000,20.756){36}{\usebox{\plotpoint}}
\put(1030,860){\usebox{\plotpoint}}
\end{picture}